\documentstyle[12pt,amssymbo,fleqn]{article}
\newtheorem{Definition}{Definition}[section]
\newtheorem{Example}[Definition]{Example}
\newtheorem{Theorem}[Definition]{Theorem}
\newtheorem{Proposition}[Definition]{Proposition}
\newtheorem{Remark}[Definition]{Remark}
\newtheorem{Lemma}[Definition]{Lemma}
\newtheorem{Corollary}[Definition]{Corollary}
\newcommand{\QPB}{quantum principal bundle }
\newcommand{\QPBS}{quantum principal bundles }
\newcommand{\QVB}{quantum vector bundle }

\newcommand{\QS}{quantum space }
\newcommand{\QSS}{quantum spaces }

\newcommand{\CS}{coordinate system }

\newcommand{\cat}[1]{\mbox{\bf{#1}}}
\newcommand{\bpro}{{\bf Proof}: }
\newcommand{\epro}{\hfill $ \Box $ \parskip0.5cm \par \noindent \parskip1.0ex }
\newcommand{\bdm}{\begin{eqnarray*}}
\newcommand{\edm}{\end{eqnarray*}}
\newcommand{\be}{\begin{equation}}
\newcommand{\ee}{\end{equation} \noindent}
\long\def\ig#1{\relax}
\ig{Thanks to Roberto Minio for this def'n.  Compare the def'n of
\comment in AMSTeX.}

\newcount \coefa
\newcount \coefb
\newcount \coefc
\newdimen\tempdimen
\newdimen\xlen
\newdimen\ylen

\newcount\tempcounta
\newcount\tempcountb
\newcount\tempcountc
\newcount\tempcountd
\newcount\tempcounte
\newcount\tempcountf
\newcount\xext
\newcount\yext
\newcount\xoff
\newcount\yoff
\newcount\gap%
\newcount\arrowtypea
\newcount\arrowtypeb
\newcount\arrowtypec
\newcount\arrowtyped
\newcount\arrowtypee
\newcount\height
\newcount\width
\newcount\xpos
\newcount\ypos
\newcount\run
\newcount\rise
\newcount\arrowlength
\newcount\halflength
\newcount\arrowtype
\newsavebox{\tempboxa}%
\newsavebox{\tempboxb}%
\newsavebox{\tempboxc}%

\catcode`@=11 
\def\settoheight#1#2{\setbox\@tempboxa\hbox{#2}#1\ht\@tempboxa\relax}%
\def\settodepth#1#2{\setbox\@tempboxa\hbox{#2}#1\dp\@tempboxa\relax}%
\let\ifnextchar=\@ifnextchar
\catcode`@=12 

\def\putbox(#1,#2)#3{\put(#1,#2){\makebox(0,0){#3}}}

\def\setsqparms[#1`#2`#3`#4;#5`#6]{%
\settripairparms[#1`#2`#3`#4`1;#6]%
\width #5
}

\def\settriparms[#1`#2`#3;#4]{\settripairparms[#1`#2`#3`1`1;#4]}%

\def\settripairparms[#1`#2`#3`#4`#5;#6]{%
\arrowtypea #1
\arrowtypeb #2
\arrowtypec #3
\arrowtyped #4
\arrowtypee #5
\height #6
\width #6
}

\def\mvector(#1,#2)#3{
\put(0,0){\vector(#1,#2){#3}}%
\put(0,0){\vector(#1,#2){30}}%
}
\def\evector(#1,#2)#3{{
\arrowlength #3
\put(0,0){\vector(#1,#2){\arrowlength}}%
\advance \arrowlength by-30
\put(0,0){\vector(#1,#2){\arrowlength}}%
}}

\def\horsize#1#2{%
\settowidth{\tempdimen}{$#2$}%
#1=\tempdimen
\divide #1 by\unitlength
}

\def\vertsize#1#2{%
\settoheight{\tempdimen}{$#2$}%
#1=\tempdimen
\settodepth{\tempdimen}{$#2$}%
\advance #1 by\tempdimen
\divide #1 by\unitlength
}

\def\vertadjust[#1`#2`#3]{%
\vertsize{\tempcounta}{#1}%
\vertsize{\tempcountb}{#2}%
\ifnum \tempcounta<\tempcountb \tempcounta=\tempcountb \fi
\divide\tempcounta by2
\vertsize{\tempcountb}{#3}%
\ifnum \tempcountb>0 \advance \tempcountb by20 \fi
\ifnum \tempcounta<\tempcountb \tempcounta=\tempcountb \fi
}

\def\horadjust[#1`#2`#3]{%
\horsize{\tempcounta}{#1}%
\horsize{\tempcountb}{#2}%
\ifnum \tempcounta<\tempcountb \tempcounta=\tempcountb \fi
\divide\tempcounta by20
\horsize{\tempcountb}{#3}%
\ifnum \tempcountb>0 \advance \tempcountb by60 \fi
\ifnum \tempcounta<\tempcountb \tempcounta=\tempcountb \fi
}

\ig{ In this procedure, #1 is the paramater that sticks out all the way,
#2 sticks out the least and #3 is a label sticking out half way.  #4 is
the amount of the offset.}

\def\sladjust[#1`#2`#3]#4{%
\tempcountc=#4
\horsize{\tempcounta}{#1}%
\divide \tempcounta by2
\horsize{\tempcountb}{#2}%
\divide \tempcountb by2
\advance \tempcountb by-\tempcountc
\ifnum \tempcounta<\tempcountb \tempcounta=\tempcountb\fi
\divide \tempcountc by2
\horsize{\tempcountb}{#3}%
\advance \tempcountb by-\tempcountc
\ifnum \tempcountb>0 \advance \tempcountb by80\fi
\ifnum \tempcounta<\tempcountb \tempcounta=\tempcountb\fi
\advance\tempcounta by20
}

\def\putvector(#1,#2)(#3,#4)#5#6{{%
\xpos=#1
\ypos=#2
\run=#3
\rise=#4
\arrowlength=#5
\arrowtype=#6
\ifnum \arrowtype<0
    \ifnum \run=0
        \advance \ypos by-\arrowlength
    \else
        \tempcounta \arrowlength
        \multiply \tempcounta by\rise
        \divide \tempcounta by\run
        \ifnum\run>0
            \advance \xpos by\arrowlength
            \advance \ypos by\tempcounta
        \else
            \advance \xpos by-\arrowlength
            \advance \ypos by-\tempcounta
        \fi
    \fi
    \multiply \arrowtype by-1
    \multiply \rise by-1
    \multiply \run by-1
\fi
\ifnum \arrowtype=1
    \put(\xpos,\ypos){\vector(\run,\rise){\arrowlength}}%
\else\ifnum \arrowtype=2
    \put(\xpos,\ypos){\mvector(\run,\rise)\arrowlength}%
\else\ifnum\arrowtype=3
    \put(\xpos,\ypos){\evector(\run,\rise){\arrowlength}}%
\fi\fi\fi
}}

\def\bfig{\begin{picture}(\xext,\yext)(\xoff,\yoff)}
\def\efig{\end{picture}}

\def\putsplitvector(#1,#2)#3#4{
\xpos #1
\ypos #2
\arrowtype #4
\halflength #3
\arrowlength #3
\gap 140
\advance \halflength by-\gap
\divide \halflength by2
\ifnum \arrowtype=1
    \put(\xpos,\ypos){\line(0,-1){\halflength}}%
    \advance\ypos by-\halflength
    \advance\ypos by-\gap
    \put(\xpos,\ypos){\vector(0,-1){\halflength}}%
\else\ifnum \arrowtype=2
    \put(\xpos,\ypos){\line(0,-1)\halflength}%
    \put(\xpos,\ypos){\vector(0,-1)3}%
    \advance\ypos by-\halflength
    \advance\ypos by-\gap
    \put(\xpos,\ypos){\vector(0,-1){\halflength}}%
\else\ifnum\arrowtype=3
    \put(\xpos,\ypos){\line(0,-1)\halflength}%
    \advance\ypos by-\halflength
    \advance\ypos by-\gap
    \put(\xpos,\ypos){\evector(0,-1){\halflength}}%
\else\ifnum \arrowtype=-1
    \advance \ypos by-\arrowlength
    \put(\xpos,\ypos){\line(0,1){\halflength}}%
    \advance\ypos by\halflength
    \advance\ypos by\gap
    \put(\xpos,\ypos){\vector(0,1){\halflength}}%
\else\ifnum \arrowtype=-2
    \advance \ypos by-\arrowlength
    \put(\xpos,\ypos){\line(0,1)\halflength}%
    \put(\xpos,\ypos){\vector(0,1)3}%
    \advance\ypos by\halflength
    \advance\ypos by\gap
    \put(\xpos,\ypos){\vector(0,1){\halflength}}%
\else\ifnum\arrowtype=-3
    \advance \ypos by-\arrowlength
    \put(\xpos,\ypos){\line(0,1)\halflength}%
    \advance\ypos by\halflength
    \advance\ypos by\gap
    \put(\xpos,\ypos){\evector(0,1){\halflength}}%
\fi\fi\fi\fi\fi\fi
}

\def\setpos(#1,#2){\xpos=#1 \ypos#2}

\def\putmorphism(#1)(#2,#3)[#4`#5`#6]#7#8#9{{%
\run #2
\rise #3
\ifnum\rise=0
  \puthmorphism(#1)[#4`#5`#6]{#7}{#8}{#9}%
\else\ifnum\run=0
  \putvmorphism(#1)[#4`#5`#6]{#7}{#8}{#9}%
\else
\setpos(#1)%
\arrowlength #7
\arrowtype #8
\ifnum\run=0
\else\ifnum\rise=0
\else
\ifnum\run>0
    \coefa=1
\else
   \coefa=-1
\fi
\ifnum\arrowtype>0
   \coefb=0
   \coefc=-1
\else
   \coefb=\coefa
   \coefc=1
   \arrowtype=-\arrowtype
\fi
\width=2
\multiply \width by\run
\divide \width by\rise
\ifnum \width<0  \width=-\width\fi
\advance\width by60
\if l#9 \width=-\width\fi
\putbox(\xpos,\ypos){$#4$}
{\multiply \coefa by\arrowlength
\advance\xpos by\coefa
\multiply \coefa by\rise
\divide \coefa by\run
\advance \ypos by\coefa
\putbox(\xpos,\ypos){$#5$} }%
{\multiply \coefa by\arrowlength
\divide \coefa by2
\advance \xpos by\coefa
\advance \xpos by\width
\multiply \coefa by\rise
\divide \coefa by\run
\advance \ypos by\coefa
\if l#9%
   \put(\xpos,\ypos){\makebox(0,0)[r]{$#6$}}%
\else\if r#9%
   \put(\xpos,\ypos){\makebox(0,0)[l]{$#6$}}%
\fi\fi }%
{\multiply \rise by-\coefc
\multiply \run by-\coefc
\multiply \coefb by\arrowlength
\advance \xpos by\coefb
\multiply \coefb by\rise
\divide \coefb by\run
\advance \ypos by\coefb
\multiply \coefc by70
\advance \ypos by\coefc
\multiply \coefc by\run
\divide \coefc by\rise
\advance \xpos by\coefc
\multiply \coefa by140
\multiply \coefa by\run
\divide \coefa by\rise
\advance \arrowlength by\coefa
\ifnum \arrowtype=1
   \put(\xpos,\ypos){\vector(\run,\rise){\arrowlength}}%
\else\ifnum\arrowtype=2
   \put(\xpos,\ypos){\mvector(\run,\rise){\arrowlength}}%
\else\ifnum\arrowtype=3
   \put(\xpos,\ypos){\evector(\run,\rise){\arrowlength}}%
\fi\fi\fi}%
\fi\fi
\fi\fi}}

\def\puthmorphism(#1,#2)[#3`#4`#5]#6#7#8{{%
\xpos #1
\ypos #2
\width #6
\arrowlength #6
\putbox(\xpos,\ypos){$#3$\vphantom{$#4$}}%
{\advance \xpos by\arrowlength
\putbox(\xpos,\ypos){\vphantom{$#3$}$#4$}}%
\horsize{\tempcounta}{#3}%
\horsize{\tempcountb}{#4}%
\divide \tempcounta by2
\divide \tempcountb by2
\advance \tempcounta by30
\advance \tempcountb by30
\advance \xpos by\tempcounta
\advance \arrowlength by-\tempcounta
\advance \arrowlength by-\tempcountb
\putvector(\xpos,\ypos)(1,0){\arrowlength}{#7}%
\divide \arrowlength by2
\advance \xpos by\arrowlength
\vertsize{\tempcounta}{#5}%
\divide\tempcounta by2
\advance \tempcounta by20
\if a#8 %
   \advance \ypos by\tempcounta
   \put(\xpos,\ypos){\makebox(0,0){$#5$}}%
\else
   \advance \ypos by-\tempcounta
   \put(\xpos,\ypos){\makebox(0,0){$#5$}}%
\fi
}}

\def\putvmorphism(#1,#2)[#3`#4`#5]#6#7#8{{%
\xpos #1
\ypos #2
\arrowlength #6
\arrowtype #7
\settowidth{\xlen}{$#5$}%
\putbox(\xpos,\ypos){$#3$}%
{\advance \ypos by-\arrowlength
\putbox(\xpos,\ypos){$#4$}}%
{\advance\arrowlength by-140
\advance \ypos by-70
\ifdim\xlen>0pt
   \if m#8%
      \putsplitvector(\xpos,\ypos){\arrowlength}{\arrowtype}%
   \else
      \putvector(\xpos,\ypos)(0,-1){\arrowlength}{\arrowtype}%
   \fi
\else
   \putvector(\xpos,\ypos)(0,-1){\arrowlength}{\arrowtype}%
\fi}%
\ifdim\xlen>0pt
   \divide \arrowlength by2
   \advance\ypos by-\arrowlength
   \if l#8%
      \advance \xpos by-40
      \put(\xpos,\ypos){\makebox(0,0)[r]{$#5$}}%
   \else\if r#8%
      \advance \xpos by40
      \put(\xpos,\ypos){\makebox(0,0)[l]{$#5$}}%
   \else
      \putbox(\xpos,\ypos){$#5$}%
   \fi\fi
\fi
}}

\def\topadjust[#1`#2`#3]{%
\yoff=10
\vertadjust[#1`#2`{#3}]%
\advance \yext by\tempcounta
\advance \yext by 10
}

\def\botadjust[#1`#2`#3]{%
\vertadjust[#1`#2`{#3}]%
\advance \yext by\tempcounta
\advance \yoff by-\tempcounta
}

\def\leftadjust[#1`#2`#3]{%
\xoff=0
\horadjust[#1`#2`{#3}]%
\advance \xext by\tempcounta
\advance \xoff by-\tempcounta
}

\def\rightadjust[#1`#2`#3]{%
\horadjust[#1`#2`{#3}]%
\advance \xext by\tempcounta
}

\def\rightsladjust[#1`#2`#3]{%
\sladjust[#1`#2`{#3}]{\width}%
\advance \xext by\tempcounta
}

\def\leftsladjust[#1`#2`#3]{%
\xoff=0
\sladjust[#1`#2`{#3}]{\width}%
\advance \xext by\tempcounta
\advance \xoff by-\tempcounta
}

\def\adjust[#1`#2;#3`#4;#5`#6;#7`#8]{%
\topadjust[#1``{#2}]
\leftadjust[#3``{#4}]
\rightadjust[#5``{#6}]
\botadjust[#7``{#8}]}

\def\putsquare(#1)[#2`#3`#4`#5;#6`#7`#8`#9]{%
\setpos(#1)
\puthmorphism(\xpos,\ypos)[#4`#5`{#9}]{\width}{\arrowtyped}b%
\advance\ypos by \height
\puthmorphism(\xpos,\ypos)[#2`#3`{#6}]{\width}{\arrowtypea}a%
\putvmorphism(\xpos,\ypos)[``{#7}]{\height}{\arrowtypeb}l%
\advance\xpos by \width
\putvmorphism(\xpos,\ypos)[``{#8}]{\height}{\arrowtypec}r%
}

\def\square[#1`#2`#3`#4;#5`#6`#7`#8]{{
\xext=\width                              
\yext=\height                             
\topadjust[#1`#2`{#5}]
\botadjust[#3`#4`{#8}]
\leftadjust[#1`#3`{#6}]
\rightadjust[#2`#4`{#7}]
\begin{picture}(\xext,\yext)(\xoff,\yoff)
\putsquare(0,0)[#1`#2`#3`#4;#5`#6`#7`{#8}]
\end{picture}
}}

\def\putptriangle(#1,#2)[#3`#4`#5;#6`#7`#8]{%
\xpos=#1 \ypos=#2
\advance\ypos by \height
\puthmorphism(\xpos,\ypos)[#3`#4`{#6}]{\height}{\arrowtypea}a%
\putvmorphism(\xpos,\ypos)[`#5`{#7}]{\height}{\arrowtypeb}l%
\advance\xpos by\height
\putmorphism(\xpos,\ypos)(-1,-1)[``{#8}]{\height}{\arrowtypec}r%
}

\def\ptriangle[#1`#2`#3;#4`#5`#6]{{
\width=\height                         
\xext=\width                           
\yext=\width                           
\topadjust[#1`#2`{#4}]
\botadjust[#3``]
\leftadjust[#1`#3`{#5}]
\rightsladjust[#2`#3`{#6}]
\begin{picture}(\xext,\yext)(\xoff,\yoff)
\putptriangle(0,0)[#1`#2`#3;#4`#5`{#6}]%
\end{picture}%
}}

\def\putqtriangle(#1,#2)[#3`#4`#5;#6`#7`#8]{%
\xpos=#1 \ypos=#2
\advance\ypos by\height
\puthmorphism(\xpos,\ypos)[#3`#4`{#6}]{\height}{\arrowtypea}a%
\putmorphism(\xpos,\ypos)(1,-1)[``{#7}]{\height}{\arrowtypeb}l%
\advance\xpos by\height
\putvmorphism(\xpos,\ypos)[`#5`{#8}]{\height}{\arrowtypec}r%
}

\def\qtriangle[#1`#2`#3;#4`#5`#6]{{
\width=\height                         
\xext=\width                           
\yext=\height                          
\topadjust[#1`#2`{#4}]
\botadjust[#3``]
\leftsladjust[#1`#3`{#5}]
\rightadjust[#2`#3`{#6}]
\begin{picture}(\xext,\yext)(\xoff,\yoff)
\putqtriangle(0,0)[#1`#2`#3;#4`#5`{#6}]%
\end{picture}%
}}

\def\putdtriangle(#1,#2)[#3`#4`#5;#6`#7`#8]{%
\xpos=#1 \ypos=#2
\puthmorphism(\xpos,\ypos)[#4`#5`{#8}]{\height}{\arrowtypec}b%
\advance\xpos by \height \advance\ypos by\height
\putmorphism(\xpos,\ypos)(-1,-1)[``{#6}]{\height}{\arrowtypea}l%
\putvmorphism(\xpos,\ypos)[#3``{#7}]{\height}{\arrowtypeb}r%
}

\def\dtriangle[#1`#2`#3;#4`#5`#6]{{
\width=\height                         
\xext=\width                           
\yext=\height                          
\topadjust[#1``]
\botadjust[#2`#3`{#6}]
\leftsladjust[#2`#1`{#4}]
\rightadjust[#1`#3`{#5}]
\begin{picture}(\xext,\yext)(\xoff,\yoff)
\putdtriangle(0,0)[#1`#2`#3;#4`#5`{#6}]%
\end{picture}%
}}

\def\putbtriangle(#1,#2)[#3`#4`#5;#6`#7`#8]{%
\xpos=#1 \ypos=#2
\puthmorphism(\xpos,\ypos)[#4`#5`{#8}]{\height}{\arrowtypec}b%
\advance\ypos by\height
\putmorphism(\xpos,\ypos)(1,-1)[``{#7}]{\height}{\arrowtypeb}r%
\putvmorphism(\xpos,\ypos)[#3``{#6}]{\height}{\arrowtypea}l%
}

\def\btriangle[#1`#2`#3;#4`#5`#6]{{
\width=\height                         
\xext=\width                           
\yext=\height                          
\topadjust[#1``]
\botadjust[#2`#3`{#6}]
\leftadjust[#1`#2`{#4}]
\rightsladjust[#3`#1`{#5}]
\begin{picture}(\xext,\yext)(\xoff,\yoff)
\putbtriangle(0,0)[#1`#2`#3;#4`#5`{#6}]%
\end{picture}%
}}

\def\putAtriangle(#1,#2)[#3`#4`#5;#6`#7`#8]{%
\xpos=#1 \ypos=#2
{\multiply \height by2
\puthmorphism(\xpos,\ypos)[#4`#5`{#8}]{\height}{\arrowtypec}b}%
\advance\xpos by\height \advance\ypos by\height
\putmorphism(\xpos,\ypos)(-1,-1)[#3``{#6}]{\height}{\arrowtypea}l%
\putmorphism(\xpos,\ypos)(1,-1)[``{#7}]{\height}{\arrowtypeb}r%
}

\def\Atriangle[#1`#2`#3;#4`#5`#6]{{
\width=\height                         
\xext=\width                           
\yext=\height                          
\topadjust[#1``]
\botadjust[#2`#3`{#6}]
\multiply \xext by2 
\leftsladjust[#2`#1`{#4}]
\rightsladjust[#3`#1`{#5}]
\begin{picture}(\xext,\yext)(\xoff,\yoff)%
\putAtriangle(0,0)[#1`#2`#3;#4`#5`{#6}]%
\end{picture}%
}}

\def\putAtrianglepair(#1,#2)[#3]{\xpos=#1 \ypos=#2%
\putAtrianglepairx[#3]}
\def\putAtrianglepairx[#1`#2`#3`#4;#5`#6`#7`#8`#9]{%
\puthmorphism(\xpos,\ypos)[#2`#3`{#8}]{\height}{\arrowtyped}b%
\advance\xpos by\height
\puthmorphism(\xpos,\ypos)[\phantom{#3}`#4`{#9}]{\height}{\arrowtypee}b%
\advance\ypos by\height
\putmorphism(\xpos,\ypos)(-1,-1)[#1``{#5}]{\height}{\arrowtypea}l%
\putvmorphism(\xpos,\ypos)[``{#6}]{\height}{\arrowtypeb}m%
\putmorphism(\xpos,\ypos)(1,-1)[``{#7}]{\height}{\arrowtypec}r%
}

\def\Atrianglepair[#1`#2`#3`#4;#5`#6`#7`#8`#9]{{%
\width=\height
\xext=\width
\yext=\height
\topadjust[#1``]%
\vertadjust[#2`#3`{#8}]
\tempcountd=\tempcounta                       
\vertadjust[#3`#4`{#9}]
\ifnum\tempcounta<\tempcountd                 
\tempcounta=\tempcountd\fi                    
\advance \yext by\tempcounta                  
\advance \yoff by-\tempcounta                 
\multiply \xext by2 
\leftsladjust[#2`#1`{#5}]
\rightsladjust[#4`#1`{#7}]%
\begin{picture}(\xext,\yext)(\xoff,\yoff)%
\putAtrianglepair(0,0)[#1`#2`#3`#4;#5`#6`#7`#8`{#9}]%
\end{picture}%
}}

\def\putVtriangle(#1,#2)[#3`#4`#5;#6`#7`#8]{%
\xpos=#1 \ypos=#2
\advance\ypos by\height
{\multiply\height by2
\puthmorphism(\xpos,\ypos)[#3`#4`{#6}]{\height}{\arrowtypea}a}%
\putmorphism(\xpos,\ypos)(1,-1)[`#5`{#7}]{\height}{\arrowtypeb}l%
\advance\xpos by\height
\advance\xpos by\height
\putmorphism(\xpos,\ypos)(-1,-1)[``{#8}]{\height}{\arrowtypec}r%
}

\def\Vtriangle[#1`#2`#3;#4`#5`#6]{{
\width=\height                         
\xext=\width                           
\yext=\height                          
\topadjust[#1`#2`{#4}]
\botadjust[#3``]
\multiply \xext by2 
\leftsladjust[#1`#3`{#5}]
\rightsladjust[#2`#3`{#6}]
\begin{picture}(\xext,\yext)(\xoff,\yoff)%
\putVtriangle(0,0)[#1`#2`#3;#4`#5`{#6}]%
\end{picture}%
}}

\def\putVtrianglepair(#1,#2)[#3]{\xpos=#1 \ypos=#2%
\putVtrianglepairx[#3]}
\def\putVtrianglepairx[#1`#2`#3`#4;#5`#6`#7`#8`#9]{%
\advance\ypos by\height
\putmorphism(\xpos,\ypos)(1,-1)[`#4`{#7}]{\height}{\arrowtypec}l%
\puthmorphism(\xpos,\ypos)[#1`#2`{#5}]{\height}{\arrowtypea}a%
\advance\xpos by\height
\puthmorphism(\xpos,\ypos)[\phantom{#2}`#3`{#6}]{\height}{\arrowtypeb}a%
\putvmorphism(\xpos,\ypos)[``{#8}]{\height}{\arrowtyped}m%
\advance\xpos by\height
\putmorphism(\xpos,\ypos)(-1,-1)[``{#9}]{\height}{\arrowtypee}r%
}

\def\Vtrianglepair[#1`#2`#3`#4;#5`#6`#7`#8`#9]{{%
\xoff=0
\yoff=2 
\xext=\height                  
\width=\height                 
\yext=\height                  
\vertadjust[#1`#2`{#5}]
\tempcountd=\tempcounta        
\vertadjust[#2`#3`{#6}]
\ifnum\tempcounta<\tempcountd
\tempcounta=\tempcountd\fi
\advance \yext by\tempcounta
\botadjust[#4``]%
\multiply \xext by2
\leftsladjust[#1`#4`{#7}]%
\rightsladjust[#3`#4`{#9}]%
\begin{picture}(\xext,\yext)(\xoff,\yoff)%
\putVtrianglepair(0,0)[#1`#2`#3`#4;#5`#6`#7`#8`{#9}]%
\end{picture}%
}}

\def\putCtriangle(#1,#2)[#3`#4`#5;#6`#7`#8]{%
\xpos=#1 \ypos=#2
\advance\ypos by\height
\putmorphism(\xpos,\ypos)(1,-1)[``{#8}]{\height}{\arrowtypec}l%
\advance\xpos by\height
\advance\ypos by\height
\putmorphism(\xpos,\ypos)(-1,-1)[#3`#4`{#6}]{\height}{\arrowtypea}l%
{\multiply\height by 2
\putvmorphism(\xpos,\ypos)[`#5`{#7}]{\height}{\arrowtypeb}r}%
}

\def\Ctriangle[#1`#2`#3;#4`#5`#6]{{
\width=\height                          
\xext=\width                            
\yext=\height                           
\multiply \yext by2 
\topadjust[#1``]
\botadjust[#3``]
\sladjust[#2`#1`{#4}]{\width}
\tempcountd=\tempcounta                 
\sladjust[#2`#3`{#6}]{\width}
\ifnum \tempcounta<\tempcountd          
\tempcounta=\tempcountd\fi              
\advance \xext by\tempcounta            
\advance \xoff by-\tempcounta           
\rightadjust[#1`#3`{#5}]
\begin{picture}(\xext,\yext)(\xoff,\yoff)%
\putCtriangle(0,0)[#1`#2`#3;#4`#5`{#6}]%
\end{picture}%
}}

\def\putDtriangle(#1,#2)[#3`#4`#5;#6`#7`#8]{%
\xpos=#1 \ypos=#2
\advance\xpos by\height \advance\ypos by\height
\putmorphism(\xpos,\ypos)(-1,-1)[``{#8}]{\height}{\arrowtypec}r%
\advance\xpos by-\height \advance\ypos by\height
\putmorphism(\xpos,\ypos)(1,-1)[`#4`{#7}]{\height}{\arrowtypeb}r%
{\multiply\height by 2
\putvmorphism(\xpos,\ypos)[#3`#5`{#6}]{\height}{\arrowtypea}l}%
}

\def\Dtriangle[#1`#2`#3;#4`#5`#6]{{
\width=\height                         
\xext=\height                          
\yext=\height                          
\multiply \yext by2 
\topadjust[#1``]
\botadjust[#3``]
\leftadjust[#1`#3`{#4}]
\sladjust[#2`#1`{#4}]{\height}
\tempcountd=\tempcountd                
\sladjust[#2`#3`{#6}]{\height}
\ifnum \tempcounta<\tempcountd         
\tempcounta=\tempcountd\fi             
\advance \xext by\tempcounta           
\begin{picture}(\xext,\yext)(\xoff,\yoff)
\putDtriangle(0,0)[#1`#2`#3;#4`#5`{#6}]%
\end{picture}%
}}

\def\setrecparms[#1`#2]{\width=#1 \height=#2}%
\def\recurse[#1`#2`#3`#4;#5`#6`#7`#8`#9]{{%
\settowidth{\tempdimen}{#1}
\ifdim\tempdimen=0pt
  \savebox{\tempboxa}{\hbox{#2}}%
  \savebox{\tempboxb}{\hbox{#4}}%
  \savebox{\tempboxc}{\hbox{#7}}%
\else
  \savebox{\tempboxa}{\hbox{$\hbox{#1}\times\hbox{#2}$}}%
  \savebox{\tempboxb}{\hbox{$\hbox{#1}\times\hbox{#4}$}}%
  \savebox{\tempboxc}{\hbox{$\hbox{#1}\times\hbox{#7}$}}%
\fi
\tempcounte=\height
\divide\tempcounte by 2
\tempcountf=\tempcounte
\advance\tempcountf by \width
\xext=\tempcountf \yext=\height
\topadjust[#2`\usebox{\tempboxa}`{#5}]%
\botadjust[#4`\usebox{\tempboxb}`{#9}]%
\sladjust[#3`#2`{#6}]{\tempcounte}%
\tempcountd=\tempcounta
\sladjust[#3`#4`{#8}]{\tempcounte}%
\ifnum \tempcounta<\tempcountd
\tempcounta=\tempcountd\fi
\advance \xext by\tempcounta
\advance \xoff by-\tempcounta
\rightadjust[\usebox{\tempboxa}`\usebox{\tempboxb}`\usebox{\tempboxc}]%
\bfig
{\settriparms[-1`1`1;\tempcounte]%
\putCtriangle(0,0)[`#3`;#6`#7`{#8}]}%
\arrowtypea=-1 \arrowtypeb=0 \arrowtypec=1 \arrowtyped=-1
\putsquare(\tempcounte,0)[#2`\usebox{\tempboxa}`#4`\usebox{\tempboxb};%
#5``\usebox{\tempboxc}`#9]%
\efig
}}

\begin{document}
\begin{titlepage}
\title{Quantum Groups on Fibre Bundles}
\author {Markus J.~Pflaum}
\date{Mathematisches Institut der Universit\"{a}t M\"{u}nchen\\
Theresienstra{\ss}e 39\\
80333 M\"{u}nchen 2, Germany\\
May 1, 1993}
\maketitle
\begin{abstract}
It is shown that the principle of locality and noncommutative geometry
can be connnected by a sheaf theoretical method. In this framework
quantum spaces are introduced and examples in mathematical physics are given.
With the language of quantum spaces noncommutative principal and
vector bundles are defined and their properties are studied.
Important constructions in the classical theory of principal fibre bundles
like associated bundles and differential calculi are carried over to the
quantum case.
At the end $q$-deformed instanton models
are introduced for every integral index.
\end{abstract}
\end{titlepage}
\tableofcontents
\section*{Introduction}
\addcontentsline{toc}{section}{Introduction}
There are two essential principles in quantum field theory,
namely symmetry and locality.
Especially in every gauge theory it must be explained what the
symmetry objects are and what locality means.

Since some time many theorists hope for noncommutative geometry \cite{con-1}
becoming the right tool to formulate quantum field theory rigorously. But now
we are in the dilemma that the language of noncommutative geometry provides
very general and powerful symmetry objects, the quantum groups, but not an
appropriate method to study local aspects.
This fundamental problem was the starting point for the present paper.

Let me explain in more detail what the principle of locality in physics says.
In the algebraic framework \cite{haa-1} a net of $C^*$-algebras on Minkowski
space describes the local structure of observables such that observable
algebras defined on spatially seperated double cones commute.
Alternativly one can require local commutation relations \cite{stw-1,itz-1}.
In any case we need mathematical methods to compare observables or fields
at different space time points or neighbourhoods. But in noncommutative
geometry there are no points respectively there is no topology on a
base space where the fields and observables are defined.
The problem becomes even more serious if we want to quantise gauge theories.
A field in a classical gauge theory is a (global) section
in a vector bundle. Usually those vector bundles are described in local
charts or in other words in local coordinates.
Now gauge transformations of the second kind change the
field locally such that the observable effects of the field stay the same.
The principle of local coordinates and gauge transformations is
mathematically welldefined in classical geometry and physics.
The appropriate language is the theory of principal fiber bundles.
But up to now it hasn't been possible to translate it into
noncommutative geometry or rigorous quantum field theory \cite{haa-1}.

Mathematics also provides arguments to connect local aspects and
noncommutative methods. Furthermore these arguments even give a hint how to
solve the above problem. Certain structures on a locally compact topological
space $M$ like differentiable or analytical ones are not characterised by
the single algebra ${\cal C}_0 (M)$ but by an appropriate sheaf on $M$.
Additiona"ly it is well known from complex geometry \cite{kak-1} that
local function algebras are in general not determined by the global one.
So it is quite natural to assume that we need a sheaf structure in the
noncommutative setting as well.
This would be very helpful also in the case where commutative function
algebras are deformed.
Then one can keep track of what happens to local algebras
of continuous, differentiable and analytical functions or sections
in a fibre bundle.

Because of these considerations we give a sheaf theoretical method
to connect locality and noncommutative geometry. Furthermore within
this method it is possible to define noncommutative principal fibre
bundles which have quantum groups as their "structure groups".

In the first section we will find an equivalent description of principal
fibre bundles in the language of sheaves. Then a definition of very general
noncommutative spaces is given in section 2. In the following noncommutative
principal bundles are defined and important objects like local coordinates,
transition functions and gauge transformations are carried over to the
noncommmutative case. By the same concept quantum vector bundles and
associated quantum vector bundles are introduced in section 4. Additionally
we study differential calculi on quantum vector bundles.
Finally $q$-deformed instanton models give interesting examples where
local gauge transformations are set up in a noncommutative language.

This paper grew out of my diploma thesis \cite{pfl-1} under the supervision of
Prof.~J.~Wess at the "Sektion Physik der Universit\"at M\"unchen".
\section{Principal Fibre Bundles}
Let us repeat the well known definition of a principal fibre bundle
(see for example \cite{dav-1,ghv-2,kas-1}).
\begin{Definition}
\label{DPFB}
Let $P,M$ be topological spaces, $G$ a topological group
and
$\, \pi : P \longrightarrow M \,$ a continous mapping.
$\, (P,M,\pi ,G) \,$ is called a principal fibre bundle with total space $P$,
basis $M$ and structure group $G$, if the following conditions are satisfied.
\begin{enumerate}
\item $\pi$ is surjective.
\item $G$ acts freely from the right on $P$.
\item The equation $\, \pi(u_1) \, = \, \pi(u_2)  \,$ for $\, u_1 , \, u_2 \in
P \,$
 is valid if and only if $\, u_1a \, = \, u_2 \,$ for an $\, a \in G \,$.
\item $P$ is locally trivial over $M$, i.~e.~there exists an open covering
 $\, {\cal U} \, = \, (U_{\iota})_{\iota \in I} \,$ of $M$ and homeomorphisms
 $\, \psi_{\iota} : \pi^{-1}(U_{\iota}) \longrightarrow
 U_{\iota} \times G \, , \: u \longmapsto \big( \pi (u) \, , \, \eta_{\iota}
(u) \big) \,$,
 such that
\begin{equation}
 \psi_{\iota} (ua) \: = \: \big( \pi (u) \, , \, \eta_{\iota} (u)a \big),
 \hspace{1cm}
 u \in \pi^{-1}(U_{\iota}), \quad a \in G.
\end{equation}
\end{enumerate}
\end{Definition}
\begin{Remark}
 The homeomorphisms $\, \psi_{\iota}$ with ${\iota} \in I \,$ are the local
trivializations
 of the principal bundle.
\end{Remark}
The conditions (i) to (iv) in the above definition are not independant from
each other.
In the next theorem we give a characterization of principal bundles where the
defining
axioms are independant. The obvious proof of the theorem is skipped.
\begin{Theorem}
\label{EDPFB}
 Let $P,M$ be topological spaces, $G$ a topological group and
 $\, \pi : P \longrightarrow M \,$ continuous.
 Assume the following two conditions to be true.
\begin{enumerate}
\item
 $\pi$ is surjective.
\item
 $P$ is locally trivial over $M$, i.~e.~there exists an open covering
 $\, {\cal U} \, = \, (U_{\iota})_{\iota \in I} \,$ of $M$ and
 homeomorphisms $\, \psi_{\iota} : \pi^{-1}(U_{\iota}) \longrightarrow
 U_{\iota} \times G \,$, such that
\begin{equation}
 \label{PRTR}
 pr_1 \circ \psi_{\iota} \, = \, \pi \! \mid_{\pi^{-1}(U_{\iota})} \, =: \,
 \pi \! \mid_{U_{\iota}},
\end{equation}
 and
\begin{eqnarray}
 \label{ATR}
 \lefteqn{ \psi_{\iota} \circ \psi_{\kappa}^{-1} \, (x,ab) }
 \nonumber \\
 & = &
 \big( x \, , \, pr_2 \big( \psi_{\iota} \circ \psi_{\kappa}^{-1}
 (x,a) \big) \, b \big),
 \hspace{1.0cm} x \in U_{\iota} \cap U_{\kappa}, \quad a,b \in G.
\end{eqnarray}
\end{enumerate}
 Then by
\begin{equation}
 \label{RA}
 u \cdot a \, := \,
 \psi_{\iota}^{-1} \big( pr_1 \big( \psi_{\iota} (u) \big) \, , \,
 pr_2 \big( \psi_{\iota} (u) \big) \, a \big) ,
 \hspace{0.7cm} u \in \pi^{-1} (U_{\iota}), \quad a \in G,
\end{equation}
 a right $G$-action on $P$ is defined.
 Furthermore $(P,M,\pi ,G)$ is a principal fibre bundle
 over $M$
 with trivialisations $\, \psi_{\iota} \, $, $\iota \in I \,$.
 On the other hand given any principal bundle $(P,M,\pi ,G)$
 with trivialisations $\, \psi_{\iota} \, $, $\iota \in I \,$
 the above conditions $(i)$ and $(ii)$ hold,
 and the right $G$-action is given locally by equation
 {\rm (\ref{RA})}.
\end{Theorem}
We introduce some notation.
\label{GD}
Let $\cal M$ (resp.~$\cal F$) be the sheaf of complex and bounded
continous functions on $M$ (resp.~$G$).
Now $\cal M$ and $\cal F$ are sheaves of Banach-$\ast $-algebras.
Let $U, \: V$ be open in $M$.
Then ${\cal P}(U)$ is the algebra of complex and bounded continous functions
on $\pi^{-1}(U)$. The canonical injections
\bdm
 i_U^V : U
 \longrightarrow V , \quad u \longmapsto u,
\edm
where $U \subset V \subset M$ open define restrictions ${\cal P}(i_U^V) =
r_U^V : {\cal P}(V) \longmapsto {\cal P} (U) \, ,
\: f \longmapsto f \! \mid_{\pi^{-1} (U)}$.
Now $\cal P$ is a sheaf of algebras on $M$, or more clearly a sheaf of
Banach-$\ast $-algebras.
The continuous function
$\, \pi : P \longrightarrow M \,$ induces a sheaf morphism
\bdm
 \varrho = \pi^* : {\cal M} \longrightarrow {\cal P}, \quad
 \pi^* (f) = f \circ \pi \! \mid_U, \hspace{1,0cm}
 f \in {\cal M}(U),
\edm
and the $G$-action on $P$ a sheaf morphism
$\, \phi : {\cal P} \longrightarrow {\cal P} \otimes {\cal F}(G) \,$
by
\begin{equation}
 \phi_U(h)(u,a) \, = \, h(ua), \hspace{1,0cm} h \in {\cal P}, \quad
 u \in \pi^{-1}(U), \quad a \in G.
\end{equation}
Finally the local trivialisations give sheaf morphisms
\begin{eqnarray*}
\begin{array}{lclcl}
 \Omega_{\iota} & : & {\cal M} \! \mid_{U_\iota} \otimes {\cal F}(G)
 & \longrightarrow & {\cal P} \! \mid_{U_\iota}  \\
 & & f \otimes g & \longmapsto  & (f \otimes g)
 \circ \psi_{\iota} \! \mid_U, \quad g \in {\cal F}_G, \; f \in {\cal M}(U),
 \; U \subset U_{\iota}.
\end{array}
\end{eqnarray*}
\begin{Proposition}
\label{SFPFB}
Let $P, M$ be locally compact topological spaces, $G$ a locally compact
topological group and $\pi : P \longrightarrow M $ continuous.
With the above definition of $\cal{P, M} , \varrho $ the quadrupel $(P,M,\pi ,
G) $
is a principal fibre bundle if and only if the following conditions
$(i)$ and $(ii)$ are satisfied.
\begin{enumerate}
\item
The sequence $\, 0 \longrightarrow {\cal M}
\stackrel{\varrho }{\longrightarrow} {\cal P} \,$ is exact.
\item
 One can find an open covering
 $\, {\cal U} = (U_{\iota})_{\iota \in I} \,$ of $M$ and continuous mappings
 $\psi_{\iota} : \pi^{-1}(U_{\iota}) \longrightarrow U_{\iota} \times G$
 with the following property. The sheaf morphisms
 $\, \Omega_{\iota} : {\cal M} \! \mid_{U_\iota} \otimes {\cal F}_G
 \longrightarrow {\cal P} \! \mid_{U_\iota} \,$, $\, \iota \in I \,$
 are isomorphisms and define sheaf morphisms
 $\, \Omega_{\kappa ,\iota} : {\cal M} \! \mid_{U_\kappa} \otimes {\cal F}_G
 \longrightarrow {\cal M} \! \mid_{U_\iota} \otimes {\cal F}_G \,$,
 $\, \iota, \kappa \in I \,$ by
 $\, \big( \Omega_{\kappa ,\iota} \big)_U \, = \,
 \big( \Omega_{\kappa} \big)_U^{-1} \circ \big( \Omega_{\iota} \big)_U \,$.
 These sheaf morphisms satisfy the equations
\begin{eqnarray}
\label{SPR}
 \lefteqn{\big( \Omega_{\iota} \big)_U (f \otimes 1) \, = \, \varrho_U(f),
 \hspace{1cm} f \in {\cal M}, \: U \subset U_{\iota},}
\end{eqnarray}
\begin{eqnarray}
\label{STR}
 \big( \big( \Omega_{\kappa ,\iota} \big)_U \otimes id \big) \circ
 (id \otimes \Delta ) \ = \ (id \otimes \Delta ) \circ
 \big( \Omega_{\kappa ,\iota} \big)_U, \quad U \subset U_{\iota} \cap
U_{\kappa},
\end{eqnarray}
where $m$ $($resp.~$\Delta )$ is the multiplication
$($resp.~comultiplication$)$ in ${\cal F}(G)$.
\end{enumerate}
\end{Proposition}
\bpro
For $\, U \subset U_{\iota} \,$ open, $\, f \in {\cal M}(U) \,$
and $\, u \in \pi^{-1}(U) \,$ wre have the equations
\begin{equation}
\label{PR1}
 \lefteqn{
 (\Omega_{\iota})_U (f \otimes 1) \, (u) \, = \,
 f \big( pr_1 \circ \psi_{\iota}(u) \big), }
\end{equation}
\begin{equation}
\label{PR2}
 \lefteqn{
 \varrho_U(f) \, (u) \, = \,  f \big( \pi (u) \big) , }
\end{equation}
and for $\, U \subset U_{\iota} \cap U_{\kappa} \,$ open, $\, f \in {\cal M}(U)
\otimes
{\cal F}(G) \,$, $\, x \in U \,$ and $\, a,b \in G \,$ the equations
\begin{equation}
\label{TR0}
\lefteqn{
 \big( \Omega_{\kappa ,\iota} \big)_U (f) \, (x,a) \, = \,
 f \big( \psi_{\iota} \circ \psi_{\kappa}^{-1}(x,a) \big) , }
\end{equation}
\begin{equation}
\label{TR1}
\lefteqn{
 \big( (id \otimes \Delta) \circ \big( \Omega_{\kappa ,\iota} \big)_U \big)
 \, (f) \, (x,a,b)
 \, = \,
 f \big( \psi_{\iota } \circ \psi_{\kappa } (x,ab) \big) , }
\end{equation}
\begin{equation}
\label{TR2}
\lefteqn{
 \big( \big( \big( \Omega_{\kappa ,\iota} \big)_U \otimes id \big) \circ
 (id \otimes \Delta) \big) \, (f) \, (x,a,b) \, = \,
 f \big( x \, , \,
 pr_2 \big( \psi_{\iota } \circ \psi_{\kappa }^{-1} \, (x,a) \big) \, b
 \big) . }
\end{equation}
First suppose $(P,M, \pi , G)$ to define a principal fibre
bundle with trivialisations
$\, \psi_{\iota} : \pi^{-1}(U_{\iota}) \longrightarrow U_{\iota} \times G \,$.
As $\pi : P \longrightarrow M$ is surjective, the sequence
$\, 0 \longrightarrow {\cal M} \stackrel{\varrho }{\longrightarrow} {\cal P}
\,$
is exact and (i) holds.
The equations (\ref{PR1}), (\ref{PR2}) and (\ref{PRTR}) imply (\ref{SPR}).
Furthermore the equations (\ref{TR1}), (\ref{TR2}) and (\ref{ATR}) give
(\ref{STR}). Altogether this proofs (ii).
\newline \noindent
Now we have to show the other implication.
Assume (i) and (ii) being true for $(P,M, \pi ,G)$.
Then the relations (\ref{PR1}), (\ref{PR2}) and (\ref{SPR}) entail
\begin{eqnarray*}
 f \big( pr_1 \circ \psi_{\iota} \, (u) \big) \, = \, f( \pi (u)),
 \hspace {2,2cm} f \in {\cal M} (U), \; u \in \pi^{-1} (u).
\end{eqnarray*}
As the continuous functions on $U$ are seperating, we get (\ref{PRTR}).
Similarly (\ref{TR1}), (\ref{TR2}) and (\ref{STR}) give the equation
\begin{eqnarray*}
 f \big( \psi_{\iota} \circ \psi_{\kappa}^{-1} \, (x,ab) \big) \, = \,
 f \big( x \, , \, pr_2 \big( \psi_{\iota} \circ \psi_{\kappa}^{-1} \, (x,a)
 \big) \, b \big)
\end{eqnarray*}
for $\, f \in {\cal M} (U) \otimes {\cal F} (G) \,$,
$\, x \in U \,$ and $\, a,b \in G \,$.
Now we have shown (\ref{ATR}) and condition (ii) in the theorem \ref{EDPFB}.
As $\, pr_1 \circ \psi_{\iota} \, = \, \pi \! \mid_{U_{\iota}} \,$ and
$\, {\cal U} = (U_{\iota})_{\iota \in I} \,$ covers $M$, $\pi$ is surjective.
That is all.
\epro
\begin{Corollary}
\label{SFPFBK}
Assume to be given a tupel $({\cal P, M},\varrho,H)$, where $\cal M$ is the
commutative
sheaf of complex bounded continuous functions on a locally compact topological
space $M$,
$\cal P$ is a sheaf of commutative $C^*$-algebras over $M$,
$\, \varrho : {\cal M} \longrightarrow {\cal P} \,$ a sheaf morphism and $H$
a commutative $($topological$)$ Hopf algebra.
$({\cal P, M},M,\varrho,H)$ can be identified with a principal fibre bundle
if the following conditions hold.
\begin{enumerate}
\item
The sequence $\, 0 \longrightarrow {\cal M}
\stackrel{\varrho }{\longrightarrow} {\cal P} \,$ is exact.
\item
There exists an open covering $\, {\cal U} = (U_{\iota})_{\iota \in I} \,$ of
$M$
and sheaf isomorphisms $\, \Omega_{\iota}  :  {\cal M} \! \mid_{U_\iota}
\otimes H
\longrightarrow {\cal P} \! \mid_{U_\iota} $, $\,  \iota \in I \,$
such that $\Omega_{\iota}$ and $\, \Omega_{\kappa ,\iota} \,$ whith
$\, \big( \Omega_{\kappa ,\iota} \big)_U \, = \,
 \big( \Omega_{\kappa} \big)_U^{-1} \circ \big( \Omega_{\iota} \big)_U,
 \quad U \subset U_{\iota} \cap U_{\kappa} \,$
satisfy the equations $(\ref{SPR})$ and $(\ref{STR})$.
\end{enumerate}
\end{Corollary}
\bpro
One can construct the locally compact topological spaces $P,M,G$ and
the continuous mappings $\pi, \psi_{\iota}$ by the Gelfand transformation.
Then the assumptions of proposition \ref{SFPFB} hold and the corollary
is shown.
\epro
\section{Quantum Spaces}
In this section we define the frame in which the principle of locality
and noncommutative geometry can be connected. We use a sheaf theoretical
language which is already well known in the commutative setting of algebraic
geometry and complex analysis.
See the appendix A for the definition of sheaves and the literature
\cite{ten-1,mam-1} for further details on sheaves.
\begin{Definition}
\label{DQS}
 Let ${\bf A}$ be subcategory of the category ${\bf Alg}$ of all
 associative algebras.
 An ${\bf A}$-\QS over a topological space $M$ is a sheaf $\cal M$ over $M$
 with objects in ${\bf A}$. The category of  ${\bf A}$-\QSS is dual
 to the category of sheaves over topological spaces and will be denoted by
 ${\bf A}$-${\bf Qs}$.
\end{Definition}
Let $\cal M$ be a sheaf over $M$,. If we consider $\cal M$ as a an object of
${\bf A}$-${\bf Qs}$, we sometimes write ${\cal M}_Q$.
Now let $\cal N$ be a sheaf over $N$ and $f : M \longrightarrow N$
a continuous mapping. A morphism ${\cal F}_Q : {\cal M}_Q \longrightarrow
{\cal N}_Q$ of sheaves over  $f$ will be written
${\cal F}_Q : {\cal M}_Q \longrightarrow {\cal N}_Q$ if regarded as a
morphism in ${\bf A}$-${\bf Qs}$.
The ${\bf A}$-\QSS over a topological space $M$ with
morphisms over the identity $id_M$ form a subcategory
$\bf A$-${\bf Qs}_M$ of $\bf A$-$\bf Qs$.

\label{EQS}
The following examples of quantum spaces show that it is possible to
include a concept of locality in noncommutative geometry.
They also comprise important objects of commutative and
noncommutative geometry.
\begin{Example}
\label{MA-i}
Let $M$ be a topological space, $A$ an object in ${\bf A}$ and $ \cal L$
the locally constant sheaf on $M$ with objects in $A$. $ \cal L$
is an ${\bf A}$-quantum space.
\end{Example}
Ringed spaces are important tools of complex analysis and algebraic
geometry \cite{har-1,kak-1}. A ringed space is simply a pair
$\big( M, {\cal O}_M \big)$, where $M$ is a topological space and
${\cal O}_M$ a sheaf of commutative rings.
\begin{Example}
\label{MA-ii}
 Ringed spaces $\big( M, {\cal O}_M \big)$ are commutative quantum spaces.
\end{Example}
Manifolds, complex spaces end schemes can also be considered as ringed spaces
or commutative quantum spaces.
To explain that let us first write down some special ringed spaces: \\
\begin{tabular}{lclcl}
 $\big( {\Bbb{R}}^n, {\cal C} \big) $, & where & ${\cal C}$ & -&
 sheaf of continuous functions on ${\Bbb{R}}^n$, $n \in {\Bbb{N}}$, \\
 $\big( {\Bbb{R}}^n, {\cal C}^r \big) $, & where & ${\cal C}^r$ & -&
 sheaf of $r$-times continuously differentiable \\
 & & & & functions on
 ${\Bbb{R}}^n$, $n \in {\Bbb{N}}$, $r \in {\Bbb{N}}^* \cup \{ \infty \}$, \\
 $\big( {\Bbb{R}}^n, {\cal C}^{\omega} \big) $, & where &
 ${\cal C}^{\omega}$ & -&
 sheaf of real analytic functions on ${\Bbb{R}}^n$, $n \in {\Bbb{N}}$, \\
 $\big( {\Bbb{C}}^n, {\cal O}_n \big) $, & where & ${\cal O}_n$ & -&
 sheaf of holomorphic functions on ${\Bbb{C}}^n$, $n \in {\Bbb{N}}$.
\end{tabular}
\begin{Example}
\label{MA-iv}
 Let $n \in {\Bbb{N}}$ and $r \in {\Bbb{N}}^* \cup \{ \infty \}$.
 A ringed space $\big( X, {\cal O}_X \big)$ is called a
\begin{enumerate}
\item
 topological manifold of dimension $n$, if
 $\big( X, {\cal O}_X \big)$ is locally isomorphic to
 $\big( {\Bbb{R}}^n, {\cal C} \big) $,
\item
 differentiable $r$-manifold of dimension $n$, if
 $\big( X, {\cal O}_X \big)$ is locally isomorphic to
 $\big( {\Bbb{R}}^n, {\cal C}^r \big) $,
\item
 real analytic manifold of dimension $n$, if
 $\big( X, {\cal O}_X \big)$ is locally isomorphic to \\
 $\big( {\Bbb{R}}^n, {\cal C}^{\omega} \big) $,
\item
 complex manifold of dimension $n$, if
 $\big( X, {\cal O}_X \big)$ is locally isomorphic to
 $\big( {\Bbb{C}}^n, {\cal O}_n \big) $,
\item
 scheme, if for every $x \in X$ there exists an open neighbourhood $U$ of $x$,
 such that $\big( U, {\cal O}_X \! \mid_U \! \big)$ is isomorphic to
 an affine scheme.
\end{enumerate}
All those spaces are quantum spaces.
\end{Example}
Supersymmetric structures (see Wess, Bagger \cite{web-1})
are our first examples of noncommutative quantum spaces.
Most easily this can be seen with the definition of superspaces
according to Manin
\cite{man-2}.
\begin{Definition}
 A superspace consists of a pair
 $\big( M, {\cal O}_M \big)$, where $M$ is a topological space and
 ${\cal O}_M$ a sheaf of supercommutative rings,
 such that all stalks $\, {\cal O}_{M,x} \,$, $\,x \in M \,$ are local.
\end{Definition}
Supermanifolds are superspaces which locally split into an even
and odd part such that the splitting is differentiable and the odd part
is a locally free module sheaf over the even part.
\begin{Example}
 Superspaces and supermanifolds are noncommutative quantum spaces.
\end{Example}
We already cite here an example of a quantum space we are going to construct
in section \ref{NCIM}.
\begin{Example}
\label{PH-iii}
 The $q$-deformed space time over the background $S^4$ is a noncommutative
 quantum space.
\end{Example}
\section{Quantum Principal Bundles}
\label{SQPBDEF}
\subsection{The Category of Quantum Principal Bundles}
Corollary \ref{SFPFBK} characterises principal bundles in the language of
sheaves
of commutative algebras. If we simply leave out the requirement for the
commutativity of the local algebras we almost get the definition of
noncommutative
principal bundles or quantum principal bundles. One further generalisation
compared with the commutative case has to be made. The reason lies in the fact
that the tensor product of a noncommutative algebra and a Hopf algebra
posses "more multiplications" than the tensor product of the
corresponding commutative objects. So we have to specify the choosen
multiplication on the local tensor products by the methods of appendix B.

Let $ M $ be a topological space and $ {\bf A }$ a subcategory of the
category of all associative $ \Bbb{C} $-algebras.
Suppose we are given the following objects:
\begin{enumerate}
\item
 a sheaf $ {\cal M} $ over $ M $ with objects in $ {\bf A} $ called the
 base quantum space,
\item
 a sheaf $ {\cal P} $ over $ M $ with objects in $ {\bf A} $ called the total
quantum space,
\item
 a sheaf morphism
 $ \varrho : {\cal M} \longrightarrow {\cal P} $ called the projection,
\item
 a Hopf algebra $ H $ called the structure quantum group,
\item
 a family of sheaf morphisms $\, \big( \Omega_{\iota} \big)_{\iota \in I} \,$,
 $\, \Omega_{\iota} \, : \, {\cal M} \! \mid_{U_\iota} \, \#_{\iota } \, H \,
  \, \longrightarrow  \, {\cal P} \! \mid_{U_\iota} \, $,
 where
 $ {\cal U} \: = \: (U_{\iota})_{\iota \in I} \: $ is an open covering
 of $ M $ and $ \#_{\iota } $ is a crossed product defined according to
 theorem \ref{CRPR} by a weak action
 $ \alpha_{\iota} : H \times
 {\cal M} \! \mid_{U_{\iota}} \longrightarrow {\cal M} \! \mid_{U_{\iota}} $
 and a normal cocycle
 $ \iota : H \times H
 \longrightarrow {\cal M} \! \big( U_{\iota} \big) $
 fulfilling the twisted module condition.
\end{enumerate}
The tupel
$\big( {\cal P,M}, \varrho , H , \big( \Omega_{\iota} \big)_{\iota \in I} \big)
$
gives the data of an $ {\bf A} $-quantum principal bundle over $M$.
Its entries can be
regarded as the noncommutative generalisations
of respectively the total space, base space, projection, structure group
and trivialisation of a classical principal bundle.
\begin{Definition}
\label{DQPB}
$\big( {\cal P,M}, \varrho , H , \big( \Omega_{\iota} \big)_{\iota \in I} \big)
$
is said to be an $ {\bf A} $-quantum principal bundle over $M$ with
coordinate system $\, \big( \Omega_{\iota} \big)_{\iota \in I} \,$, if the
following conditions are fulfilled.
\begin{enumerate}
\item
 The sequence
 $\, 0 \longrightarrow {\cal M}
 \stackrel{\varrho }{\longrightarrow} {\cal P} \,$
 is exact.
\item
 The algebras $ {\cal M} (U) $ and $ {\cal P}(U) $ are unitary
 for $ U \subset U_{\iota} $ open.
\item
 Let the sheaf morphisms
 $\, \Omega_{\kappa , \iota }
 \, : \, {\cal M} \! \mid_{U_{\iota} \cap U_{\kappa}}
 \#_{\iota } \, H \, \longrightarrow \,
 {\cal M} \! \mid_{U_{\iota} \cap U_{\kappa}} \#_{\kappa } \, H \,$ be defined
 by $\, \Big( \Omega_{\kappa , \iota } \Big)_U \, =
 \, \Big( \Omega_{\kappa } \Big)_U^{-1}
 \circ \, \Big( \Omega_{\iota } \Big)_U \,$,
 where $\, U \subset U_{\iota} \cap U_{\kappa} \,$ open.
 Then the the following equations are valid:
\begin{equation}
 \label{DQPB1}
  \left( \Omega_{\iota} \right)_U (f \#_{\iota } 1) \, = \,
  \varrho_U(f),  \mbox{ }
  \hspace{1cm} f \in {\cal M}(U), \quad U \subset U_{\iota}
\end{equation}
\begin{equation}
 \label{DQPB2}
 \big( \big( \Omega_{\kappa , \iota } \big)_U \otimes id \big) \circ
 ( id \otimes \Delta ) \, = \, ( id \otimes \Delta ) \circ
 \big( \Omega_{\kappa , \iota } \big)_U, \mbox{ }
 \hspace{1cm} U \subset U_{\iota} \cap U_{\kappa}.
\end{equation}
\end{enumerate}
\end{Definition}
Suppose we are given a second
$ {\bf A} $-quantum principal bundle
$\big( {\cal P,M}, \varrho , H , \big( \tilde{\Omega}_{\kappa} \big)_{\kappa
\in K} \big) $
over $M$ with \CS $ \big( \tilde{\Omega}_{\kappa} \big)_{\kappa \in K} $
being defined on the open covering $ \big( V_{\kappa} \big)_{\kappa \in K} $
of $M$. The two $ {\bf A} $-quantum principal bundles with \CS are
equivalent, if
for $ U \subset U_{\iota} \cap U_{\kappa} $ open,
$ \iota \in I $, $\kappa \in J$
the sheaf morphisms
\begin{equation}
\begin{array}{lclc}
 \tilde{\Omega}_{\kappa , \iota } \, : \,
 {\cal M} \! \mid_{U_{\iota} \cap U_{\kappa}}
 \, \#_{\iota } \, H & \longrightarrow &
 {\cal M} \! \mid_{U_{\iota} \cap U_{\kappa}} \, \#_{\kappa } \, H , & \;
 \big( \tilde{\Omega}_{\kappa , \iota } \big)_U  \, = \,
 \big( \tilde{\Omega}_{\kappa } \big)_U ^{-1} \circ \big( \Omega_{\iota }
\big)_U ,
\end{array}
\end{equation}
satisfy the equation
\begin{equation}
\big( \big( \tilde{\Omega}_{\kappa , \iota } \big)_U \otimes id \big) \circ
 ( id \otimes \Delta ) \: = \: ( id \otimes \Delta ) \circ
 \big( \tilde{\Omega}_{\kappa , \iota } \big)_U.
\end{equation}
This relation is an equivalence relation in the class of all
${\bf A}$-quantum
principal bundles over $M$ with coordinate system.
\begin{Definition}
An equivalence class of ${\bf A}$-quantum principal bundles over $M$ with \CS
is called an ${\bf A}$-\QPB over $M$.
\end{Definition}
\begin{Remark}
If no misunderstandings can arise we will not distinguish betweeen quantum
principle bundles with coordinate system and their equivalence classes, the
quantum principal bundles.
\end{Remark}
\begin{Remark}
 The algebras ${\cal M} \! \mid_{U_{\iota} \cap U_{\kappa}}
 \, \#_{\iota } \, H $ are socalled Hopf Galois extensions $($see
 {\rm \cite{blm-1}} for further details$)$.
 Therefore one can interpret \QPBS as sheaves which locally look like
 appropriate Hopf Galois extensions.
\end{Remark}
We would like to regard the quantum principle bundles as objects of
a certain category.
The following definition provides the necessary morphisms of this
category.
\begin{Definition}
\label{DMQPB}
Let
$\big( {\cal P,M}, \varrho , H , \big( \Omega_{\iota} \big)_{\iota \in J} \big)
$
$($resp.~$\big( {\cal N}, {\cal Q},
\tilde{\varrho} , \tilde{H} , \big( \tilde{\Omega}_{\kappa} \big)_{\kappa \in
K} \big) $$)$
be an ${\bf A}$-QPB over $M$ $($resp.~over $N$$)$, where the \CS
$\big( \Omega_{\iota} \big)_{\iota \in J} $
$($resp.~$\big( \tilde{\Omega}_{\kappa} \big)_{\kappa \in K} $$)$
is defined on the open covering $\big( U_{\iota} \big)_{\iota \in J} $
$($resp.~$\big( V_{\kappa} \big)_{\kappa \in K} $$)$
of $M$ $($resp.~N$)$. A morphism of ${\bf A}$-quantum principal bundles
\[
({\cal P,M} ,\varrho,H,\big( U_{\iota} \big)_{\iota \in J} )
\longrightarrow
({\cal Q,N},\tilde{\varrho},\tilde{H},\big( V_{\kappa} \big)_{\kappa \in K} )
\] \noindent
consists of a tupel $ ({\cal R},{\cal F},f, h) $
such that the relations $(i)$ to $(iv)$ are satisfied.
\begin{enumerate}
\item
 $ f : M \longrightarrow N $ is a continuous mapping.
\item
 $ {\cal R} : {\cal Q} \longrightarrow {\cal P} $ and
 $ {\cal F} : {\cal N} \longrightarrow {\cal M} $ are
 morphismen of \cat{A}-sheaves over $ f $ such that the diagram
\begin{center}
\xext=1400 \yext=500
\adjust[`\tilde{\varrho};`\varrho;`\cal R;`\cal F]
\begin{picture}(\xext,\yext)(\xoff,\yoff)
\putmorphism(0,0)(1,0)[0`\phantom{\cal M}`]{700}{1}{b}
\putmorphism(0,500)(1,0)[0`\phantom{\cal N}`]{700}{1}{a}
\setsqparms[1`1`1`1;700`500]
\putsquare(700,0)[\cal N`\cal Q`\cal M`\cal P;\tilde{ \varrho }`\cal F`\cal
R`\varrho]
\end{picture}
\end{center}
 commutes.
\item
$ h : \tilde{H} \longrightarrow H $ is a morphism of Hopf algebras.
\item
 Let the mapping ${\cal T}_{\iota, \kappa, V}$ with
 $ \iota \in I $, $ \kappa \in J $, $ V \subset V_{\kappa} $
 open and $ U \: = \: f^{-1} (V) \cap U_{\iota} $ be defined by
\begin{eqnarray*}
\lefteqn{ {\cal T}_{\iota, \kappa, V} \: : \: {\cal N} (V) \#_{\kappa}
\tilde{H}
 \: \stackrel{ (\tilde{\Omega}_{\iota} )_V }{ \longrightarrow } \:
 {\cal Q} (V) \stackrel{\tilde{\cal R}_V}{\longrightarrow} } \\
 & & \stackrel{\tilde{\cal R}_V}{\longrightarrow} \:
 {\cal P} \big( f^{-1} (V) \big) \:
 \stackrel{ r_U^{f^{-1} (V) } }{ \longrightarrow } \:
 {\cal P} (U) \: \stackrel{ (\tilde{\Omega}_{\iota} )_V^{-1} }{ \longrightarrow
} \:
 {\cal M} (U) \#_{\iota} H
\end{eqnarray*}
 Then we have
\begin{eqnarray}
\lefteqn{
 \big( id \otimes \Delta_H \big) \circ {\cal T}_{\iota, \kappa, V}
 \: = \:
 \big( {\cal T}_{\iota, \kappa, V} \otimes h \big) \circ
 \big( id \otimes \Delta_{\tilde{H}} \big) }.
\end{eqnarray}
\end{enumerate}
\end{Definition}
By a standard calculation we get the following theorem.
\begin{Theorem}
 The  \cat{A}-quantum principal bundles and their morphisms $($ definition
 {\rm \ref{DMQPB}}$)$ form a category \cat{A-Qpb}.
 The \cat{A}-\QPBS  over $M$ $($resp.~the \cat{A}-\QPBS  over $M$
 with basis $\cal M )$ together with the morphisms $(f,{\cal R, F}, h)$
 of the form $ f \, = \, id_M $ $($resp.~$ f \, = \, id_M $ and
 $ {\cal F} \, = \, id_{\cal M} )$ form a subcategory being denoted by
 $ {\cat{A-Qpb}}_M$ $($resp.~${\cat{A-Qpb}}_{\cal M})$.
\end{Theorem}
\subsection{Coaction of the Structure Quantum Group}
The structure quantum group $H$ can be regarded as a gauge quantum group.
In analogy with the commutative case $H$ should (co-)act on the quantum bundle
or
in other words should provide noncommutative gauge transformations of the first
kind.
Starting from the example of commutative principle bundles we will
show how to define this coaction and derive some fundamental results about it.

Let the sheaves $ {\cal M} $, $\cal P$ and the Hopf algebra $H$
be given by a principal bundle $ (P,M,\pi,G) $.
The next question is what kind of
$H$-coaction the $G$-action on $P$ induces.
To give an answer define for all $ U \subset M $ open a homomorphism
$\phi_U : {\cal P}(U) \longrightarrow {\cal P}(U) \otimes  H  $
by
\begin{equation}
 \label{DRHCO}
 \phi_U (f) (u,a) \: = \: f(ua), \hspace{1cm}  f \in {\cal P}(U), \quad
 u \in \pi^{-1}(U), \quad a \in G.
\end{equation}
As $\, u(ab) \: = \: (ua)b \,$ for $ u \in \pi^{-1}(U) $,
$\, a,b \in G \,$ (\ref{DRHCO}) entails
\begin{eqnarray}
 \big( (id \otimes \Delta) \circ \phi_U \big) \, (f) \,
 (u,a,b) & = & f(u(ab)) \nonumber \\
 & = & f((ua)b) \nonumber \\
 & = & \big( (\phi_U \otimes id) \circ \phi_U \big) \, (f) \, (u,a,b),
\end{eqnarray}
that is
\begin{equation}
 (id \otimes \Delta) \circ \phi_U \: = \: (\phi_U \otimes id) \circ
 \phi_U.
\end{equation}
A similar consideration using $ ue \: = \: u$ for $ u \in \pi^{-1}(U) $ proofs
\begin{equation}
 (id \otimes \varepsilon ) \circ \phi_U \: = \: id.
\end{equation}
Therefore $ \phi_U $ gives $ {\cal P}(U) $ the structure of an
$H$-right-comodule
and provides for a sheaf-morphism
$\phi : {\cal P} \longrightarrow {\cal P} \otimes  H  $.
If $ U \subset U_{\iota} $, we can express $ \phi_U $ directly by
the local trivialisations $\Omega_{\iota}$ of the sheaf ${\cal P}$.
Using (\ref{RA}) in theorem \ref{EDPFB}
as well as (\ref{DRHCO}) one gets the relation
\begin{eqnarray}
 \lefteqn{
 \phi_U (f) (u,a) } \nonumber \\
 & = & f \left( \psi^{-1}_{\iota} \left( pr_1 \left( \psi_{\iota} (u) \right),
 pr_2 \left( \psi_{\iota} (u) \right) a \right) \right) \nonumber \\
 & = & \big( \left( \Omega_{\iota} \otimes id \right)
 \circ (id \otimes \Delta)
 \circ \Omega_{\iota}^{-1} \, (f) \big) \, (u,a),
\end{eqnarray}
that is
\begin{equation}
\label{TRRCO}
\phi_U \: = \: (\Omega_{\iota} \otimes id) \circ (id  \otimes \Delta)
\circ \Omega_{\iota}^{-1}.
\end{equation}
Equation (\ref{TRRCO}) will be now be used to define a sheaf-morphism
\[
 \phi : {\cal P} \longrightarrow {\cal P} \otimes  H
\] \noindent
in the general case of an arbitrary quantum principal bundle
$({\cal P,M},\varrho,H, (\Omega_{\iota})_{\iota \in I})$.
Let us show that by equation (\ref{TRRCO}) $\phi$ is welldefined even in the
noncommutative
setting. We first have to proof
\begin{equation}
 \big( \big( \Omega_{\iota} \otimes id \big) \circ (id  \otimes \Delta) \circ
 \Omega_{\iota}^{-1} \big) (f)
 \: = \:
 \big( \big( \Omega_{\kappa} \otimes id \big) \circ (id  \otimes \Delta) \circ
 \Omega_{\kappa}^{-1} \big) (f)
\end{equation}
for all $\, U \subset U_{\iota} \cap U_{\kappa} \,$ open and $\, f \in {\cal
P}(U) \,$.
But this is a consequence from
\begin{eqnarray}
 \lefteqn{
 \big( \big( \Omega_{\iota} \otimes id \big) \circ \left( id
 \otimes \Delta \right) \circ \Omega_{\iota}^{-1} \big) (f) } \nonumber \\
 & = & \big( \big( \big( \Omega_{\kappa} \circ \Omega_{\kappa, \iota} \big)
 \otimes id \big) \circ \left( id \otimes  \Delta \right) \circ
 \big( \Omega_{\iota, \kappa} \circ \Omega_{\kappa}^{-1} \big) \big) \, (f)
\nonumber \\
 & = & \big( \big( \Omega_{\kappa} \otimes id \big) \circ
 ( id \otimes \Delta ) \circ \Omega_{\kappa, \iota} \circ
 \Omega_{\iota, \kappa} \circ \Omega_{\kappa}^{-1} \big) \, (f) \nonumber \\
 & = & \big( \big( \Omega_{\kappa} \otimes id \big) \circ
 ( id \otimes \Delta ) \circ \Omega_{\kappa}^{-1} \big) \, (f).
\end{eqnarray}
In the second step let $ U \subset M $ be open and $ f \in {\cal P}(U) $. Then
for all
$ \iota \in I $ the homorphism
$\, \phi_{U_{\iota} \cap U} \big( r^U_{U_{\iota} \cap U}(f) \big) \,$
is defined, and for all $\iota , \kappa \in I $
\begin{eqnarray}
 r^{U_{\iota} \cap U}_{U_{\iota} \cap U_{\kappa} \cap U} \circ
 \phi_{U_{\iota} \cap U} \, \big( r^U_{U_{\iota} \cap U} (f) \big)
 & = & \phi_{U_{\iota} \cap U_{\kappa} \cap U} \,
 \big( r^{U}_{U_{\iota} \cap U_{\kappa} \cap U} (f) \big) \nonumber \\
 & = & r^{U_{\kappa} \cap U}_{U_{\iota} \cap U_{\kappa} \cap U} \circ
 \phi_{U_{\kappa} \cap U} \, \big( r^U_{U_{\kappa} \cap U} (f) \big).
\end{eqnarray}
is true.
As $ {\cal P} \otimes  H  $ is a sheaf, one gets a
$ \phi_U(f) \in {\cal P}(U) \otimes  H  $ with
\begin{equation}
 r^{U}_{U_{\iota} \cap U} \circ \phi_U (f) \: = \:
 \big( \big( \Omega_{\iota} \otimes id \big) \circ (id \otimes \Delta) \circ
 \Omega_{\iota}^{-1} \big) \, \big( r^U_{U_{\iota} \cap U} (f) \big).
\end{equation}
Now the next theorem is evident.
\begin{Theorem}
\label{DNRCO}
 Let
 $({\cal P,M},\varrho,H,(\Omega_{\iota})_{\iota \in I})$ be an
 $\bf A$-\QPB. Then there exists a uniquely defined sheaf-morphism
 $ \phi : {\cal P} \longrightarrow {\cal P} \otimes  H  $ fulfilling
\begin{equation}
 \label{QRCO}
 \phi_U (f) \: = \: \big( \big( \Omega_{\iota} \otimes id\big) \circ (id
\otimes \Delta)
 \circ \Omega_{\iota}^{-1} \big) \, (f) , \hspace{1,5cm}
 f \in {\cal P}(U), \quad U \subset U_{\iota}.
\end{equation}
 If $({\cal P,M},\varrho,H,(\Omega_{\iota})_{\iota \in I}))$ is given by a
 commutative principal bundle $(P,M,\pi ,G) $,
 the relation
\begin{equation}
 \phi_U (f) (u,a) \: = \: f (ua), \hspace{1cm} f \in {\cal P}(U), \quad
 u \in \pi^{-1}(U), \quad a \in G
\end{equation}
 is true.
\end{Theorem}
\begin{Corollary}
 $ {\cal P}(U) $ is an $H$-right-comodule with coaction $ \phi_U $, that is the
 following relations hold:
\begin{equation}
 \label{EHCO1}
 (id \otimes \Delta) \circ \phi_U \: = \: (\phi_U \otimes id) \circ \phi_U,
\end{equation}
\begin{equation}
 \label{EHCO2}
 (id \otimes \varepsilon ) \circ \phi_U \: = \: id.
\end{equation}
\end{Corollary}
\bpro
As $ \phi $ is a sheaf-morphism, it suffices to show (\ref{EHCO1})
and (\ref{EHCO2}) only locally for $ U \subset U_{\iota} $ open.
We have:
\begin{eqnarray}
 \lefteqn{
 \big( \phi_U \otimes id \big) \circ \phi_U } \nonumber \\
 & = & \big( \Omega_{\iota} \otimes id \otimes id \big) \circ
 (id \otimes \Delta \otimes id) \circ \big( \Omega_{\iota}^{-1} \otimes id
\big)
 \circ \nonumber \\
 & & \big( \Omega_{\iota} \otimes id \big) \circ
 (id \otimes \Delta) \circ \Omega_{\iota}^{-1} \nonumber \\
 & = & \big( \Omega_{\iota} \otimes id \otimes id \big) \circ
 (id \otimes \Delta \otimes id) \circ
 (id \otimes \Delta) \circ \Omega_{\iota}^{-1} \nonumber \\
 & = & \big( \Omega_{\iota} \otimes id \otimes id \big) \circ
 (id \otimes id \otimes \Delta) \circ
 (id \otimes \Delta) \circ \Omega_{\iota}^{-1} \nonumber \\
 & = & (id \otimes \Delta) \circ \big( \Omega_{\iota} \otimes id \big) \circ
 (id \otimes \Delta) \circ \Omega_{\iota}^{-1} \nonumber \\
 & = & (id \otimes \Delta) \circ \phi_U ,
\end{eqnarray}
\begin{eqnarray}
 \lefteqn{
 (id \otimes \varepsilon) \circ \phi_U } \nonumber \\
 & = & (id \otimes \varepsilon ) \circ \big(\Omega_{\iota} \otimes id \big)
 \circ (id \otimes \Delta) \circ \Omega_{\iota}^{-1} \nonumber \\
 & = & \big( \Omega_{\iota} \otimes id\big) \circ (id \otimes id
 \otimes \varepsilon) \circ (id \otimes \Delta)
 \circ \Omega_{\iota}^{-1} \nonumber \\
 & = & \Omega_{\iota} \circ \Omega_{\iota}^{-1} \nonumber \\
 & = & id .
\end{eqnarray}
Quod erat demonstrandum.
\epro
\begin{Remark}
 The last corollary justifies to call $\phi$ a noncommutative gauge
 transformation of the first kind.
\end{Remark}
For the moment let us suppose again ${\cal P,M}, \varrho ,H $ being defined
by a classical principal bundle $(P,M,\pi,G)$.
The relation $\, \pi (ua) \: = \: \pi (u) \,$ for $ u \in P, \; a \in G $
implies
\begin{equation}
 \phi_U  \circ \varrho_U (f) \: = \: \varrho_U (f) \otimes 1, \hspace{1cm}
 f \in {\cal M}(U),
\end{equation}
because the equations
\begin{equation}
 \phi_U \circ \varrho_U (f) (u,a) \: = \: \phi_U (f \circ \pi ) \, (u,a)
 \: = \: f ( \pi (ua) )
\end{equation}
\begin{equation}
 \varrho_U (f) \otimes 1 (u,a) \: = \: f \circ \pi (u)
\end{equation}
are true. An analogous result holds in the noncommmutative case.
\begin{Theorem}
 Let $\, U \subset M \,$ be open. Then $\, h \in {\cal P}(U) \,$
 satisfies the equation
\begin{equation}
 \phi_U (h) \: = \: h \otimes 1
\end{equation}
 if and only if $\, h \, = \, \varrho_U(f) $ for a $\, f \in {\cal M}(U) \,$.
\end{Theorem}
\bpro
 As $ \varrho $ and $ \phi $ are sheaf-morphism it suffices to assume
 $\, U \subset U_{\iota} \,$.
 Let us first suppose that $\, h \, = \, \varrho_U(f) \,$ for a
 $\, f \in {\cal M}(U) \,$. Then we get with (\ref{DQPB1}):
\begin{eqnarray}
 \lefteqn{
 \phi_U \circ \varrho (f) } \nonumber \\
 & = & \big( \big( \Omega_{\iota} \otimes id \big) \circ (id \otimes \Delta )
\circ
 \Omega_{\iota}^{-1} \big) \, \big( \varrho_U (f) \big) \nonumber \\
 & = & \big( \big( \Omega_{\iota} \otimes id \big) \circ ( id \otimes \Delta )
 \big) \, (f \#_{\iota} 1) \nonumber \\
 & = & \big( \Omega_{\iota} \otimes id \big) \, (f \#_{\iota} 1 \otimes 1)
\nonumber \\
 & = & \varrho_U (f) \otimes 1.
\end{eqnarray}
 This gives one direction of the assertion.
 Now assume $\, \phi_U (h) \, = \, h \otimes 1$ for a $\, h \in {\cal P}(U)
\,$.
 Then equation (\ref{QRCO}) implies:
\begin{eqnarray}
 \lefteqn{
 (id \otimes \varepsilon \otimes id) \, \big( \Omega_{\iota}^{-1} (h)
 \otimes 1 \big) } \nonumber \\
 & = & \big( (id \otimes \varepsilon \otimes id) \circ (id \otimes \Delta)
 \circ \Omega_{\iota}^{-1} \big) \, (h) \nonumber \\
 & = & \Omega_{\iota}^{-1} (h) .
\end{eqnarray}
 As $\, (id \otimes \varepsilon \otimes id) \,
 \big( \Omega_{\iota}^{-1} (h) \otimes 1 \big) \in {\cal M}(U) \, \#_{\iota} \,
 H  \,$,
 the relations (\ref{DQPB1}) and $\, f \, := \, (id \otimes \varepsilon) \circ
 \Omega_{\iota}^{-1} (h) \,$ entail
 $\, h \, = \, \varrho_U (f) $, which gives the other direction.
\epro
\subsection{Transition Functions}
In the following we will derive some basic properties of the
local coordinate changes
$\, \Omega_{\kappa, \iota} \,$. Define the linear mappings
$\, \tau_{\iota, \kappa} :  H
\longrightarrow {\cal M} (U_{\iota} \cap U_{\kappa}) \,$, $\, \iota, \kappa
\in I \,$ by
\begin{equation}
\label{QTRF}
 \tau_{\iota, \kappa} (g) \: := \: (id \otimes \varepsilon )
 \circ \Omega_{\kappa, \iota} (1 \#_{\iota} g),
 \hspace{0.7cm} g \in  H  .
\end{equation}
Then the $\, \Omega_{\kappa, \iota} \,$ can be written in the form:
\begin{eqnarray}
\lefteqn{
 \Omega_{\kappa, \iota}  ( f \#_{\iota} g ) } \nonumber \\
 & = & \Omega_{\kappa, \iota}  ( f \#_{\iota} 1 ) \cdot
 \Omega_{\kappa, \iota}  ( 1 \#_{\iota} g ) \nonumber \\
 & = & ( f \#_{\kappa} 1 ) \cdot \big( ( id \otimes \varepsilon \otimes id )
 \circ ( id \otimes \Delta ) \circ \Omega_{\kappa, \iota} \big)
 ( 1 \#_{\iota} g ) \nonumber \\
 & = & ( f \#_{\kappa} 1 ) \cdot \big( ( id \otimes \varepsilon \otimes id )
 \circ ( \Omega_{\kappa, \iota} \otimes id ) \circ  (id \otimes \Delta )
 \big) ( 1 \#_{\iota} g ) \nonumber \\
 & = & \sum \limits_{ (g) } \, ( f \#_{\kappa} 1 ) \cdot \big( \big(
 r^{U_{\iota} \cap U_{\kappa} }_{U} \circ  \tau_{\iota, \kappa} \big)
 \big( g_{(1)} \big) \, \#_{\kappa} \, g_{(2)} \big) \nonumber \\
\label{NCCCH1}
 & = &
 \sum \limits_{ (g) } \, \big( f  \cdot \big(
 r^{U_{\iota} \cap U_{\kappa} }_{U} \circ \tau_{\iota, \kappa} \big)
 \big( g_{(1)} \big) \big) \, \#_{\kappa} \, g_{(2)},
\end{eqnarray}
where
$\, f \in {\cal M}(U) \,$,
$\, U \subset U_{\iota} \cap U_{\kappa}\,$,
and $\, g \in  H \,$, or in the form
\begin{equation}
\label{NCCCH}
 \Omega_{\kappa, \iota} \: = \: \big( ( m \otimes id ) \circ
 \big( id \otimes \big( r^{U_{\iota} \cap U_{\kappa} }_{U} \circ
 \tau_{\iota, \kappa} \big)
 \otimes id \big) \circ (id \otimes \Delta) \big) .
\end{equation}
Let us show that the linear mappings $\tau_{\iota, \kappa}$ can be
considered as a generalisation of the transition functions in classical
geometry.
Suppose the \QPB $({\cal P,M},\varrho,H,(\Omega_{\iota})_{\iota \in I} )$ is
given by a principal bundle
$(P,M,\pi,G)$ with trivialisations
$\, \psi_{\iota} : \pi^{-1}(U_{\iota}) \longrightarrow U_{\iota} \times G \,$.
By the definition of the $\tau_{\iota , \kappa} $ it is obvious that
they have the form
\bdm
\begin{array}{lcl}
 \tau_{\iota,\kappa} \, : \, H \, \longrightarrow \, {\cal M} ( U_{\iota}
 \cap U_{\kappa} ) , & \quad & g \, \longmapsto \, g \circ \eta_{\iota,
\kappa}.
\end{array}
\edm
where the $\eta_{\iota,\kappa}$ are the classical transition functions defined
by
\begin{equation}
 \psi_{\iota} \circ \psi^{-1}_{\kappa} (x,a) \: = \:
 (x,\eta_{\iota, \kappa}(x)a).
\end{equation}
In the commutative case the $\tau_{\iota, \kappa}$ are morphisms of algebras,
whereas in general they are only linear mappings between algebras.

The $\tau_{\iota, \kappa} $ are not independant from each other but fulfill
certain conditions we will derive in the sequel. Let us first give an important
definition.
\begin{Definition}
\label{COCY}
 Let $ { \cal U } \: = \: (U_{\iota})_{\iota \in I } $ be an open
 covering of $M$, $H$ a Hopf algebra and
 $\cal M $ an $ \bf A $-quantum space over $M$.
 Further let
 $(\tau_{\iota , \kappa})_{(\iota , \kappa ) \in I \times I } $
 be a family of linear mappings
 $\, \tau_{\iota , \kappa } : H \, \longrightarrow \, {\cal M}
 ( U_{\iota} \cap U_{\kappa} ) \,$
 satisfying the following conditions.
\begin{enumerate}
\item
\label{COCY0}
 $ \quad \tau_{\iota,\iota} (1) \: = \: 1 $,
\item
\label{COCY1}
 $ \quad r_U^{U_{\iota} \cap U_{\lambda} } \circ \tau_{\iota , \lambda} \: = \:
 \big( r_U^{U_{\iota} \cap U_{\kappa} } \circ \tau_{\iota, \kappa} \big) *
 \big( r_U^{U_{\kappa} \cap U_{\lambda} } \circ \tau_{\kappa, \lambda} \big) $,
\item
\label{COCY2}
 $ \quad \tau_{\iota , \iota} (g) \: = \: \varepsilon (g) \cdot 1, \hspace{2cm}
 g \in  H $,
\end{enumerate}
where $\, U \, = \, U_{\iota} \cap  U_{\kappa} \cap U_{\lambda} \,$
and the convolution product $*$ is to be formed in the convolution algebra
$Hom( H ,{\cal M}(U))$.
Then $\,  ( \tau_{\iota, \kappa})_{(\iota, \kappa) \in
I \times I} \,$ is called an $H$-cocycle in $\cal M$.\footnote{The cocycles
defined
here are different to the ones in theorem \ref{CRPR}. The context always
makes clear which kind of cocycles are meant so that no confusions can arise.}
\end{Definition}
Now the defining equation (\ref{QTRF}) implies
\begin{equation}
  \tau_{\iota,\kappa} (1) \: = \: 1 .
\end{equation}
Then we have for
$U \subset U_{\iota} \cap U_{\kappa} \cap U_{\lambda} $ open because of
(\ref{NCCCH})
and the definition of the $\Omega_{\iota, \kappa} $:
\begin{eqnarray}
 \lefteqn{
 (m \otimes id) \circ \big( id \otimes \big(
 r_U^{U_{\iota} \cap U_{\lambda} } \circ
 \tau_{\iota , \lambda} \big) \otimes id \big) \circ (id \otimes \Delta)}
 \nonumber \\
 & = & \Omega_{\lambda, \iota} \nonumber \\
 & = & \Omega_{\lambda, \kappa} \circ \Omega_{\kappa, \iota} \nonumber \\
 & = & (m \otimes id) \circ \big( id \otimes
 \big( r_U^{U_{\kappa} \cap U_{\lambda} }
 \circ \tau_{\kappa , \lambda} \big) \otimes id \big) \circ
 (id \otimes \Delta) \circ \nonumber \\
 & & (m \otimes id) \circ \big( id \otimes \big(
 r_U^{U_{\iota} \cap U_{\kappa} } \circ
 \tau_{\iota , \kappa} \big) \otimes id \big) \circ (id \otimes \Delta)
\nonumber \\
 & = & (m \otimes id) \circ (m \otimes id \otimes id) \circ \nonumber \\
 & & \big( id \otimes
 \big( r_U^{U_{\iota} \cap U_{\kappa} } \circ \tau_{\iota , \kappa} \big)
 \otimes
 \big( r_U^{U_{\kappa} \cap U_{\lambda} } \circ \tau_{\kappa , \lambda} \big)
 \otimes
 id \big) \circ \nonumber \\
 & & (id \otimes id \otimes \Delta) \circ  (id \otimes \Delta) \nonumber \\
 & = & (m \otimes id) \circ (id \otimes m \otimes id) \circ \nonumber \\
 & & \big( id \otimes
 \big( r_U^{U_{\iota} \cap U_{\kappa} } \circ \tau_{\iota , \kappa} \big)
 \otimes
 \big( r_U^{U_{\kappa} \cap U_{\lambda} } \circ \tau_{\kappa , \lambda} \big)
 \otimes
 id \big) \circ \nonumber \\
 & & (id \otimes \Delta \otimes id) \circ  (id \otimes \Delta) .
\end{eqnarray}
This relation entails for $g \in H$:
\begin{eqnarray}
 \lefteqn{
 \big( r_U^{U_{\iota} \cap U_{\lambda} } \circ \tau_{\iota , \lambda}
 \big) \, (g) } \nonumber \\
 & = & \big( (id \otimes \varepsilon) \circ (m \otimes id) \circ \big( id
 \otimes \big(
 r_U^{U_{\iota} \cap U_{\lambda} } \circ \tau_{\iota , \lambda} \big)
 \otimes id \big) \circ (id \otimes \Delta) \big) \, (1 \otimes g) \nonumber \\
 & = & \big( (id \otimes \varepsilon) \circ
 (m \otimes id) \circ (id \otimes m \otimes id) \circ \nonumber \\
 & & \big( id \otimes
 \big( r_U^{U_{\iota} \cap U_{\kappa} } \circ \tau_{\iota , \kappa} \big)
 \otimes \big(
 r_U^{U_{\kappa} \cap U_{\lambda} } \circ \tau_{\kappa , \lambda} \big)
 \otimes id \big) \circ \nonumber \\
 & & (id \otimes \Delta \otimes id) \circ  (id \otimes \Delta) \big) \,
 (1 \otimes g) \nonumber \\
 & = & \big( m \circ \big(
 \big( r_U^{U_{\iota} \cap U_{\kappa} } \circ \tau_{\iota , \kappa} \big)
 \otimes \big(
 r_U^{U_{\kappa} \cap U_{\lambda} } \circ \tau_{\kappa , \lambda} \big) \big)
 \circ \Delta \big) \, (g).
\end{eqnarray}
Finally (\ref{DQPB2}) gives for $\, \iota = \kappa \, $ and
$\, g \in H \,$ the equation:
\begin{equation}
 \tau_{\iota , \iota} (g) \: = \: \varepsilon (g) \cdot 1.
\end{equation}
Let us subsume the last results in a proposition.
\begin{Proposition}
\label{CTRFU}
 The transition functions $\, \tau_{\iota , \kappa} : H \, \longrightarrow \,
 {\cal M} (U_{\iota} \cap U_{\kappa}) \, $ of a \QPB
 $({\cal P,M},\varrho,H,(\Omega_{\iota})_{\iota \in I} )$
 form an $H$-cocycle in ${\cal M}$.
\end{Proposition}
The transition functions $ \tau_{\iota , \kappa} $ characterise the (quantum)
principal
bundle in the commutative as well as in the noncommutative case.
We will show how to construct a \QPB out of a family
$(\tau_{\iota , \kappa})_{(\iota , \kappa ) \in I \times I } $ of
transition functions fulfilling the cocycle conditions in definition
\ref{COCY}.
Let us further suppose we are given a family
$(\alpha_{\iota } )_{\iota \in I} $
of weak actions
$\, \alpha_{\iota} : {\cal M}(U_{\iota}) \times  H \, \longrightarrow \,
 {\cal M}(U_{\iota}) \,$
and a family $(\iota )_{\iota \in I} $ of normal cocycles
$\, \iota : H \times H \, \longrightarrow \, {\cal M} \big( U_{\iota} \big) \,$
fulfilling the twisted module condition\footnote{See the appendix for further
details on
weak actions and normal cocycles.}
According to appendix B, theorem \ref{CRPR}
the crossed products $\, {\cal M}(U) \#_{\iota} H \,$
exist for all $\, U \subset U_{\iota} \,$ open.
It is assumed now that
the linear mappings
\bdm
\begin{array}{lclcl}
 \lefteqn{ (m \otimes id) \circ \left( id \otimes \tau_{\iota , \kappa}
 \otimes id \right)
 \circ (id \otimes \Delta) : } \\
 & & {\cal M}(U_{\iota} \cap U_{\kappa}) \#_{\iota}   H &
 \longrightarrow &
 {\cal M}(U_{\iota} \cap U_{\kappa}) \#_{\kappa} H
\end{array}
\edm
are morphisms of algebras with unity.
\newline \noindent
To construct the desired \QPB we consider the algebras
\begin{eqnarray}
\label{AlgPN}
\begin{array}{lcl}
 {\cal P}_0 (U) & = & \bigoplus \limits_{ \iota \in I } \;
 {\cal M} ( U_{\iota} \cap U ) \#_{\iota}  H
\end{array}
\end{eqnarray}
for all $\, U \subset M \,$ open and their subalgebras
\begin{eqnarray}
\label{AlgP}
\begin{array}{lcl}
 {\cal P } (U) & = &
 \Big{\{} \sum \limits_{ \iota \in I } \: f_{\iota } \in
 {\cal P }_0 (U)  \, : \,
 \forall \iota, \kappa \in I  \:
 \big( r_{ U_{\iota } \cap U_{\kappa } \cap U }^{ U_{\iota } \cap U } \otimes
 id \big) \, \big( f_{\iota } \big) \: = \\
 & & \big( ( m \otimes id )
 \circ \big( id \otimes \big(
 r_{ U_{\iota } \cap U_{\kappa } \cap U }^{ U_{\iota } \cap U_{\kappa } }
 \circ \tau_{\iota , \kappa } \big) \otimes id \big) \circ \\
 & & \circ ( id \otimes \Delta )
 \circ \big( r_{ U_{\iota } \cap U_{\kappa } \cap U }^{ U_{\kappa } \cap U }
 \otimes id \big) \big) \, \big( f_{\kappa} \big) \, \Big{\}}.
\end{array}
\end{eqnarray}
Obviously $\, U \longrightarrow {\cal P}_0 (U) \,$ defines a sheaf
$ {\cal P}_0 $ on $ M $ and $\, U \longrightarrow {\cal P} (U) \,$
a subsheaf $\, \cal P $.
The next lemma helps to characterise the sheaf $ \cal P $.
\begin{Lemma}
\label{CHSH}
 Suppose $ \sum \limits_{ \iota \in I } \: f_{\iota } \in {\cal P}_0 (U) $ and
 $ U \subset U_{\kappa} $ open.
 Then $ \sum \limits_{ \iota \in I } \: f_{\iota } \in {\cal P} (U) $
 if and only if for all $ \iota \in I $ the equation
\begin{eqnarray}
\label{TRANS}
\begin{array}{lcl}
 \big( r_{ U_{\iota } \cap U_{\kappa } \cap U }^{ U_{\iota } \cap U } \otimes
 id \big) \, ( f_{\iota } ) & = &
 \big( ( m \otimes id ) \circ \big( id \otimes \big(
 r_{ U_{\iota } \cap U_{\kappa } \cap U }^{ U_{\iota } \cap U_{\kappa } }
 \circ
 \tau_{\iota , \kappa } \big)
 \otimes id \big) \circ \\
 & & \circ ( id \otimes \Delta ) \circ
 \big( r_{ U_{\iota } \cap U_{\kappa } \cap U }^{ U_{\kappa } \cap U }
 \otimes id \big) \big) \, ( f_{\kappa} )
\end{array}
\end{eqnarray}
is satisfied.
\end{Lemma}
\bpro
Let $ \sum \limits_{ \iota \in I } \: f_{\iota } \in {\cal P}_0 (U) $
fulfill the relation (\ref{TRANS}).
According to definition \ref{COCY} (2) we have for all
$ \iota , \lambda , \mu \in I $:
\begin{eqnarray}
 \label{COCYMU}
 \lefteqn{
 \big( ( m \otimes id ) \circ \big( id \otimes \big(
 r_{ U_{\iota } \cap U_{\lambda } \cap U }^{ U_{\iota } \cap U_{\lambda } }
 \circ \tau_{\iota , \lambda } \big) \otimes id \big)
 \circ ( id \otimes \Delta ) \big)
 \circ } \nonumber \\
 & & \big( ( m \otimes id ) \circ \big( id \otimes \big(
 r_{ U_{\lambda } \cap U_{\mu } \cap U }^{ U_{\lambda } \cap U_{\mu } }
 \circ \tau_{\lambda , \mu } \big) \otimes id \big)
 \circ ( id \otimes \Delta ) \big) \nonumber \\
 & = & \big( ( m \otimes id ) \circ ( id \otimes m \otimes id )
 \circ \nonumber \\
 & & \big( id \otimes \big(
 r_{ U_{\iota } \cap U_{\lambda } \cap U }^{ U_{\iota } \cap U_{\lambda } }
 \circ \tau_{\iota , \lambda } \big) \otimes
 \big( r_{ U_{\lambda } \cap U_{\mu } \cap U }^{ U_{\lambda } \cap U_{\mu } }
 \circ \tau_{\lambda , \mu } \big) \otimes id \big) \circ \nonumber \\
 & & ( id \otimes \Delta \otimes id ) \circ
 ( id \otimes \Delta ) \big) \nonumber \\
 & = & \big( ( m \otimes id ) \circ \big( id \otimes \big(
 r_{ U_{\iota } \cap U_{\mu } \cap U }^{ U_{\iota } \cap U_{\mu } }
 \circ \tau_{\iota , \mu } \big) \otimes id \big)
 \circ ( id \otimes \Delta ) \big) .
\end{eqnarray}
This equation as well as (\ref{TRANS}) und \ref{COCY} (2) imply
\begin{eqnarray}
 \lefteqn{
 \big( r_{ U_{\kappa } \cap U_{\lambda } \cap U }^{ U_{\kappa } \cap U }
\otimes
 id \big) \, \big( f_{\kappa } \big) } \nonumber \\
 & = & \big( ( m \otimes id ) \circ \big( id \otimes \big(
 r_{ U_{\kappa } \cap U_{\lambda } \cap U }^{ U_{\kappa } \cap U_{\lambda } }
\circ
 \tau_{\kappa , \lambda } \big) \otimes id \big)
 \circ ( id \otimes \Delta ) \circ \nonumber \\
 & & \big( r_{ U_{\kappa } \cap U_{\lambda } \cap U }^{ U_{\lambda } \cap U }
 \otimes id \big) \big) \, \big( f_{\kappa} \big) .
\end{eqnarray}
With (\ref{COCYMU}) one derives the relation
\begin{eqnarray}
 \lefteqn{
 \big( r_{ U_{\iota } \cap U_{\lambda } \cap U }^{ U_{\iota } \cap U } \otimes
 id \big) \, \big( f_{\iota } \big) } \nonumber \\
 & = &
 \big( ( m \otimes id ) \circ \big( id \otimes \big(
 r_{ U_{\iota } \cap U_{\lambda } \cap U }^{ U_{\iota } \cap U_{\kappa } }
\circ
 \tau_{\iota , \kappa } \big)
 \otimes id \big) \circ ( id \otimes \Delta ) \circ \nonumber \\
 & & \big( r_{ U_{\iota } \cap U_{\lambda } \cap U }^{ U_{\kappa } \cap U }
 \otimes id \big) \big) \, \big( f_{\kappa} \big) \nonumber \\
 & = & \big( ( m \otimes id ) \circ \big( id \otimes \big(
 r_{ U_{\iota } \cap U_{\lambda } \cap U }^{ U_{\iota } \cap U_{\kappa } }
\circ
 \tau_{\iota , \kappa } \big) \otimes id \big) \circ ( id \otimes \Delta )
 \circ \nonumber \\
 & & ( m \otimes id ) \circ \big( id \otimes \big(
 r_{ U_{\iota } \cap U_{\lambda } \cap U }^{ U_{\iota } \cap U_{\kappa } }
\circ
 \tau_{\kappa , \lambda } \big) \otimes id \big) \circ ( id \otimes \Delta )
 \circ \nonumber \\
 & & \big( r_{U_{\iota } \cap U_{\lambda } \cap U }^{U_{\lambda } \cap U }
 \otimes id \big) \big) \, \big( f_{\lambda } \big) \nonumber \\
 & = & \big( ( m \otimes id ) \circ \big( id \otimes \big(
 r_{ U_{\iota } \cap U_{\lambda } \cap U }^{ U_{\iota } \cap U_{\lambda } }
\circ
 \tau_{\iota , \lambda } \big) \otimes id \big) \circ ( id \otimes \Delta )
 \circ \nonumber \\
 & & \big( r_{ U_{\iota } \cap U_{\lambda } \cap U }^{ U_{\lambda }
 \cap U } \otimes id \big) \big) \, \big( f_{\lambda } \big).
\end{eqnarray}
Therefore $ \sum \limits_{ \iota \in I } \: f_{ \iota }  \in {\cal P}(U) $ and
one part of the assertation is proven. The other one is trivial.
\epro
We have to supply sheaf-morphisms
$\, \varrho : {\cal M } \longrightarrow {\cal P } \,$ and
$\, \Omega_{\iota} : {\cal M } \! \mid_{U_{\iota }} \#_{\iota} {\cal F }_G
\longrightarrow {\cal P } \! \mid_{U_{\iota }} \,$.
As for all $\, f \in {\cal M} (U) \,$, $\, U \in M \, $ open the sum
$ \sum \limits_{ \iota \in I }
\: r_{U_{\iota } \cap U }^{ U } (f) \#_{\iota} 1 $  lies in
$\, {\cal P } (U) \,$, we can set
\bdm
\begin{array}{lclclclcl}
 \varrho & : & {\cal M } & \longrightarrow & {\cal P } , & & & & \\
 \varrho_U & : & {\cal M } (U) & \longrightarrow & {\cal P } (U) , & \quad &
 f & \longmapsto & \sum \limits_{ \iota \in I }
 \: r_{U_{\iota } \cap U }^{ U } (f) \#_{\iota} 1 .
\end{array}
\edm
The mapping
$\, \Omega_{\iota} : {\cal M } \! \mid_{U_{\iota }} \#_{\iota} {\cal F }_G
\, \longrightarrow \, {\cal P } \! \mid_{U_{\iota }} \,$ shall be
given by
\bdm
\begin{array}{lclclclclc}
 \Omega_{\iota} & : & {\cal M } \! \mid_{U_{\iota }} \#_{\iota} {\cal F }_G &
 \longrightarrow & {\cal P } \! \mid_{U_{\iota }} & & & & & \\
 (\Omega_{\iota})_U & : & {\cal M } (U) \#_{\iota} {\cal F }_G &
 \longrightarrow & {\cal P } (U) , & \quad &
 f & \longmapsto & \sum \limits_{ \iota \in I } \: f_{\iota } , &
 U \subset U_{\iota },
\end{array}
\edm
where
\begin{eqnarray}
 \lefteqn{
 f_{\kappa } \: = \: \big( ( m \otimes id ) \circ } \nonumber \\
 \label{DSUM}
 & & \circ \big( id \otimes \big(
 r_{ U_{\iota } \cap U_{\kappa } \cap U }^{U_{\iota } \cap U_{\kappa } }
 \circ \tau_{\iota , \kappa } \big) \otimes id \big)
 \circ ( id \otimes \Delta )
 \circ \big( r_{ U_{\iota } \cap U_{\kappa } \cap U }^{ U_{\iota } \cap U }
 \otimes id \big) \big) \, ( f ).
\end{eqnarray}
We have in particular $\, f_{\iota } \: = \: f \,$. Now lemma \ref{CHSH}
shows that $ \Omega_{\iota } $ is welldefined.
By the definition of $\varrho$ and $\Omega_{\iota}$
it is clear that the equation (\ref{DQPB1})
$\, {( \Omega_{\iota } ) }_U (f \otimes 1 ) \, = \, \varrho_U (f) \,$
for $\, f \in {\cal M} (U) \,$, $\, \quad U \subset U_{\iota } \,$ is
satisfied.
If we can proof $ \Omega_{\iota } $ being bijective, our considerations
show that $ \varrho_U $ is injective for $ U \subset U_{\iota } $.
This will give the exactness of the sequence
\bdm
 0 \longrightarrow {\cal M} \stackrel{\varrho }{\longrightarrow} {\cal P} .
\edm
Therefore it has to be proven that $ \Omega_{\iota } $ is an isomorphism
which satisfies (\ref{DQPB2}).
Define for $ U \subset U_{\iota } $ open:
\bdm
\begin{array}{lclclclcl}
 \Xi_{\iota } & : & {\cal P} \! \mid_{U_{\iota }} & \longrightarrow &
 {\cal M} \! \mid_{U_{\iota }} \#_{\iota} H & & & &\\
 (\Xi_{\iota })_U & : & {\cal P}(U) & \longrightarrow &
 {\cal M}(U) \#_{\iota}  H ,& \quad &
 \sum \limits_{ \kappa \in I } \: f_{\kappa } & \longmapsto & f_{\iota }.
\end{array}
\edm
Then it is easy to see
\begin{equation}
 \Xi_{\iota } \circ \Omega_{\iota } \: = \: id \quad
 \mbox{und} \quad
 \Omega_{\iota } \circ \Xi_{\iota } \: = \: id ,
\end{equation}
that is $\Omega_{\iota}$ is a sheaf-isomorphism with inverse $\Xi_{\iota}$.
Further we get (\ref{DQPB2})
\begin{equation}
( \Omega_{\iota} \otimes id) \circ (id \otimes \Delta) \: = \:
(id \otimes \Delta) \circ \Omega_{\iota},
\end{equation}
which follows from the definition of $ \Omega_{\iota}$ and the
coassociativity in $H$. Now we can state the final theorem.
\begin{Theorem}
\label{CQPBTRF}
 Let $\, {\cal U } \: = \: ( U_{\iota } )_{\iota \in I } \,$ be an open
 covering of $M$, $H$ a Hopf algebra, ${\cal M}$ an $\bf A $-quantum space
 over $M$, and $ (\tau_{\iota , \kappa })_{(\iota , \kappa)
 \in I \times I} $ an $H$-cocycle in $\cal M$.
 Further let $\, \alpha_{\iota} : {\cal M}(U_{\iota}) \times  H \,
 \longrightarrow \, {\cal M}(U_{\iota}) \,$
 be weak actions and
 $\, \iota : H \times H \, \longrightarrow
 \, {\cal M} \big( U_{\iota} \big) \,$
 normal cocycles fulfilling the twisted module condition. The linear
 mappings
\bdm
\begin{array}{lclcl}
 \lefteqn{ (m \otimes id) \circ \left( id \otimes \tau_{\iota , \kappa}
 \otimes id \right)
 \circ (id \otimes \Delta) : } \\
 & & {\cal M}(U_{\iota} \cap U_{\kappa}) \#_{\iota} H &
 \longrightarrow &
 {\cal M}(U_{\iota} \cap U_{\kappa}) \#_{\kappa} H
\end{array}
\edm
are supposed to be morphisms of algebras with unity.
Then there exists an
$\bf A $-\QPB $({\cal P,M},\varrho,H,(\Omega_{\iota})_{\iota \in I})$
over $M$ uniquely defined up to isomorphism
such that the $ \tau_{\iota ,\kappa } $
are its transition functions or in other words such that
\begin{eqnarray}
\label{CCCTRF}
\begin{array}{lcl}
 \big( \Omega_{\kappa, \iota} \big)_U & = & ( m \otimes id ) \circ
 \big( id \otimes \big(
 r_{ U_{\iota } \cap U_{\kappa } \cap U }^{U_{\iota } \cap U_{\kappa } }
 \circ \tau_{\iota , \kappa } \big) \otimes id \big) \circ \\
 & & \circ ( id \otimes \Delta )
 \circ \big( r_{ U_{\iota } \cap U_{\kappa } \cap U }^{ U_{\iota } \cap U }
 \otimes id \big)
\end{array}
\end{eqnarray}
is satisfied for $\, U \subset U_{\iota} \cap U_{\kappa} \,$ open.
\end{Theorem}
\bpro
Most of the theorem has been proven above, but we still have to show
(\ref{CCCTRF}). Let $\, U \subset U_{\iota } \cap U_{\kappa } \,$ be open,
$\, f \in {\cal M } (U) \#_{\iota} H \,$,
$\, g \in {\cal M } (U) \#_{\kappa} H \,$ and
$\, g \, = \, \big( \big( \Omega_{\kappa, \iota } \big)_U \big) \, (f) \,$.
Then (\ref{DSUM}) entails
\begin{eqnarray}
 \lefteqn{
 g \: = \: g_{\kappa } \: = \: pr_{\kappa } \big( ( \Omega_{\kappa } )_U
 (g) \big) \: = \: pr_{\kappa } \big( ( \Omega_{\iota } )_U (f) \big)
 \: = \: f_{\kappa } \: = \: \big( ( m \otimes id ) \circ } \nonumber \\
 & & \big( id \otimes \big(
 r_{ U_{\iota } \cap U_{\kappa } \cap U }^{U_{\iota } \cap U_{\kappa } }
 \circ \tau_{\iota , \kappa } \big) \otimes id \big)
 \circ ( id \otimes \Delta )
 \circ \big( r_{ U_{\iota } \cap U_{\kappa } \cap U }^{ U_{\iota } \cap U }
 \otimes id \big) \big) \, ( f ) ,
\end{eqnarray}
 which gives the desired equation. The statement about the uniqueness
 of the \QPB up to isomorphism is clear by definition.
\epro
\section{Quantum Vector Bundles}
\subsection{Definition and Examples}
We can also translate vector bundles in the language of quantum spaces. As
typical
fibres we use quadratic algebras which according to Manin \cite{man-1} are
considered as the
noncommutative linear spaces. Like in the case of quantum groups
the multiplication on the tensor products serving as the local
trivialisations has to be defined by the methods in appendix B.

Suppose we are given the following objects:
\begin{enumerate}
\item
 a sheaf ${\cal M}$ over a topological space $M$ with objects in the category
$\bf A$ called
 the base quantum space,
\item
 a sheaf ${\cal V}$ over $M$ with objects in $\bf A$ called the total quantum
space,
\item
 a sheaf morphism $\, \varrho : {\cal M} \longmapsto {\cal V} $
 called the projection,
\item
 a quadratic algebra $A$ called the typical fibre,
\item
 a Hopf-algebra $H$ called the structure quantum group,
\item
 a coaction $\, \varphi : A \longmapsto H \otimes A \,$,
\item
 a family $\, ( \Gamma_{\iota} )_{\iota \in I} \,$ of
 sheaf morphisms $\, \Gamma_{\iota} : {\cal M} \! \mid_{U_\iota} \,
 \#_{\iota } \, A  \longrightarrow  {\cal V} \! \mid_{U_\iota} \,$, where
 $\, {\cal U} \, = \, (U_{\iota})_{\iota \in I} \, $ is an open covering of $M$
and
 $\#_{\iota}$ is a crossed product which is given according
 to theorem \ref{SMPRKA} by a weak action
 $\, \alpha_{\iota} : H \times {\cal M} \! \mid_{U_{\iota}} \longrightarrow
 {\cal M} \! \mid_{U_{\iota}} \,$, a normal cocycle
 $\, \iota : H \times H \longrightarrow {\cal M}
 \big( U_{\iota} \big) \,$ fulfilling the twisted module condition and
 the coaction $ \varphi $.
\end{enumerate}
$( {\cal V,M},\varrho,A,H,\varphi,(\Gamma_{\iota})_{\iota \in I})$ gives the
data
of an $\bf A$-quantum vector bundle over $M$.
\begin{Definition}
\label{DQVB}
The tupel $( {\cal V,M},\varrho,A,H,\varphi,(\Gamma_{\iota})_{\iota \in I})$
is said to be an $\bf A$-quantum vector bundle with coordinate system
$\, (\Gamma_{\iota})_{\iota \in I} \,$, if the following conditions hold:
\begin{enumerate}
\item
 The sequence $\, 0 \longrightarrow {\cal M}
 \stackrel{\varrho }{\longrightarrow} {\cal V} \,$ is exact.
\item
 The algebras ${\cal M}(U)$ and ${\cal V}(U)$ are unitary for
 $\, U \subset U_{\iota} \,$ open.
\item
 Let the sheaf morphisms
 $\, \Gamma_{\kappa , \iota } : {\cal M} \! \mid_{U_{\iota} \cap U_{\kappa}}
 \, \#_{\iota } \, A  \longrightarrow {\cal M} \! \mid_{U_{\iota}
 \cap U_{\kappa}} \, \#_{\kappa } \, A \,$ be defined by
 $\, \big( \Gamma_{\kappa , \iota } \big)_U \, = \,
 \big( \Gamma_{\kappa } \big)_U^{-1} \circ \big( \Gamma_{\iota } \big)_U \,$,
 where $\, U \subset U_{\iota} \cap U_{\kappa} \,$ open.
 Then one can find linear mappings
 $\,\tau_{\iota, \kappa}  :  H  \longrightarrow {\cal M}(U_{\iota} \cap
U_{\kappa}) \,$,
 $\, {\iota},{\kappa} \in I \,$ such that
 the following equations hold:
\begin{equation}
\label{DQVB1}
  \lefteqn{\left( \Gamma_{\iota} \right)_U (f \#_{\iota } 1) \, = \,
  \varrho_U(f) \hspace{1cm} f \in {\cal M}, \: U \subset U_{\iota}}
\end{equation}
\begin{eqnarray}
\label{DQVB2}
 \lefteqn{
 \big( \Gamma_{\kappa , \iota } \big)_U  \, = \,
 (m \otimes id) \circ ( id \otimes \big( r^{U_{\iota} \cap U_{\kappa}}_U
 \circ
 \tau_{ \iota, \kappa } \big) \otimes id ) \circ (id \otimes \varphi)} \\
\nonumber & & \hfill U \subset U_{\iota} \cap U_{\kappa}.
\end{eqnarray}
\end{enumerate}
\end{Definition}
\begin{Remark}
Equation $(\ref{DQVB2})$ can also be written in the form
\begin{equation}
 \Gamma_{\kappa, \iota} (f \#_{\iota} g) \, = \,
 \sum \limits_{ (g) } \: f \, \tau_{\iota,\kappa}(g_{(-1)}) \otimes g_{(0)},
\end{equation}
 where $ f \in {\cal M}(U) $, $ g \in A $ and $ U \subset
 U_{\iota} \cap U_{\kappa} \cap U_{\lambda} $.
\end{Remark}
\begin{Remark}
The tupel
$ ({\cal V,M},\varrho ,A ,H, \varphi,(\Gamma_{\iota})_{\iota \in I} ) $
should better be defined as a \QVB with coordinate system
similarly like in the case of quantum principal
bundles $($see definition $\ref{DQPB})$.
Quantum vector bundles were equivalence classes of quantum vector bundles
with coordinate system. But this procedure does not give new aspects, and the
technical details are analog to the ones in the definition of
qauntum principal bundles.
\end{Remark}
The transition functions $\tau_{\iota, \kappa}$ are not independant from each
other.
\begin{Proposition}
The linear mappings
\begin{eqnarray*}
\begin{array}{lclcl}
 \tau_{\iota, \kappa} & : & H & \longrightarrow &
 {\cal M}(U_{\iota} \cap U_{\kappa}),
 \hspace{1cm} {\iota},{\kappa} \in I
\end{array}
\end{eqnarray*}
form an $H$-cocycle in $\cal M$ over $(U_{\iota})_{\iota \in I}$.
\end{Proposition}
\bpro
 This can be shown exactly like in proposition \ref{CTRFU}.
\epro
We state the next theorem but postpone the proof till we introduce
noncommutative
associated bundles.
\begin{Theorem}
\label{KQVB}
 Let
 ${ \big( {\tau}_{(\iota ,\kappa)} \big) }_{(\iota ,\kappa) \in I \times I} $
 be an $H$-cocycle in $ \cal M $, and
 $\  \varphi : A \longrightarrow H \otimes A \,$ a left coaction on the
 quadratic algebra $A$. Further suppose that the mappings
 $\, \alpha_{\iota} : H \times {\cal M} \! \mid_{U_{\iota}} \longrightarrow
 {\cal M} \! \mid_{U_{\iota}} \,$ are actions and the cocycles
 $\, \iota : H \times H \longrightarrow {\cal M}
 \big( U_{\iota} \big) \,$ are trivial that means
 $\, \iota (h,l) \, = \, \epsilon (h) \, \epsilon (l) \,$ for $\, 1 \quad h,l
\in H \,$.
 Then there exists a \QVB which has $A$ as its typical fibre, $H$ as its
structure
 group and the $\tau_{\iota , \kappa}$ as transition functions.
\end{Theorem}
Classical vector bundles are natural examples of quantum vector bundles
as will be shown in the following.

Let $ \pi : E \longrightarrow M $
be a real vector bundle of dimension $n$ over the topological
space $M$, where the structure group $ G \subset Gl(n, {\Bbb{R}}) $ is compact.
Now define the following objects
\footnote{
 Here we use a slight topological generalisation of our concept,
 but dont want to go deeper in the subject at the moment.}:
\begin{enumerate}
\item
 $ \cal M $ is the sheaf of continuous bounded ${\Bbb{C}}$-functions on $M$.
\item $\cal V$ is the sheaf on $M$ defined by
\bdm
\begin{array}{lcll}
 U & \longrightarrow & {\cal V} (U), & \hspace{0.6cm} U \subset M
 \mbox{ offen}, \\
 i^U_V & \longrightarrow & {\cal V} \big(  i^U_V \big),  &
 \hspace{0.6cm} V \subset U \subset M \mbox{ open}.
\end{array}
\edm
 Here $ {\cal V} (U) $ is the algebra of complex continuous bounded
 functions on $\pi^{-1} (U) $ and ${\cal V} \big( i^U_V \big) $
 the restriction from $ \pi^{-1} (U) $ to $ \pi^{-1} (V) $.
\item
 $\varrho$ is the sheaf morphism
\bdm
\begin{array}{lclcllcl}
 \pi^* & : & {\cal M} & \longrightarrow & {\cal V}, & & & \\
 \pi^*_U & : & {\cal M}(U) & \longrightarrow & {\cal V}(U), &
 f & \longmapsto & f \circ \pi \! \mid_U, \hspace{0.7cm} U \subset M.
\end{array}
\edm
\item
 $A$ is the quadratic algebra of complex polynomials in n variables
 $x_1 \, , ... , \, x_n \,$, where the $\, x_i \,$
 are the coordinate projections of ${\Bbb{R}}^n $.
 Further let $\tilde{A} $ be the $*$-Fr\'echet algebra of complex
 continuous functions on ${\Bbb{R}}^n$. Then $A$ lies densly in $\tilde{A}$.
\item
 $G$ gives the (topological) Hopf algebra $H$ of continuous functions on $G$.
\item
 $ \varphi$ is dual to the action $\, G \times {\Bbb{R}}^n
 \longrightarrow {\Bbb{R}}^n \,$, that means $\varphi$ is the coaction
\bdm
\begin{array}{lclclcl}
 \tilde{A} & \longrightarrow & H \otimes {\tilde{A}} & & & & \\
 f & \longmapsto & \sum \limits_{(f)} \, f_{(-1)} \otimes f_{(0)}
 & = & ( \, (a,v) & \longmapsto & f(av) \, ).
\end{array}
\edm
\item
 Let
 $\big( U_{\iota} \big)_{\iota \in I} $ be an open covering
 of $ M $ such that trivialisations
 $\, \psi_{\iota} : E \! \mid_{U_{\iota}} \longrightarrow U_{\iota} \times
 {\Bbb{R}}^n \,$ exist. These induce sheaf isomorphisms
\begin{eqnarray*}
\begin{array}{lclcl}
 \Gamma_{\iota} & : & {\cal M} \! \mid_{U_\iota} \, \#_{\iota } \, A &
 \longrightarrow & {\cal V} \! \mid_{U_\iota} , \\
 & & f \otimes g & \longmapsto & (f \otimes g) \circ \psi_{\iota} \!
 \mid_{U_{\iota}},
\end{array}
\end{eqnarray*}
where $ g \in \tilde{A}$, $f \in {\cal M}(U)$, and $ U \subset U_{\iota}$ open.
\end{enumerate}
Obviously the above defined objects give rise to the following example.
\begin{Example}
The tupel $( {\cal V,M},\varrho, A , H , \varphi ,
(\Gamma_{\iota})_{\iota \in I}) $ is a quantum vector bundle over $M$.
\end{Example}
\begin{Example}
 Let $\cal M$ be an arbitrary quantum space over $M$, $H$ a Hopf algebra,
 and $A$ a quadratic $H$-left comodule algebra with coaction $\varphi$.
 Then
\bdm
\begin{array}{lcl}
 {\cal V} & = &  {\cal M} \otimes A,
\end{array}
\edm
\vspace{-1,45cm} \\
\bdm
\begin{array}{lclcl}
 \varrho & : & {\cal M} & \longrightarrow & {\cal V}, \\
 & & f & \longmapsto & f \otimes 1,
\end{array}
\edm
 gives a trivial
 quantum vector bundle $({\cal V,M},\varrho,A,H,\varphi, (id))$.
\end{Example}
\subsection{Associated Quantum Vector Bundles}
One of the most import tools in the geometry of fibre bundles are the
associated vector bundles. They are used in physics as well. More
precisely do material fields
live in vector bundles which are associated to a principal bundle describing
the gauge transformations. Because of their importance in geometry and physics
we want to translate associated vector bundles to the quantum language.
To find the right definition we will first examine the classical analogon
and then dualise the classical objects and relations.

In classical geometry one first forms the cartesian product
$P \times V$, where $P$ is a principal bundle over a topological
space $M$ and $V$ a vector space on which the structure group $G$ of the
principal bundle acts from the left.
Now
\begin{equation}
\begin{array}{lcl}
 (P \times V) \times G & \longrightarrow & P \times V \\
 \big( (a,v) \, , \,  g \big) &
 \longmapsto & (ag \, , \, g^{-1} \, v), \hspace{0.7cm}
 a \in P, \quad v \in V, \quad g \in G.
\end{array}
\end{equation}
defines a right $G$-action on $P \times V$.

In the noncommutative case we have a quantum principal bundle
${\cal P}$ over $\cal M$ with coaction
$ \phi : {\cal P \longrightarrow P} \otimes H $,
a quadratic algebra $A$ and a left coaction
$\varphi : A \longrightarrow H \otimes A $. The left coaction is supposed
to be a morphism of algebras.
Now one can construct a morphism of sheaves of complex vector spaces
$\, \psi :  A \otimes {\cal P}  \longrightarrow  A \otimes {\cal P} \otimes H
\,$
by
\begin{equation}
 f \otimes g \: \longmapsto \: \psi_U ( f \otimes g) \, = \,
 \sum \limits_{(f),(g)} f_{(0)} \otimes
 g_{(0)} \otimes \big( S^{-1} f_{(-1)} \big) \, g_{(1)},
\end{equation}
where $ f\in A $, $ g \in {\cal P}(U) $ and $ U \subset M $ open.
Furthermore we used the notation
\begin{equation}
 \lefteqn{
 \phi (g) \: = \: \sum \limits_{(g)} g_{(0)} \otimes g_{(1)} \quad \mbox{ and }
 \varphi (f) \: = \: \sum \limits_{(f)} f_{(-1)} \otimes f_{(0)} } .
\end{equation}
The $\psi$ is the noncommutative analogon to the above
$G$-action on $P \times V$.
\begin{Lemma}
 $ \psi_U : A \otimes {\cal P}(U)
 \longrightarrow A \otimes {\cal P}(U) \otimes H $
 is a right $H$-coaction for all $U \subset M$ open, that is
\begin{eqnarray}
\label{ARKW1}
 \lefteqn{ (id \otimes \Delta) \circ \psi_U \: = \: (\psi_U \otimes id )
 \circ \psi_U } \\
\label{ARKW2}
 \lefteqn{ ( id \otimes id \otimes \epsilon) \circ \psi_U \: = \: id. }
\end{eqnarray}
\end{Lemma}
\bpro
Let us first show equation (\ref{ARKW1}). On the one hand the relation
\begin{eqnarray}
\lefteqn{ ( \psi_U \otimes id ) \circ \psi_U \; (f \otimes g) } \nonumber \\
 & = & ( \psi_U \otimes id) \sum \limits_{(f),(g)}
 f_{(0)} \otimes g_{(0)} \otimes \big( S^{-1} f_{(-1)} \big) \, g_{(1)}
\nonumber \\
 & = & \sum \limits_{(f),(g)}
 f_{(0)} \otimes g_{(0)} \otimes \big( S^{-1} f_{(-1)} \big) \, g_{(1)}
 \otimes \big( S^{-1} f_{(-2)} \big) \, g_{(2)},
\end{eqnarray}
is true. On the other hand we have with the flip operator $\tau$
\begin{eqnarray}
 \lefteqn{ \Delta \big( S^{-1} f_{(-1)} \, g_{(1)} \big) } \nonumber \\
 & = & \Delta \big( S^{-1} f_{(-1)} \big) \cdot \Delta \big( g_{(1)} \big)
\nonumber \\
 & = & \big( \big( S^{-1} \otimes S^{-1} \big) \circ \tau \circ
       \Delta \big( f_{(-1)} \big) \big) \cdot \Delta
       \big( g_{(1)} \big) \nonumber \\
 & = & \big( S^{-1} f_{(-1)} \otimes S^{-1} f_{(-2)} \big) \cdot
        \big( g_{(1)} \otimes g_{(2)} \big).
\end{eqnarray}
Altogether this gives
\begin{eqnarray}
\lefteqn{ (\psi_U \otimes id ) \circ \psi_U \; (f \otimes g) }  \nonumber \\
 & = & \sum \limits_{(f),(g)}
       f_{(0)} \otimes g_{(0)} \otimes \Delta \big( S^{-1} f_{(-1)} \,
       g_{(1)} \big) \nonumber \\
 & = & ( id \otimes \Delta ) \psi_U \; (f \otimes g).
\end{eqnarray}
The relation (\ref{ARKW2}) is a consequence of
\begin{eqnarray}
\lefteqn{ ( id \otimes id \otimes \epsilon) \circ \psi_U \;
 (f \otimes g) } \nonumber  \\
 & = & ( \epsilon \circ S^{-1} \otimes id \otimes id
       \otimes \epsilon ) \circ ( \varphi   \otimes \phi ) \;
       (f \otimes g) \nonumber \\
 & = & ( \epsilon \otimes id \otimes id
       \otimes \epsilon ) \circ ( \varphi   \otimes \phi ) \;
       (f \otimes g) \nonumber \\
 & = & f \otimes g.
\end{eqnarray}
This proofs the lemma.
\epro
In the theory of commutative fibre bundles one defines an equivalence relation
$ \sim $ on $ P \times V $ by
\begin{equation}
 (a,v) \: \sim (b,w) \quad \Leftrightarrow \quad
 (a,v) \, = \, (bg,g^{-1}w), \quad g \in G.
\end{equation}
The equivalence classes of this equivalence relation form
a vector bundle $E$ over $M$, the associated vector bundle.
We want to dualise this. A function $\, f \in {\cal C} (P \times V) \,$
is said to be lifted by a function $\bar{f} \in {\cal C} (E) $, if
$\, \bar{f} \circ p \, = \, f \,$ for the
canonical projection $\, p : P \times V \longrightarrow E \,$.
$f$ can be lifted or regarded as a function on the vector bundle $E$
if and only if for all $ a \in \cal P$, $v \in V$ and $g \in G$
\begin{equation}
 f(ag,g^{-1}v) \: = \: f(a,v).
\end{equation}
In the language of \QPB this means that the element
\begin{equation}
 f \: = \: \sum \limits_{ {i}} {f}_{ {i}} \otimes
 f'_{ {i}} \: \in \:
 A \otimes {\cal P}(U), \quad U \subset M
\end{equation}
can be regarded as an element of the associated \QPB if and only if
\begin{equation}
 \psi_U \Big( \sum \limits_{ {i}} \, {f}_{ {i}} \otimes f'_{ {i}} \Big)
 \: = \:
 {f}_{ {i}} \otimes f'_{ {i}} \otimes 1.
\end{equation}
Therefore we define for all $U \subset M $ open
\bdm
 \lefteqn{ {\cal V} (U) } \\
 & = & \Big{\{}
  \sum \limits_{ {i} } \, {f}_{ {i}} \otimes f'_{ {i}}  \in
  A \otimes {\cal P}(U) \: : \:
  \psi_U \Big( \sum \limits_{ {i} }  \,
  {f}_{ {i} } \otimes f'_{ {i}} \Big) \: = \:
  {f}_{ {i}} \otimes f'_{ {i}} \otimes 1 \Big{\}}.
\edm
\begin{Remark}
 ${\cal V} (U)$ is the cotensor product of $A$ and ${\cal P}(U)$ over $H$,
 in signs
\begin{equation}
 {\cal V} (U) \: = \: A \, \Box_{H} \, {\cal P}(U).
\end{equation}
\end{Remark}
\begin{Theorem}
 Let us set
\bdm
\begin{array}{lcc}
 {\cal V}(U) & = & A \, \Box_{H} \, {\cal P}(U), \\
 {\cal V}(i^U_V) & = & id_A \otimes r^U_V
\end{array}
\edm
 for $V \subset U \subset M$ open. This gives a sheaf of associative algebras
 over $M$, where the multiplication
 $\, m  :  {\cal V}(U) \otimes {\cal V}(U)  \longrightarrow  {\cal V}(U) \,$
 is defined by
 $\, f \cdot g  \: = \:
 \sum \limits_{ {i}, {j} } \, {f}_{ {i}} {g}_{ {j}}
 \otimes f'_{ {i}} g'_{ {j}}, \,$ with
 $\, f = \Big( \sum \limits_{ {i} } \, {f}_{ {i}}
 \otimes f'_{ {i}} \Big), \,$
 $\, g = \Big( \sum \limits_{ {j} } \, {g}_{ {j}}
 \otimes g'_{ {j}} \Big) \, \in {\cal V}(U) \,$.
 So $ {\cal V} $ becomes a quantum space.
\end{Theorem}
\bpro
It is obvious that ${\cal V}$ is a subsheaf of the sheaf $A \otimes {\cal P}$
of complex vector spaces. Therefore we only have to show
\begin{equation}
\lefteqn{
 \psi_U \Big(
 \sum \limits_{ {i}, {j} } \, {f}_{ {i}} {g}_{ {j}}
 \otimes f'_{ {i}} g'_{ {j}}
 \Big) \: = \:
 \sum \limits_{ {i}, {j} } \, {f}_{ {i}} {g}_{ {j}}
 \otimes f'_{ {i}} g'_{{j}} \otimes 1 }.
\end{equation}
We get with the universal property of the tensor product:
\begin{eqnarray}
\lefteqn{
 \psi_U \Big(
 \sum \limits_{ {i}, {j} } \, {f}_{ {i}} {g}_{ {j}}
 \otimes f'_{ {i}} g'_{ {j}}
 \Big) } \nonumber \\
 & = &
 \sum \limits_{{i}, {j}} \,
 \sum \limits_{ ({f}_{ {i}}), ({g}_{ {j}}),
 (f'_{ {i}}), (g'_{ {j}}) } \,
 {f}_{ {i} \, (0)} \, {g}_{ {j} \, (0)}
 \otimes f'_{ {i} \, (0)} \, g'_{ {j} \, (0)} \otimes \nonumber \\
 &  & \hspace{4.8cm}
 \otimes
 \big( S^{-1} {g}_{ {j} \, (-1)} \big) \,
 \big( S^{-1} {f}_{ {i} \, (-1)} \big) \,
 f'_{ {i} \, (1)} \, g'_{ {j} \, (1)} \nonumber \\
 & = &
 \sum \limits_{{j}} \,
 \sum \limits_{ ({g}_{ {j}}), (g'_{ {j}}) } \,
 \Big(
 \sum \limits_{{i}} \,
 \sum \limits_{ ({f}_{ {i}}), (f'_{ {i}}) } \,
 {f}_{ {i} \, (0)} \, {g}_{ {j} \, (0)}
 \otimes f'_{ {i} \, (0)} \, g'_{ {j} \, (0)} \otimes \nonumber \\
 &  & \hspace{4.8cm}
 \otimes
 \big( S^{-1} {g}_{ {j} \, (-1)} \big) \,
 \big( \big( S^{-1} {f}_{ {i} \, (-1)} \big) \,
 f'_{ {i} \, (1)} \big) \, g'_{ {j} \, (1)} \nonumber \\
 & = &
 \sum \limits_{{j}}
 \sum \limits_{ ({g}_{ {j}}), (g'_{ {j}}) } \, \Big(
 \sum \limits_{{i}} \,
 {{f}_{ {i}}} \, {g}_{ {j} \, (0)}
 \otimes f'_{ {i}} \, g'_{ {j} \, (0)} \otimes
 \big( S^{-1} {g}_{ {j} \, (-1)} \big)
 \, g'_{ {j} \, (1)} \Big) \nonumber \\
 & = &
 \sum \limits_{ {i}, {j} } \, {f}_{ {i}} {g}_{ {j}}
 \otimes f'_{ {i}} g'_{ {j}} \otimes 1.
\end{eqnarray}
\mbox{  }
\epro
The quantum space $\cal V$ will turn out to be a quantum vector bundle.
We are going to proof this and want to find appropriate
local trivialisations.
Let $ {\cal U} = (U_{\iota})_{\iota \in I} $ be an open covering of $M$
such that the given quantum principal bundle
$({\cal P,M},\varrho,H,(\Omega_{\iota})_{\iota \in I})$
is locally trivial over $\cal U$ and the $\tau_{\iota, \kappa}$ are
transition functions. According to proposition \ref{CTRFU} the family
$ (\tau_{\iota, \kappa})_{\iota \in I} $ is an $H$-cocycle in $\cal M$.
Now the morphisms of sheaves with values in the category of
complex vector spaces
\bdm
\begin{array}{lclcl}
 \Gamma_{\iota} & : & {\cal M} \! \mid_{U_\iota} \, \#_{\iota } \, A &
 \longrightarrow & A \otimes {\cal P} \! \mid_{U_\iota} \\
 & & f \otimes g & \longmapsto & \big( \tau \circ \big(
 \Omega_{\iota} \otimes id\big) \circ (id \otimes \varphi) \big) \,
 ( f \otimes g ) = \\
 & & & & = \:
 \sum \limits_{(g)} \, g_{(0)} \otimes \big( \Omega_{\iota} \, \big( f \#
 g_{(-1)} \big) \big)
\end{array}
\edm
can be defined, where the crossed product
$ {\cal M} \! \mid_{U_\iota} \, \#_{\iota } \, A $
is given by the weak action
$\, \alpha_{\iota} :  H \times {\cal M} \! \mid_{U_\iota}
\longrightarrow {\cal M} \! \mid_{U_\iota} \,$, the normal cocycle
$\iota : H \times H \longrightarrow {\cal M} \big( U_{\iota} \big)$
and the left $H$-coaction $\, \varphi : A \longrightarrow H \otimes A \,$
in the sense of theorem \ref{SMPRKA}.
The $ \Gamma_{\iota} $ are the local trivialisations we are looking
for. Before we can proof this, some general statements about
the $\Gamma_{\iota}$ have to be made.
\begin{Lemma}
\label{LeGa}
 For all $U \subset M$ open
\begin{enumerate}
\item
 $ \big( \Gamma_{\iota} \big)_U $ is a morphism of algebras with unity,
\item
 $Im \big( \Gamma_{\iota} \big)_U \, \subset \, {\cal V}(U). $
\end{enumerate}
\end{Lemma}
\bpro
We only show the lemma for the case of a trivial cocycle
$\, \iota : H \times H \longrightarrow {\cal M} \big( U_{\iota} \big) \,$.
The general case goes with the same argument but requires
a lot more writing.
Let $f$, $f' \in {\cal M}(U) $ and $g$, $g' \in A$.
Then i) is a consequence of the following two equations:
\begin{eqnarray}
\lefteqn{
 \Gamma_{\iota} \big( (f \otimes g) (f' \otimes g') \big) } \nonumber \\
 & = & \sum \limits_{(g)}
 \Gamma_{\iota} \big( f \, g_{(-1)}  f'\otimes g_{(0)} \, g' \big) \nonumber \\
 & = & \tau \Big(
 \sum \limits_{(g)} \, \big(
 \Omega_{\iota} \big( f \, g_{(-2)} \, f' \# g_{(-1)} \, g'_{(-1)} \big)
 \otimes g_{(0)} \, g'_{(0)} \big) \Big) \nonumber\\
 & = & \tau \Big(
 \sum \limits_{(g)} \, \big(
 \Omega_{\iota} \big( f \# g_{(-1)} \big)
 \cdot \Omega_{\iota} \big( f' \# g'_{(-1)} \big)
 \otimes g_{(0)} \, g'_{(0)} \big) \Big) \nonumber \\
 & = & \tau \Big(
 \sum \limits_{(g)} \, \big(
 \Omega_{\iota} \big( f \# g_{(-1)} \big) \otimes g_{(0)} \big)
 \cdot
 \big( \Omega_{\iota} \big( f' \# g'_{(-1)} \big) \otimes g'_{(0)} \big)
 \Big) \nonumber \\
 & = & \Gamma_{\iota} (f \otimes g) \cdot \Gamma_{\iota} ( f'\otimes g' ),
\end{eqnarray}
\begin{eqnarray}
\lefteqn{ \Gamma_{\iota} \, ( 1 \otimes 1 ) \: = \: ( 1 \otimes 1 ). }
\end{eqnarray}
As the inverse $S^{-1}$ of the antipode $S$ satisfies the relation
\begin{equation}
\label{IANT}
 \sum \limits_{(g)} \, S^{-1} \big( g_{(2)} \big) \cdot g_{(1)} \, = \,
 \epsilon (g) \, 1
\end{equation}
one gets
\begin{eqnarray}
\lefteqn{
 \psi \circ \Gamma_{\iota} (f \otimes g) } \nonumber \\
 & = &
 \sum \limits_{(g)} \, \psi \big( g_{(0)} \otimes \Omega_{\iota}
 \big( f \# g_{(-1)} \big) \big) \nonumber \\
 & = &
 \sum \limits_{(g)} \, g_{(0)} \otimes \Omega_{\iota}
 \big( f \# g_{(-3)} \big) \otimes S^{-1} \big( g_{(-1)}
 \big) \cdot g_{(-2)} \nonumber \\
 & = &
 \sum \limits_{(g)} \, g_{(0)} \otimes \Omega_{\iota}
 \big( f \# g_{(-2)} \big) \otimes \epsilon \big( g_{(-1)}
 \big) 1 \nonumber \\
 & = &
 \sum \limits_{(g)} \, g_{(0)} \otimes \Omega_{\iota}
 \big( f \# g_{(-1)} \big) \otimes 1 \nonumber \\
 & = &
 \psi ( f \otimes g) \otimes 1.
\end{eqnarray}
This entails ii).
\epro
Now $\Gamma_{\iota}$ can be considered as a mapping
$\, \Gamma_{\iota} : {\cal M} \! \mid_{U_\iota} \, \#_{\iota } \, A
  \longrightarrow   {\cal V} \! \mid_{U_\iota} \,$.
Next we need an inverse of this $ \Gamma_{\iota} $. Define
\bdm
\begin{array}{lclcl}
 \zeta_{\iota} & : & A \otimes {\cal P} \! \mid_{U_{\iota}} & \longrightarrow &
 {\cal M} \! \mid_{U_{\iota}} \#_{\iota}  A \\
 & & ( f \otimes g) & \longmapsto &
 \big( ( id \otimes \epsilon ) \circ \Omega^{-1}_{\iota} \big)
 (g) \otimes f = \\
 & & & & = \: ( id \otimes \epsilon \otimes id) \circ
 \big( \Omega^{-1}_{\iota} \otimes id \big) \circ \tau \, (f \otimes g).
\end{array}
\edm
Then the equation
\begin{eqnarray}
\lefteqn{
 \zeta_{\iota} \circ \Gamma_{\iota} \, (f \otimes g) } \nonumber \\
 & = &
 ( id \otimes \epsilon \otimes id ) \circ \big( \Omega_{\iota} \otimes
 id \big) \circ \big( \Omega \otimes id \big) \circ
 ( id \otimes \varphi ) \: ( f \otimes g ) \nonumber \\
 & = &
 ( id \otimes \epsilon \otimes id ) \circ ( id \otimes \varphi )
 \: ( f \otimes g ) \nonumber \\
 & = &
 f \otimes g,
\end{eqnarray}
is true and therefore
\begin{equation}
\label{LII}
 \zeta_{\iota} \circ \Gamma_{\iota} \: = \: id.
\end{equation}
If we further set $\, P_{\iota} \, = \,
\Gamma_{\iota} \circ \zeta_{\iota}  :  A \otimes {\cal P}
\! \mid_{U_{\iota}}  \longrightarrow  A \otimes {\cal P}
\! \mid_{U_{\iota}} \,$, equation (\ref{LII}) implies
\begin{equation}
\label{PRGA}
 P_{\iota} \, = \, P_{\iota} \circ P_{\iota} ,
\end{equation}
that means $\big( P_{\iota} \big)_U$ is a projection onto
$Im \big( \Gamma_{\iota} \big)_U $.
Finally only $\, Im \big( P_{\iota} \big)_U \, = \, {\cal V} (U) \,$
has to be shown for $U \subset U_{\iota} $ open.
\newline \noindent
{}From now on we must assume
$\, \iota (h,l) = \epsilon (h) \, \epsilon (l) \, 1 \,$ for all
$\, h,l \in H \,$.
\begin{Lemma}
\label{HBIM}
 $A \otimes {\cal P} (U)$, $U \subset U_{\iota} $ is an $H$-bimodule, if the
 left and right action are defined as follows:
\begin{enumerate}
\item
$
\begin{array}{ccl}
 H \times \big( A \otimes {\cal P} (U) \big) & \longrightarrow &
 A \otimes {\cal P} (U) \\
 (h,f \otimes g) & \longmapsto & h \cdot (f \otimes g) \: = \:
 f \otimes \big( \Omega_{\iota} ( 1 \# h ) \, g \big),
\end{array}
$
\item
$
\begin{array}{ccl}
 \big( A \otimes {\cal P} (U) \big) \times H & \longrightarrow &
 A \otimes {\cal P} (U) \\
 (f \otimes g,h) & \longmapsto & (f \otimes g) \cdot h \: = \:
 f \otimes \big( g \, \Omega_{\iota} ( 1 \# h ) \big).
\end{array}
$
\end{enumerate}
\end{Lemma}
\bpro
 The proof is done by an easy calculation.
\epro
Now we have all the necessary tools to show the main proposition.
\begin{Proposition}
For all $ \iota \in I $
\bdm
\begin{array}{lclcl}
 \Gamma_{\iota} & : & {\cal M} \! \mid_{U_\iota} \, \#_{\iota } \, A &
 \longrightarrow & {\cal V} \! \mid_{U_\iota} \\
 & & f \otimes g & \longmapsto &
 \sum \limits_{(g)} \, g_{(0)} \otimes \big( \Omega_{\iota} \, \big( f \#
 g_{(-1)} \big) \big)
\end{array}
\edm
is an isomorphism of sheaves of algebras.
The restriction of the sheaf morphism
\bdm
\begin{array}{lclcl}
 \zeta_{\iota} & : & A \otimes {\cal P} \! \mid_{U_{\iota}} & \longrightarrow &
 {\cal M} \! \mid_{U_{\iota}} \#_{\iota}  A \\
 & & ( f \otimes g) & \longmapsto &
 \big( ( id \otimes \epsilon ) \circ \Omega^{-1}_{\iota} \big)
 (g) \otimes f
\end{array}
\edm
to $ \, {\cal V} \! \mid_{U_{\iota}} $ gives the inverse of $\Gamma_{\iota}$.
Furthermore for all $\iota, \kappa \in I$
\begin{eqnarray}
\lefteqn{
 \big( \Gamma_{\kappa}^{-1} \big)_U \circ \big( \Gamma_{\iota} \big)_U
 \: ( f \otimes g ) } \nonumber \\
 & = & \big( \zeta_{\kappa} \big)_U \circ  \big( \Gamma_{\iota} \big)_U
 \: (f \otimes g ) \nonumber \\
\label{QVBW}
 & = & \sum \limits_{(g)} \, f \cdot \tau_{\iota,\kappa} \big(
 g_{(-1)} \big) \otimes g_{(0)}, \hspace{0.7cm}
 f \otimes g  \in {\cal M} \#_{\iota} A.
\end{eqnarray}
\end{Proposition}
\bpro
 Because of equation (\ref{PRGA}) it suffices to show
\begin{equation}
 Im \big( P_{\iota} \big)_U \: = \: {\cal V} (U), \hspace{0.7cm} U \subset
 U_{\iota}.
\end{equation}
for the proof of the first part of the proposition.
The elements$\, f \in A $, $g \in {\cal M}(U) $, $ h \in H $
satisfy the equation
\begin{equation}
\label{QVBWi}
 \psi \, \big( f \otimes \Omega_{\iota} ( g \# h ) \big) \: = \:
 \sum \limits_{(f),(h)} \, f_{(0)} \otimes \Omega_{\iota} \big( g \#
 h_{(1)} \big) \otimes \big( S^{-1} \, f_{(-1)} \big) \, h_{(2)}
\end{equation}
according to the definition of $ \psi $. As also
\begin{eqnarray}
\lefteqn{
 P_{\iota} \,  \big( f \otimes \Omega_{\iota} ( g \# h ) \big) } \nonumber \\
 & = & \Gamma_{\iota } ( \epsilon (h) \, g \otimes f ) \nonumber \\
 & = & \sum \limits_{(f)} f_{(0)} \otimes \big( \epsilon (h) \,
 \Omega_{\iota} \big( g \# f_{-1)} \big) \big).
\end{eqnarray}
is true, equation (\ref{IANT}) and the right action
in lemma \ref{HBIM} imply the following relation:
\begin{eqnarray}
\lefteqn{
 \sum \limits_{(f),(h)} \,
 P_{\iota} \, \big( f_{(0)}  \otimes \Omega_{\iota} \big( g \# h_{(1)}
 \big) \big) \cdot \big( \big( S^{-1} \, f_{(-1)} \big) \, h_{(2)} \big)
 } \nonumber \\
 & = &
 \sum \limits_{(f)} \,
 \big( f_{(0)} \otimes \Omega_{\iota} \big( g \# f_{(-1)}
 \big) \big) \cdot \big( \big( S^{-1} \, f_{(-2)} \big) \, h \big)
 \nonumber \\
 & = &
 \sum \limits_{(f)} \,
 f_{(0)} \otimes \big( \Omega_{\iota} \big( g \# f_{(-1)}
 \big) \: \Omega_{\iota} \big( 1 \# \big( S^{-1} \, f_{(-2)} \big) \big)
 \: \Omega_{\iota} \big( 1 \# h \big) \big)
 \nonumber \\
 & = &
 \sum \limits_{(f)} \,
 f_{(0)} \otimes \big( \Omega_{\iota} \big( g \# \big( f_{(-1)}
 \, \big( S^{-1} \, f_{(-2)} \big) \big) \big)
 \: \Omega_{\iota} \big( 1 \# h \big) \big)
 \nonumber \\
 & = &
 \sum \limits_{(f)} \,
 f_{(0)} \otimes \big( \Omega_{\iota} \big( g \#
 \epsilon \big( f_{(-1)} \big)
 \, 1 \big)
 \, \Omega_{\iota} \big( 1 \# h \big) \big)
 \nonumber \\
 & = &
 \label{QVBWii}
 f \otimes \Omega_{\iota} \big( g \# h \big).
\end{eqnarray}
For $\, \sum \limits_{i} \, f_i \otimes g_i \in A \otimes {\cal P}(U) \,$
assume $\, \psi \, \Big(  \sum \limits_{i} \, f_i \otimes g_i \Big) \, = \,
  \sum \limits_{j} \, f'_j \otimes g'_j \otimes h_j \,$.
As $ \Omega_{\iota} $ is a sheaf morphism (\ref{QVBWi}) and (\ref{QVBWii})
entail $ P_{\iota} $ having the property
\begin{equation}
\label{QVBWiii}
 \sum \limits_{j} \, P_{\iota} \big( f'_j \otimes g'_j \big) \cdot h_j
 \, = \,
 \sum \limits_{i} \, f_i \otimes g_i.
\end{equation}
So if $\, \sum \limits_{i} \, f_i \otimes g_i \in {\cal V}(U) \,$,
equation (\ref{QVBWiii}) gives
\begin{equation}
 \sum \limits_{i} \, P_{\iota} \big( f_i \otimes g_i \big)
 \, = \,
 \sum \limits_{j} \, P_{\iota} \big( f'_j \otimes g'_j \big) \cdot h_j
 \, = \,
 \sum \limits_{i} \, f_i \otimes g_i .
\end{equation}
Therefore $ Im \big( P_{\iota} \big)_U = {\cal V} (U) $.

The equation (\ref{QVBW}) is a direct consequence of the
definition of the $\Gamma_{\iota}$, $\zeta_{\iota}$ and the transition
functions $\tau_{\iota, \kappa}$.
Explicitly the relation (\ref{NCCCH1}) gives
\begin{eqnarray}
\lefteqn{
 \big( \zeta_{\kappa} \big)_U \circ \big( \Gamma_{\iota} \big)_U
 \: ( f \otimes g ) } \nonumber \\
 & = & \sum \limits_{(g)} \, ( id \otimes \epsilon ) \circ
 \Omega_{\kappa}^{-1} \circ \Omega_{\iota} \: \big( f \otimes g_{(-1)}
 \big) \otimes g_{(0)} \nonumber \\
 & = & \sum \limits_{(g)} \, f \cdot  \tau_{\iota, \kappa}
 \big( g_{(-1)} \big) \otimes g_{(0)}.
\end{eqnarray}
Quod erat demonstrandum.
\epro
Now we get the desired result.
\begin{Corollary}
\label{KAQVB}
 Let $A$ be a quadratic algebra, $ \varphi $ a left $H$-coaction and
 ${\cal P} $ an $\bf A$-quantum principal bundle over $\cal M$, where
 the trivialisations are defined by actions
 $\, \alpha_{\iota} :  H \times {\cal M} \! \mid_{U_\iota}
 \longrightarrow {\cal M} \! \mid_{U_\iota} \,$
 and trivial cocycles
 $\, \iota : H \times H \longrightarrow {\cal M} \big( U_{\iota} \big) \,$.
 Then the tupel
 $( {\cal V,M},\tilde{\varrho},A,H,\varphi,(\Gamma_{\iota})_{\iota \in I} )$,
 gives an ${\bf A}$-quantum vector bundle with transition functions
 $\, \tau_{\iota, \kappa} $, $\iota, \kappa \in I \,$,
 where ${\cal V}$, $\Gamma_{\iota}$
 are constructed like above, and $\tilde{\varrho}$
 is defined by
\bdm
\begin{array}{lclcl}
 \tilde{\varrho} & : & {\cal M} & \longrightarrow & {\cal V} \\
 & & f & \longmapsto & 1 \otimes \varrho \, (f).
\end{array}
\edm
\end{Corollary}
\bpro
It only has to be shown that $ \tilde{\varrho} $ is welldefined.
Now
\begin{equation}
 \phi ( \varrho \, (f) ) \: = \: \varrho \, (f) \otimes 1 ,
 \hspace{0.7cm} f \in {\cal M} (U), \quad U \subset M
\end{equation}
implies
\begin{equation}
 \psi ( 1 \otimes \varrho \, (f) ) \: = \: 1 \otimes \varrho \, (f) \otimes 1,
\end{equation}
and that proofs $\tilde{\varrho}$ being welldefined.
\epro
Now the proof of theorem \ref{KQVB} can be given.
$\vspace{0.5cm} \\ \bf Proof$ (Theorem \ref{KQVB})$\bf :$
For the $H$-cocycle
$\, \big( \tau_{\iota,\kappa} \big)_{(\iota, \kappa) \in I \times I} \,$
construct a \QPB according to theorem \ref{CQPBTRF}. Corollary \ref{KAQVB}
provides the \QVB we are looking for.
\epro
\subsection{Differential Calculus}
In this section we want to define a differential calculus and connections
on quantum principal and quantum vector bundles.
\newline \noindent
First we generalise the concept of a differential calculus over an algebra
\cite{wor-1,wez-1} to quantum spaces.

\begin{Definition}
 Let $\cal M$ be a quantum space, $\cal D$ a bimodule sheaf over $\cal M$ and
 $\, d : {\cal M} \longrightarrow {\cal D} \,$
 a sheaf morphism.
 $({\cal D}, d)$ is said to be a differential calculus over ${\cal M}$,
 if for all $f,g \in {\cal M} (U)$, $U \subset M$ the
 following conditions are satisfied:
\begin{enumerate}
\item
 $ d(fg) \, = \, (df) \, g \, + \, f \, (dg)$.
\item
 Every element $\omega \in {\cal D} (U)$ has the form
\bdm
 \omega \, = \, \sum \limits_{k = 1}^{n} \, f_k \, dg_k,
\edm
 where $\, f_k, g_k \in {\cal M} (U)$, $\, k = 1,...,n \, $, $ \, n \in
 {\Bbb{N}}$.
\end{enumerate}
\end{Definition}
If we assume to have a \QPB
$0 \longrightarrow {\cal M} \stackrel{\varrho}{\longrightarrow} {\cal P}$
and differential calculi over ${\cal M}$ and ${\cal P}$, the question
is in which sense the differential calculi fit together.
\begin{Definition}
 Let
 $0 \longrightarrow {\cal M} \stackrel{\varrho}{\longrightarrow} {\cal P}$
 be a quantum principal bundle, $({\cal D}',d')$ a differential
 calculus over ${\cal M}$ and $({\cal D},d)$ a differential calculus
 over ${\cal P}$.
 $({\cal D}, d)$ is said to induce $({\cal D}', d')$,
 if there exists a sheaf morphism
 $\varrho_* : {\cal D}'  \longrightarrow {\cal D} $
 over $\varrho $ such that the diagram
\begin{center}
\xext=1400 \yext=500
\adjust[`\varrho;`{{\varrho}_*};`d;`{d'}]
\begin{picture}(\xext,\yext)(\xoff,\yoff)
\putmorphism(0,0)(1,0)[0`\phantom{\cal M}`]{700}{1}{b}
\putmorphism(0,500)(1,0)[0`\phantom{\cal N}`]{700}{1}{a}
\setsqparms[1`1`1`1;700`500]
\putsquare(700,0)[\cal M`\cal P`{{\cal D}'}`\cal D; \varrho
`{d'}`d`{{\varrho}_*}]
\end{picture}
\end{center}
 commutes with exact horizontal lines.
\end{Definition}
The morphism $\varrho$ determines the morphism $\varrho_*$
in a certain sense. To see this choose
$f_k, f'_k \in {\cal M}(U)$, $\, U \subset M \,$ open, $\, k = 1,...,n \,$.
Then
\begin{equation}
\label{DDiffk}
 \varrho_* \, \Big( \sum \limits_{k = 1}^{n} \, f_k \, d' f'_k \Big) \: = \:
 \sum \limits_{k = 1}^{n} \, \varrho \big( f_k \big) \, d  \varrho \big( f'_k
 \big) .
\end{equation}
This equation can be taken to define a morphism $\varrho_*$.
\begin{Theorem}
 Let $({\cal D}, d)$ be a differential calculus over a \QPB ${\cal P}$.
 If the pair $( {\cal D}',d')$ is defined by
\begin{eqnarray*}
\begin{array}{lcl}
 {\cal D}' (U) & = & \Big\{ \sum \limits_{k = 1}^{n} \,
 \varrho \big( f_k \big) \, d \varrho \big( f'_k \big) \, \in {\cal D} (U) \, :
\\
 & & f_k, f'_k \in {\cal M} (U) \, , \; k =  1,...,n \, , \; n \in {\Bbb{N}}
 \Big\} \\
 d' & = & d \! \mid_{\varrho ({\cal M}) } \circ \varrho,
\end{array}
\end{eqnarray*}
one receives a differential calculus $( {\cal D}',d')$ over $\cal M$,
which is induced by the morphism
\bdm
\begin{array}{lclcl}
 \varrho_* & : & {\cal D}' & \longrightarrow & {\cal D} \\
 \big( \varrho_* \big)_U & : & {\cal D}'(U) & \longrightarrow & {\cal D}(U) \\
 & & \sum \limits_{k = 1}^{n} \, \varrho \big( f_k \big) \, d \varrho
 \big( f'_k \big) & \longmapsto &
 \sum \limits_{k = 1}^{n} \, \varrho \big( f_k \big) \, d \varrho \big( f'_k
 \big).
\end{array}
\edm
\end{Theorem}
\bpro
Obvious by the above considerations.
\epro
In the following let $(U_{\iota})_{\iota \in I}$
be a trivialisation covering of ${\cal P}$.
Further assume the cocycles $\, \iota : H \times H
\longrightarrow {\cal M} (U_{\iota}) \,$ being trivial and the
$\, \alpha_{\iota} : H \times {\cal M} \! \mid_{U_{\iota}}
\longrightarrow {\cal M}_{U_{\iota}} \,$ being actions.

For an open $U \subset U_{\iota}$ the equations
$\, {\cal D}_{\iota} \, = \, {\cal D} \! \mid_{U_{\iota}} \,$ and
$\, d_{\iota} \big( f \#_{\iota} g \big) \, = \,
d \circ {\Omega}_{\iota} \big( f \#_{\iota} g \big) \,$, $\,
f \in {\cal M} (U) \,$, $\, g \in H$
define a differential calculus on
${\cal M} \! \mid_{U_{\iota}} \#_{\iota} H $ such that
$\, d_{\iota} \big( f \#_{\iota} 1 \big) \, = \, d' (f) \,$ for
$\, f \in {\cal M} (U) $ and
\begin{equation}
 d_{\iota} \big( f \#_{\iota} g \big) \: = \: d' (f) \cdot \Omega_{\iota} \big(
 1 \#_{\iota} g \big) \, + \, \varrho (f) \cdot d \Omega \big( 1 \#_{\iota} g
\big).
\end{equation}
Now we would like to care about the inverse problem
and suppose to be given a differential calculus $(D_H,d)$
on the Hopf algebra $H$ and a differential calculus
$({\cal D}'_{\cal M},d')$ on $\cal M$.
Assuming the following consistency condition we will construct
a differential calculus $({\cal D}_{\cal P},d)$ on ${\cal P}$.
\begin{itemize}
\item
 (Consistency Condition) \\
 The actions
 $\, \alpha_{\iota} : H \times {\cal M} \! \mid_{U_{\iota}} \longrightarrow
 {\cal M}_{U_{\iota}} \,$ satisfy the equation
\begin{equation}
 \sum \limits_{(g),1 \leq k \leq n} \, \big( g_{(1)} \cdot_{\iota} f_k \big) \,
 d' \big( g_{(2)} \cdot_{\iota} f'_k \big) \: = \: 0
\end{equation}
 for all $ g \in H$, if
 $\,  \sum \limits_{k = 1}^{n} \, f_k d' f'_k \, = 0 \, $
 with
 $\, f_k, f'_k \in {\cal M}(U)\, $, $\, U \subset U_{\iota}$ open.
\end{itemize}
This consistency condition guarantees the existence of an action
\bdm
\begin{array}{lclcl}
 \alpha  & : & H \times {\cal D}_{\cal M} \! \mid_{U_{\iota}}
 & \longrightarrow & {\cal D}_{\cal M} \! \mid_{U_{\iota}} \\
 & & \Big( g, \sum \limits_{k = 1}^{n} \, f_k \, d' f'_k \Big)
 & \longmapsto &
 \sum \limits_{(g),1 \leq k \leq n} \,\big(  g_{(1)} \cdot_{\iota}
 f_k \big) \, d' \big( g_{(2)} \cdot_{\iota} f'_k \big).
\end{array}
\edm
\begin{Lemma}
 Define for $U \subset M$ open:
\bdm
\begin{array}{lcl}
 {\cal D}^0_{\cal P} (U) & = & \bigoplus \limits_{\iota \in I} \,
 \big( {\cal D}^0_{\cal P} (U) \big)_{\iota} \\
 \big( {\cal D}^0_{\cal P} (U) \big)_{\iota} & = &
 {\cal D}_{\cal M} (U_{\iota} \cap U) \otimes H \, \oplus \,
 {\cal M} (U_{\iota} \cap U) \otimes D_H.
\end{array}
\edm
Then ${\cal D}^0_{\cal P} (U)$ is a bimodule over
\bdm
\begin{array}{lcl}
 {\cal P}_0 (U) & = & \bigoplus \limits_{\iota \in I} \,
 {\cal M} (U_{\iota} \cap U) \#_{\iota} H.
\end{array}
\edm
The left and right actions are given by
\begin{eqnarray}
\lefteqn{
 (f \otimes g) \, (a \otimes g'+ f'\otimes b) } \nonumber \\
 & = & \sum \limits_{(g)} \,
 f \, \big( g_{(1)} \cdot a \big) \otimes g_{(2)} \, g'
 + f \, \big( g_{(1)} \cdot f' \big) \otimes
 \big( g_{(2)} \cdot b \big) \, ,
\end{eqnarray}
\vspace{-0,2cm}
\begin{eqnarray}
\lefteqn{
 (a \otimes g + f \otimes b) \, (f' \otimes g') } \nonumber \\
 & = & \sum \limits_{(g),(b)} \,
 a \, \big( g_{(1)} \cdot f'\big) \otimes g_{(2)} \, g'
 + f \, \big( b_{(-1)} \cdot f' \big) \otimes
 \big( b_{(0)} \cdot g' \big) \, ,
\end{eqnarray}
where $\, f,f'\in {\cal M} (U_{\iota} \cap U) \,$, $\, g,g'\in H \,$,
$\, a \in {\cal D}_{\cal M} (U_{\iota} \cap U) \,$ and $\, b \in D_H$.
\end{Lemma}
\bpro
 The following two equations
\begin{eqnarray}
\lefteqn{
 \big( ( f'' \otimes  g'' ) (f \otimes g) \big) \, ( a \otimes g'
 + f' \otimes b) } \nonumber \\
 & = & \sum \limits_{(g'')} \big( f'' \, \big( g_{(1)}'' \cdot f \big)
 \otimes g_{(2)}'' \, g
 \big) \, ( a \otimes g' + f' \otimes b ) \nonumber \\
 & = & \sum \limits_{(g),(g'')}  f'' \, \big( g_{(1)}'' \cdot f \big) \,
 \big( \big( g_{(2)}'' \, g_{(1)} \big)
 \cdot a \big) \otimes g_{(3)}'' \, g_{(2)} \, g'
 \, + \nonumber \\
 & & + \, f'' \, \big( g_{(1)}'' \cdot f \big) \, \big( \big(
 g_{(2)}'' \, g_{(1)} \big) \cdot f' \big) \otimes
 \big( \big( g_{(3)}'' \, g_{(2)} \big) \cdot b \big) \, ,
\end{eqnarray}
\begin{eqnarray}
\lefteqn{
 ( f'' \otimes  g'' ) \, \big( (f \otimes g) ( a \otimes g'
 + f' \otimes b) \big) } \nonumber \\
 & = & ( f'' \otimes  g'' ) \, \Big( \sum \limits_{(g)} \,
 f \, \big( g_{(1)} \cdot a \big) \otimes g_{(2)} \, g'
 + f \, \big( g_{(1)} \cdot f' \big) \otimes
 \big( g_{(2)} \cdot b \big) \Big) \nonumber \\
 & = & \sum \limits_{(g),(g'')} \, f'' \, \big( g_{(1)}''
 \cdot \big( f \, \big( g_{(1)}
 \cdot a \big) \big) \big) \otimes g_{(2)}'' \, g_{(2)} \, g' \, + \nonumber \\
 & & + \, f'' \, \big( g_{(1)}'' \cdot
 \big( f \, \big( g_{(1)}
 \cdot f' \big) \big) \big) \otimes \big( \big( g_{(2)}'' \, g_{(2)}
 \big) \cdot b \big)
 \nonumber \\
 & = & \sum \limits_{(g),(g'')}  f'' \, \big( g_{(1)}'' \cdot f \big) \,
 \big( \big( g_{(2)}'' \, g_{(1)} \big) \cdot a \big)
 \otimes g_{(3)}'' \, g_{(2)} \, g'
 \, + \nonumber \\
 & & + \, f'' \, \big( g_{(1)}'' \cdot f \big) \, \big( \big(
 g_{(2)}'' \, g_{(1)} \big) \cdot f' \big) \otimes
 \big( \big( g_{(3)}'' \, g_{(2)} \big) \cdot b \big) \, ,
\end{eqnarray}
 proof
\begin{eqnarray}
\lefteqn{
 \big( ( f'' \otimes  g'' ) (f \otimes g) \big) \, ( a \otimes g'
 + f' \otimes b) } \nonumber \\
 & = &
 ( f'' \otimes  g'' ) \, \big( (f \otimes g) ( a \otimes g'
 + f' \otimes b) \big).
\end{eqnarray}
 The rest is shown similarly.
\epro
In the next step we will provide the derivation operator $d$.
\bdm
\begin{array}{lclcllcl}
 d & : & {\cal P}^0 & \longrightarrow & {\cal D}_{\cal P}^0 (U), & d & = &
 \sum \limits_{\iota \in I} \, d_{\iota} \\
 d_{\iota } & : & {\cal M} (U_{\iota } \cap U) \#_{\iota } H & \longrightarrow
 & \big( {\cal D }_{\cal P }^0 (U) \big)_{\iota } & & & \\
 & & f \# g & \longmapsto & d' f \otimes g \, + \, f \otimes d g. & & &
\end{array}
\edm
The operator $d$ the Leibniz rule:
\begin{eqnarray}
\lefteqn{ d_{\iota } \big( ( f \otimes g ) ( f' \otimes g' ) \big) }
 \nonumber \\
 & = & \sum \limits_{(g)} \, d_{\iota } \big( f \, \big( g_{(1)}
 \cdot f' \big) \otimes
 g_{(2)} \, g' \big) \nonumber \\
 & = & \sum \limits_{(g)} \, \big( d' \big( f \, \big( g_{(1)}
 \cdot f' \big) \big)
 \big) \otimes g_{(2)} \, g' \big) +
 f \, \big( g_{(1)} \cdot f' \big) \otimes d
 \big(  g_{(2)} \, g' \big) \nonumber \\
 & = & \sum \limits_{(g)} \, \big( \big( d' f \big) \, \big( g_{(1)}
 \cdot f' \big) + f \, d' \big( g_{(1)} \cdot f' \big) \big)
 \otimes g_{(2)} \, g' \, + \nonumber \\
 & & + \, f \, \big( g_{(1)} \cdot f' \big)
 \otimes \big( \big( d g_{(2)} \big) \, g' +
 g_{(2)} \, \big( d g' \big) \big) ,
\end{eqnarray}
\begin{eqnarray}
\lefteqn{
 \big( d_{\iota} ( f \otimes g ) \big) \, ( f' \otimes g' ) \, + \,
 ( f \otimes g ) \, \big( d_{\iota} ( f' \otimes g' ) \big) } \nonumber \\
 & = & \sum \limits_{(g)} \, \big( d' f \otimes g +
 f \otimes d g \big) \, ( f' \otimes g' ) \, + \, ( f \otimes g ) \,
 \big( d' f' \otimes g' + f' \otimes d g' \big) \nonumber \\
 & = & \sum \limits_{(g)} \, \big( d' f \big) \, \big( g_{(1)}
 \cdot f' \big) \otimes g_{(2)} \, g'  \, + \,
 f \, \big( g_{(1)} \cdot f' \big)
 \otimes \big( d g_{(2)} \big) \, g' \, + \nonumber \\
 & & f \, d' \big( g_{(1)} \cdot f' \big) \otimes g_{(2)} \, g' \, + \,
 f \, \big( g_{(1)} \cdot f' \big) \otimes g_{(2)} \, \big( d g' \big) .
\end{eqnarray}
Now we have to show that every
$\, e \, = \, \sum \limits_{k = 1}^{n} \, a_k \, d' b_k \otimes c_k \, + \,
 a_k' \otimes b_k' \, d c_k' \,$
with $ a_k, a_k' , b_k \in {\cal M} (U_{\iota} \cap U) $, $\, c_k , b_k' ,
c_k' \in H $ has the form
\begin{equation}
 e \, = \, \sum \limits_{l = 1}^{m} \, f_l \, d f_l'
\end{equation}
with $\, f_l , f_l' \in {\cal M} (U \cap U_{\iota}) \#_{\iota} H \,$.
But this is an easy consequence from
\begin{eqnarray}
\begin{array}{lcl}
 e & = & \sum \limits_{k = 1}^{n} \, \sum \limits_{(c_k)} \,
 \big( a_k \otimes c_{k \, (2)} \big) \, d \big( \big(
 S^{-1} c_{k \, (1)} \cdot b_k \big) \otimes 1 \big) \, + \\
 & & + \big( a_k \otimes b_{k}' \big) \, d \big( 1 \otimes c_k'\big),
\end{array}
\end{eqnarray}
where we have used the antipode being bijective.

The ${\cal P} (U)$-bimodule ${\cal D}_{\cal P}(U)$
has to be defined. According to equation (\ref{AlgP}) the algebra
${\cal P} (U)$ can be regarded as a subalgebra of ${\cal P}_0 (U)$.
Therefore we set
\begin{equation}
 {\cal D}_{\cal P} (U) \: = \:
 _{\cal P} Mod \Big\{ \sum \limits_{\iota \in I}
 \, d_{\iota} f_{\iota} \: : \: \sum \limits_{\iota} \, f_{\iota}
 \in {\cal P} (U) \Big\}  \quad \subset \: {\cal D }_{\cal P}^0 (U),
\end{equation}
which means that ${\cal D}_{\cal P} (U)$ is the ${\cal P}(U)$-left module
generated by $\sum \limits_{\iota \in I} \, d_{\iota} f_{\iota}$.
Because of the Leibniz rule ${\cal D}_{\cal P} (U)$ is a
${\cal P}(U)$-right module as  well.
Furthermore the relation
\begin{equation}
\label{DRos}
 \varrho_* \Big( \sum \limits_{k = 1}^{n} \, f_k \, d' f'_k \Big) \: = \:
 \sum \limits_{k = 1}^{n} \, \sum \limits_{\iota \in I} \,
 \big( r_{U_{\iota} \cap U}^U (f'_k) \, d'  r_{U_{\iota} \cap U}^U (f_k)
 \big) \otimes 1 \, + \, 0,
\end{equation}
whith $\, f_k , f_k' \in {\cal M}(U)$,
$\, 1 \leq k \leq n$, $\, n \in {\Bbb{N}}$ gives a welldefined
sheaf morphism
$\, \varrho_* : {\cal D}' \longrightarrow {\cal D} \,$.
Then
\begin{equation}
 \varrho_* \Big( \sum \limits_{k = 1}^{n} \, f_k \, d' f_k' \Big) \, = \,
 \sum \limits_{k = 1}^{n} \, \big( r_{U_{\iota} \cap U}^U (f_k) \, \# \, 1
\big)
 \, d \big( r_{U_{\iota} \cap U}^U (f_k') \, \# \, 1 \big),
\end{equation}
and the diagram
\begin{center}
\xext=1400 \yext=500
\adjust[`\varrho;`{{\varrho}_*};`d;`{d'}]
\begin{picture}(\xext,\yext)(\xoff,\yoff)
\putmorphism(0,0)(1,0)[0`\phantom{\cal M}`]{700}{1}{b}
\putmorphism(0,500)(1,0)[0`\phantom{\cal N}`]{700}{1}{a}
\setsqparms[1`1`1`1;700`500]
\putsquare(700,0)[\cal M`\cal P`{{\cal D}'}`\cal D; \varrho
`{d'}`d`{{\varrho}_*}]
\end{picture}
\end{center}
commutes.
It still has to be shown the lower sequence being exact.
But this is obvious remembering equation (\ref{DRos}) and the
properties of the restriction mappings.
\newline \noindent
We subsume our results in the following theorem.
\begin{Theorem}
 Let $H$ be a Hopf algebra with differential calculus $(D_H,d)$.
 Further let $({\cal D}_{\cal M}',d')$ be a differential calculus
 on the base quantum space $\cal M$.
 Then every \QPB ${\cal P} $ which fulfills the consistency condition
 has a differential calculus $({\cal D}_{\cal P}, d)$ induced
 by $({\cal D}_{\cal M}',d')$.
\end{Theorem}
Finally the concept of connections on a quantum principal
bundle shall be explained.
It is still assumed the differential calculi $({\cal D}_{\cal M}',d')$
and $({\cal D}_{\cal P}, d)$ being induced by
$\, \varrho : {\cal D}_{\cal M}'  \longrightarrow  {\cal D}_{\cal P} \,$.
Denotate by $A_{\cal M} (\cal P)$ (resp.~${\cal A}_{\cal M} (\cal P)$)
the basis subalgebra (resp.~basis subsheaf) of differential
forms on $\cal P$, i.~e.
\begin{equation}
\begin{array}{lclccclcl}
 A_{\cal M} ({\cal P}) & = & Im \big( (\varrho_*)_M \big)
 & \quad & \mbox{and} & \quad &
 {\cal A}_{\cal M} ({\cal P}) & = & Im (\varrho_*).
\end{array}
\end{equation}
Then
$\, {\cal A}_h \, = \,
_{\cal P} Mod \big( {\cal A}_{\cal M} ({\cal P}) \big) \,$
is the sheaf of horizontal differential forms on ${\cal P}$ or in
other words is the ${\cal P}$-submodule sheaf generated
by ${\cal A}_{\cal M} ({\cal P}) $ .
The motivation for this comes out of commutative geometry, where the basis
subalgebra of differential forms and the sheaf of horizontal forms are defined
in an analog way (compare Greub, Halperin, Vanstone \cite{ghv-2}).
Now if the sheaf ${\cal A}_h$ has a complementary sheaf ${\cal A}_v$
in ${\cal D}_{\cal P} $, we say that ${\cal A}_v$ defines a connection
on $\cal P$.
\begin{Definition}
 A quantum principal bundle is said to have a connection,
 if the sequence
\[
 0 \longrightarrow {\cal A}_h \longrightarrow
 {\cal D}_{\cal P} \longrightarrow  {\cal D}_{\cal P} / \, {\cal A}_h
 \longrightarrow 0
\] \noindent
 splits.
 In that case the connection is a module sheaf ${\cal A}_v$ such that
 $\, {\cal D}_{\cal P} \, = \, {\cal A}_h \, \oplus \, {\cal A}_v \,$
 gives the splitting.
\end{Definition}
\begin{Remark}
 Locally always connections exist, but it is not obvious whether they
 can be glued together. In the classical case of principal bundles
 over paracompact manifolds this is possible as one has an appropriate
 partition of unity.
\end{Remark}
\section{Noncommutative Instanton Models}
\label{NCIM}
\subsection{A $q$-deformed Space Time}
In the following we will construct a quantum space ${\cal M}$,
which is a deformation of the sheaf of continuous functions on the $4$-sphere.
This quantum space will be important for the definition of the noncommutative
instanton models and can be regarded as a noncommutative space time
over the classical euclidean background $S^4$.
\newline \noindent
Let us first introduce some notation.
\[
\begin{array}{lcll}
 M & = & S^4 & (\mbox{space time}) \\
 \tilde{M} & = & S^3 \times [-1,1] & (\mbox{enlarged space time}) \\
 NP & = & (0,0,0,0,1) & (\mbox{north pole of } S^4) \\
 SP & = & (0,0,0,0,-1) & (\mbox{south pole of } S^4) \\
 U_1 & = & S^4 \setminus \{ SP \} & (\mbox{northern hemisphere of } S^4) \\
 U_2 & = & S^4 \setminus \{ NP \} & (\mbox{southern hemisphere of } S^4) \\
\end{array}
\] \noindent
Let $ \cal R $ be the locally constant sheaf on $ S^3 $ with objects in the
algebra
$ SU_q(2) $, $ \cal Z $ be the sheaf of continuous, bounded and complex valued
functions
on the interval $ [-1,1] $ such that the condition
\begin{equation}
 f(x) \, = \, 0
\end{equation}
for $\, f \in {\cal Z}(U) \,$, $\, U \subset [-1,1] \,$ open and
$\, x \in \{ -1,1 \} \,$ is fulfilled.

Before we start with the explicit and somewhat technical construction let us
give some
motivation and interpretation. The sheaf $ \cal R $ is being regarded as the
$q$-deformed spatial part of the classical background space time $M = S^4$,
the sheaf $\cal Z$ as the undeformed time part. It is not possible to directly
build a sheaf $\cal M$ out of $\cal R $ and $ \cal Z $.
First the tensor product of $ \cal R $ and $ \cal Z $ is formed which gives
a sheaf $ \tilde{\cal M} $ on the cylinder $ \tilde{M} $.
In the next step $ S^3 \times \{ -1 \} $ (resp.~$ S^3 \times \{ 1 \} $) are
glued
together to the north pole (resp.~south pole) of $ \tilde{M} $.
The gluing is carried over to the local algebras of the sheaf $ \tilde{\cal M}
$,
where it gives the desired $ {\cal M} $.
Altogether one could say that $\cal M$ is $q$-deformed in the horizontal
direction and undeformed in the vertical direction.

Now for the construction of $ \tilde{\cal M} $ take
$ W,W ' \subset S^3 $ and $ V,V ' \subset [-1,1] $ open. Define
\begin{eqnarray}
\begin{array}{lclcl}
\label{EQQSiii}
 \tilde{\cal M}( V \times W ) & := &
 {\cal R}(V) \otimes {\cal Z}(W), & & \\
 \tilde{\cal M} (i^{V \times W}_{V ' \times W '}) & := &
 r^{V \times W}_{V ' \times W '} & := &
 r^V_{V '} \otimes r^W_{W '}.
\end{array}
\end{eqnarray}
As the $V \times W$ form a basis of the topology
of $\tilde{M}$, we get a uniquely defined sheaf $ \tilde{\cal M} $
on $ \tilde{M} = S^3 \times [-1,1] $. Let us now precise what we mean
by "gluing".
Mathematically this is nothing else than the projection
\begin{eqnarray*}
\label{EQQSi}
\begin{array}{lclcl}
 \pi & : & S^3 \times [-1,1] & \longrightarrow & S^4, \\
     &   & \big( (y_1,...,y_4),r \big) & \longmapsto &
     \big( \sqrt{1-r^{2}} \, y_1,..., \sqrt{1-r^{2}} \, y_4, r \big).
\end{array}
\end{eqnarray*}
This projection has an inverse on $ S^3 \times ]-1,1[ $, namely
\begin{eqnarray*}
\begin{array}{lclcl}
 \psi & : & S^4 \setminus \{ NP,SP \} & \longrightarrow &
  S^{3} \, \times \, ]-1,1[, \\
 & & (x_1,...,x_5) & \longmapsto &
 \Big( \frac{1}{\sqrt{1-x_{5}^{2}}} \, (x_1,...,x_4),x_5 \Big).
\end{array}
\end{eqnarray*}
This mirrors the fact that the $4$-sphere without the north and south pole is
topologically equivalent to the cylinder $\, S^3 \, \times \, ]-1,1[ \,$.
Now set
\begin{equation}
\label{EQQSii}
 {\cal M}(U) \: := \: \tilde{\cal M} \big( \pi^{-1}(U) \big) \: = \:
 \tilde{\cal M} \big( \psi(U) \big)
\end{equation}
for $\, U \subset S^4 \setminus \{ NP,SP \} = U_1 \cap U_2 \,$. To find the
right definition
for $\, {\cal M} (U) \,$, if $ NP \in U$ or $ SP \in U$, let us first
have a look at
the sheaf of continuous bounded functions on the sphere $S^4$.
Each such function $f$ defined on the northern hemisphere $U_1$ can uniquely be
written in the
form
\begin{equation}
\label{DST}
 f \, = \, f_{red} + z,
\end{equation}
where $\, f_{red} \in {\cal C}(U) \,$, $\, f(NP) = 0 \,$ and $\, z \in
{\Bbb{C}} \,$.
Furthermore the set of continuous bounded and complex valued functions on $U_1$
whose
value at the north pole vanish can be identified with the (topological) tensor
product
$\, {\cal C}(S^3) \otimes {\cal Z} (]-1,1[) \,$. Therefore the set of complex
continuous
bounded functions on the northern hemisphere $U_1$ is isomorphic to
$\, \big( {\cal C} (S^3) \otimes {\cal Z} ( ] \! -1,1[ ) \big) \oplus {\Bbb{C}}
\,$.
$U_2$ satisfies an analogous result.
The above considerations suggest to set
\begin{eqnarray}
\begin{array}{lclcl}
\label{EQQSiv}
 {\cal M} (U_1) & := &
 \tilde{\cal M} \big( S^3 \, \times \: ]{-1},1] \big)
 \oplus {\Bbb{C}}
 & = &
 \tilde{\cal M} \big( \pi^{-1} (U_1) \big)
 \oplus {\Bbb{C}}, \\
 {\cal M} (U_2) & := &
 \tilde{\cal M} \big( S^3 \times [-1,1[ \big)
 \oplus {\Bbb{C}}
 & = &
 \tilde{\cal M} \big( \pi^{-1} (U_2) \big)
 \oplus {\Bbb{C}}.
\end{array}
\end{eqnarray}
Furthermore we define
\begin{eqnarray}
\label{EQQSv}
\begin{array}{lcl}
 {\cal M} (U) & := & \tilde{\cal M} \big( \pi^{-1} (U) \big)
 \oplus {\Bbb{C}} ,
\end{array}
\end{eqnarray}
where $ U \subset M $ and either $ NP \in U $ or $ SP \in U $.
The restriction morphisms
$\, r^U_V : {\cal M} (U) \longrightarrow {\cal M}(V) \,$, $\, V \subset U
\subset M \,$
are given by the following definition.
\begin{itemize}
\item
 Let $ U,V \subset M \setminus \{ NP,SP \} $ and
 $ f \in {\cal M} (U) = \tilde{\cal M} \big( \psi(U) \big) $. Then define
\begin{equation}
\label{EQQSvi}
 r^U_V (f) \, := \, r^{\psi(U)}_{\psi(V)}(f).
\end{equation}
\item
 Let $U \subset M $ and $ V \subset U $, where  either $ NP \in U $ or
 $ SP \in U $, and $ f \in {\cal M}(U) $.
 With the representation $\, f = f_{red} + z \,$,
 $\, f_{red} \in \tilde{\cal M} \big( \pi^{-1} (U) \big) \,$,
 $\, z \in {\Bbb{C}} \,$ define
\begin{equation}
\label{EQQSvii}
 r^U_V (f) \, := \, r^{\psi(U)}_{\psi(V)}(f_{red}) + z.
\end{equation}
\end{itemize}
For a basis of the topology of $M$ we have defined local algebras
$ {\cal M} (U) $ and restriction morphisms such that
over the basis of the topology the sheaf axioms are satisfied.
Therefore we get a sheaf $\cal M$ over $M$,
which we call the $q$-deformed space time over the background $S^4$.
\begin{Proposition}
\label{PQDSPT}
 The $q$-deformed space time is a quantum space.
\end{Proposition}
\subsection{The $q$-deformed Instanton Models}
The base quantum space of the $q$-deformed instanton models is the
quantum space in proposition \ref{PQDSPT}.
\newline \noindent
Let us introduce some more notation:
\begin{eqnarray*}
\begin{array}{lclcl}
 V_a & = & \{ x \in S^4 \, : \, x_5 > a \} & \mbox{mit} & -1 \leq a < 1, \\
 _b V & = & \{ x \in S^4 \, : \, x_5 < b \} & \mbox{mit} & -1 < b \leq 1.
\end{array}
\end{eqnarray*}
Now we can define the mappings
$\, \tau^1_{\iota,\kappa} : SU_q(2) \longrightarrow {\cal M}
 \big( U_\iota \cap U_\kappa \big) \,$ for $\, \iota,\kappa = 1,2 \,$ by
\bdm
\begin{array}{lcl}
 \tau_{1,1}^1 (h) & = & \epsilon (h) \, 1, \\
 \tau_{2,2}^1 (h) & = & \epsilon (h) \, 1, \\
 \tau_{1,2}^1 (h) & = & h \otimes 1, \\
 \tau_{2,1}^1 (h) & = & S(h) \otimes 1,
\end{array}
\edm
where we used $h \in SU_q (2) $ and the relation
$\, {\cal M}(U_1 \cap U_2) \, = \, SU_q (2) \otimes {\cal Z}( ] -1, 1 [ ) \,$
(see (\ref{EQQSii}) and (\ref{EQQSiii})).
The following lemma is obvious.
\begin{Lemma}
 $\big( \tau^1_{\iota ,{\kappa}} \big)_{1 \leq \iota ,{\kappa} \leq 2}$ is a
 $SU_q (2)$-cocycle in $\cal M$ over
 $\big( U_{\iota} \big)_{{\iota} = 1,2 }$.
\end{Lemma}
\begin{Remark}
 If $\, q \, = \, 1$ $($that means in the commutative case$)$
 we can write for
 $\, h \in SU_1 (2) \, = \, {\cal F} (SU(2)) \,$, $\, x \in U_1 \cap U_2 \,$:
\bdm
\begin{array}{lclcl}
 \tau_{1,1}^1 (h) \, (x) & = & h (1) & = & h \big( \eta_{1,1} (x) \big), \\
 \tau_{2,2}^1 (h) \, (x) & = & h (1) & = & h \big( \eta_{2,2} (x) \big), \\
 \tau_{1,2}^1 (h) \, (x) & = & h \big( \eta_{1,2} (x) \big), & &     \\
 \tau_{2,1}^1 (h) \, (x) & = & h \big( \eta_{2,1} (x) \big) & = &
 h \big( \eta_{1,2}^{-1} (x) \big),
\end{array}
\edm
 where the $\, \eta_{\iota , \kappa} : U_{\iota} \cap U_{\kappa}
 \longrightarrow SU(2) \,$, $ \, {\iota}, {\kappa} = 1,2 \, \,$ are
 classical transition functions given by
\bdm
\begin{array}{lclcl}
 (x_1,...,x_5) & \longmapsto &
 \eta_{1,1} (x_1,...,x_5) & = & 1, \\
 (x_1,...,x_5) & \longmapsto &
 \eta_{2,2} (x_1,...,x_5) & = & 1, \\
 (x_1,...,x_5) & \stackrel{\psi}{\longmapsto} &
 \lefteqn{ (y_1,...,y_5) } \\
 & \longmapsto & \eta_{1,2} (x_1,...,x_5) & = &
 \left(
\begin{array}{cc}
 y_1+iy_2 & -y_3+iy_4 \\
 y_3+iy_4 & y_1-iy_2
\end{array}
\right),  \\
 (x_1,...,x_5) & \longmapsto &
 \eta_{2,1} (x_1,...,x_5) & = & \eta_{1,2}^{-1} (x_1,...,x_5).
\end{array}
\edm
\end{Remark}
In the following we need $SU_q (2)$-actions on the quantum space
${\cal M} (U_{\iota}) $, ${\iota} = 1,2$.
The quantum group $SU_q (2)$ acts trivially on ${\cal M}(U_2)$:
\bdm
\begin{array}{lclcll}
 \big( \alpha_2^1 \big)_U & : & SU_q (2) \times {\cal M} (U) &
 \longrightarrow & {\cal M} (U), & \hspace{0.5cm} U \subset U_2
 \mbox{ open} \\
 & & (h,f) & \longmapsto & h \cdot_2 f \, = \, \epsilon (h) \, f . &
\end{array}
\edm
The family $\, \big( \big( \alpha_2^1 \big)_U \big)_{U \subset U_2}\,$
gives a sheaf morphism
$\, \alpha_2^1 : SU_q (2) \times {\cal M} \! \mid_{U_2}
\longrightarrow {\cal M} \! \mid_{U_2} \,$. Now define functions
\bdm
\begin{array}{lclcll}
 \big( \alpha_1^1 \big)_U & : &  SU_q (2) \times {\cal M}
 (U) & \longrightarrow & {\cal M} (U), & \hspace{0.5cm} U \subset U_1
 \mbox{ open} \\
 & & (h,f) & \longmapsto & h \cdot_1 f &
\end{array}
\edm
by the following procedure:
\begin{itemize}
\item
 If $U \, = \, \pi ( \, V \times ] a,b [ \, ) $ with $V \subset S^3$ open,
 $-1 \leq a < b \leq 1$, equations (\ref{EQQSiii}) and (\ref{EQQSii})
 proof $\, {\cal M} (U) \, = \, SU_q (2) \otimes {\cal Z} ( ] a,b [ ) \,$.
 In this case define for $ h,f  \in SU_q (2) $,
 $r \in {\cal Z} ( ] a,b [ )$:
\begin{eqnarray}
\lefteqn{
 h \cdot_1 (f \otimes r) } \nonumber \\
 & = & \big( \alpha ^1_1 \big)_U (h, f \otimes r) \nonumber \\
 & = & \sum \limits_{(h)} \, h_{(1)} \, f \, \big( S \, h_{(2)} \big) \otimes
r.
\end{eqnarray}
\item
 If $ U \, = \, V_c$ with $c > -1$, equation (\ref{EQQSv}) implies
 $\, {\cal M} (U) \, = \,  SU_q (2) \otimes {\cal Z}
 ( ] c,1 [ ) \oplus {\Bbb{C}} \,$.
 In that case define for $ h,f \in SU_q (2) $,
 $ r \in {\cal Z} ( ] c,1 [ )$, $ z \in {\Bbb{C}}$:
\begin{eqnarray}
\lefteqn{
 h \cdot_1 (f \otimes r + z) } \nonumber \\
 & = & \big( \alpha ^1_1 \big)_U (h, f \otimes r + z)  \nonumber \\
 & = & \sum \limits_{(h)} \, h_{(1)} \, f \, \big( S \, h_{(2)} \big) \otimes
 r + z.
\end{eqnarray}
\end{itemize}
The mappings $\big( \alpha_1^1 \big)_U$ are $ SU_q (2)$-actions on
${\cal M}(U)$. Furthermore the diagram
\begin{center}
\setsqparms[1`1`1`1;1000`500]
\square[SU_q(2) \otimes {\cal M}(U)`{\cal M}(U)`SU_q(2) \otimes {\cal
M}(\tilde{U})`{\cal M}(\tilde{U});
 \big( \alpha^1_1 \big)_U`id \otimes i^U_{\tilde{U}}`i^U_{\tilde{U}}`\big(
\alpha^1_1 \big)_U]
\end{center}
commutes for all $\tilde{U} \subset U$ open with
$\, U, \tilde{U} \in {\cal B}$ and
\bdm
 \lefteqn{ {\cal B} \: = \: \{ \pi ( \, V \times ] a,b [  \, ), \,
 V_c \in Pot (M) : } \\
 & & V \subset S^3
 \mbox{ open }, \; -1 \leq a < b \leq 1, \; c > -1 \, \}.
\edm
As ${\cal B}$ is a basis of the open sets in $U$, we get a sheaf morphism
$\, \alpha_1^1 : SU_q (2) \times {\cal M} \! \mid_{U_1}
\longrightarrow {\cal M} \! \mid_{U_1} \,$,
whose components are actions.

Now we can construct the smash products
$\, {\cal M} \! \mid_{U_1} \#_1 SU_q (2) \,$ and
$\, {\cal M} \! \mid_{U_2} \#_2 SU_q (2) \,$ with the actions
$\, \alpha_1^1 \,$ and $\, \alpha_2^1 \,$.
Next we will show that the noncommutative coordinate changes
\bdm
\lefteqn{ \Omega_{{\kappa},\iota }^1 \: = \: (m\otimes id) \circ
 (id \otimes \tau_{\iota ,{\kappa}}^1 \otimes id) \circ (id \otimes \Delta): }
\\
 & & {\cal M} (U_1 \cap U_2 ) \#_{\iota} SU_q  (2) \: \longrightarrow \:
 {\cal M} (U_1 \cap U_2 ) \#_{\kappa} SU_q  (2)
\edm
are morphisms of algebras. Let $\iota = 1$, $\, {\kappa} = 2$.
Then $\Omega_{{\kappa},\iota}^1$ has the form
\begin{equation}
 \Omega_{{\kappa},\iota}^1 ( (f \otimes r) \# h ) \: = \:
 \sum \limits_{(h)} \,
 \big( f \cdot h_{(1)} \otimes r \big) \# h_{(2)},
\end{equation}
where $h,f \in SU_q (2) $ and $r \in {\cal Z} ( ] -1,1[ ) $.
But that is exactly the form the morphism
$\Phi$ in theorem \ref{BCMii} has, if we let $u_1, \, u_2 \in
Hom \big( SU_q(2),{\cal M}(U) \big) $ be defined by:
\bdm
\begin{array}{lcll}
 u_1(h) & = & h \otimes 1, & \hspace{1cm} h \in SU_q (2) \\
 u_2(h) & = & \epsilon (h) 1, & \hspace{1cm} h \in SU_q (2).
\end{array}
\edm

The quantum space $\cal M$, the $SU_q (2)$-cocycle
$\big( \tau^1_{\iota ,{\kappa}} \big)_{1 \leq \iota ,{\kappa} \leq 2}$ and
the actions
$ \, \alpha_{\iota} $, $\, \iota = 1,2 \,$ fulfill the conditions
of theorem \ref{CQPBTRF}. Therefore we get a unique \QPB
\bdm
 \big( {\cal P}^1, {\cal M} , {\varrho}^1 , SU_q (2) , (U_{\iota})_{\iota
 \in \{ 1,2 \} } \big)
\edm
with transition functions
$\big( \tau^1_{\iota ,{\kappa}} \big)_{1 \leq \iota ,{\kappa} \leq 2}$ and
call it the instanton model for the index $k = 1$.
In the case $q = 1$ we get the undeformed instanton model with index $1$.

The instanton model for the index $k = 0$ is the trivial one:
\bdm
 \big( {\cal P}^0, {\cal M}, M, {\varrho}^0 , SU_q (2) \big),
\edm
that means $\, {\cal P}^0 \, = \, {\cal M} \otimes SU_q (2) \,$ and
\bdm
\begin{array}{lclcl}
 \varrho^0 & : & {\cal M} & \longrightarrow & {\cal P}^0, \\
 & & f & \longmapsto & \varrho^0 (f) \, = \, f \otimes 1, \quad f \in {\cal M}
(U),
 \quad U \subset M .
\end{array}
\edm
Let us subsume the results in a theorem.
\begin{Theorem}
 The $q$-deformed instanton models
\bdm
 \big( {\cal P}^k, {\cal M}, M, {\varrho}^k , SU_q (2) \big)
\edm
 with index $k = 0,1$ are noncommutative quantum principal bundles,
 which turn into the classical ones for $q = 1 $.
\end{Theorem}
But there exist $q$-deformed instanton models for all $k \in {\Bbb{Z}}$.
To see that consider the mappings:
\bdm
 u_1^k, \, u_2^k \, \in Hom \big( SU_q (2) , {\cal M} (U_1 \cap U_2 ) \big),
 \quad k \in {\Bbb{Z}},
\edm
which are defined for all $n \in {\Bbb{N}}$ by
\bdm
\begin{array}{lclcl}
 u_1^n (h) & = & \sum \limits_{(h)} \, h_{(1)} \cdot ... \cdot h_{(n)}
 \otimes 1, & \quad & h \in SU_q (2) \\
 u_2^n (h) & = & \epsilon (h) \, 1 \otimes 1 , & \quad & h \in SU_q (2) \\
 u_1^{-n} (h) & = & u_2^n  (h) , & \quad & h \in SU_q (2) \\
 u_2^{-n} (h) & = & u_1^n \circ S \, (h) & \quad & h \in SU_q (2).
\end{array}
\edm
The for all $ k \in {\Bbb{Z}}$:
\begin{equation}
 u_1^k * u_2^{-k} \: = \: 1 \: = \: u_2^k * u_1^{-k}.
\end{equation}
Let us construct actions
$\, \alpha_{\iota}^k : SU_q (2) \times {\cal M}  \! \mid_{U_{\iota}}
\longrightarrow {\cal M}  \! \mid_{U_{\iota}} \,$
with the $u_{\iota}^k$ and the following recipe:
\begin{itemize}
\item
 If $ U \, = \, V_c$, $c > -1 $, ${\cal M}(U)$ has the form
 $ SU_q (2) \otimes {\cal Z} ( ]c,1]) \oplus {\Bbb{C}} $ and we define for
 $ h,f  \in SU_q (2) $,
 $r \in {\cal Z} ( ] c,1 ] )$ and $ z \in {\Bbb{C}}$:
\begin{eqnarray}
\lefteqn{
 h \cdot_1 (f \otimes r + z) } \nonumber \\
 & = & \big( \alpha^k_1 \big)_U (h, f \otimes r + z) \nonumber \\
 & = & \sum \limits_{(h)} \, u_1^k \big( h_{(1)} \big)
 \, ( f \otimes r ) \, u_2^{-k} \big( h_{(2)} \big) +z.
\end{eqnarray}
\item
 If $U \, = \, _d V$, $d < 1$, ${\cal M}(U)$ has the form
 $ SU_q (2) \otimes {\cal Z} ( [-1,d[ ) \oplus {\Bbb{C}} $
 and we define for $ h,f  \in SU_q (2) $,
 $r \in {\cal Z} ( ] c,1 ] )$ and $ z \in {\Bbb{C}}$:
\begin{eqnarray}
\lefteqn{
 h \cdot_2 (f \otimes r + z) } \nonumber \\
 & = & \big( \alpha^k_2 \big)_U (h, f \otimes r + z) \nonumber \\
 & = & \sum \limits_{(h)} \, u_2^k \big( h_{(1)} \big)
 \, ( f \otimes r ) \, u_1^{-k} \big( h_{(2)} \big) +z.
\end{eqnarray}
\item
 But if $U \, = \, \pi ( \, V \times ] a,b [ \, ) $ with
 $V \subset S^3$ open,
 $-1 \leq a < b \leq 1$, ${\cal M}(U)$ has the form
 $\, SU_q (2) \otimes {\cal Z} ( ] a,b [ ) \,$ and we set for
 $ h,f  \in SU_q (2) $ and
 $r \in {\cal Z} ( ] a,b [ )$:
\begin{eqnarray}
\lefteqn{
 h \cdot_1 (f \otimes r) } \nonumber \\
 & = & \big( \alpha ^k_1 \big)_U (h, f \otimes r) \nonumber \\
 & = & \sum \limits_{(h)} \, u_1^k \big( h_{(1)} \big) \, ( f \otimes r ) \,
 u_2^{-k} \big( h_{(2)} \big) ,
\end{eqnarray}
\begin{eqnarray}
\lefteqn{
 h \cdot_2 (f \otimes r) } \nonumber \\
 & = & \big( \alpha ^k_2 \big)_U (h, f \otimes r) \nonumber \\
 & = & \sum \limits_{(h)} \, u_2^k \big( h_{(1)} \big) \, ( f \otimes r ) \,
 u_1^{-k} \big( h_{(2)} \big) .
\end{eqnarray}
\end{itemize}
Altogether this provides sheaf morphisms
\bdm
\begin{array}{lclcl}
 \alpha_1^k & : & SU_q (2) \times {\cal M} \! \mid_{U_1} & \longrightarrow &
 {\cal M} \! \mid_{U_1}, \\
 \alpha_2^k & : & SU_q (2) \times {\cal M} \! \mid_{U_2} & \longrightarrow &
 {\cal M} \! \mid_{U_2},
\end{array}
\edm
whose components $\big( \alpha_{\iota}^k \big)_U$, $\, U \subset U_{\iota} $,
${\iota} = 1,2$ are weak actions by theorem \ref{BCMi}. Furthermore
theorem \ref{BCMi}
gives normal cocycles $\, \iota: SU_q(2) \times SU_q(2) \longrightarrow
{\cal M}(U_{\iota}) \,$ and crossed products $\, {\cal M} (U_{\iota})
\#_{\iota} SU_q(2) \,$.

Finally we have to find transition functions which lead to quantum principal
bundles. Theorem \ref{BCMii} in the appendix gives a hint. If we define
$\, \tau_{\iota, \kappa}^k : SU_q (2) \longrightarrow {\cal M}
(U_{\iota} \cap U_{\kappa} ) \,$ for $\iota, \kappa = 1,2$ by
\bdm
\begin{array}{lclcl}
 \tau_{1,1}^k (h) & = & \epsilon (h) 1, & \quad & h \in SU_q (2), \\
 \tau_{2,2}^k (h) & = & \epsilon (h) 1, & \quad & h \in SU_q (2), \\
 \tau_{1,2}^k (h) & = & \sum \limits_{(h)} \, u_1^k \big(  h_{(1)} \big) \,
 u_1^{-k} \big( h_{(2)} \big) , & \quad & h \in SU_q (2), \\
 \tau_{2,1}^k (h) & = & \sum \limits_{(h)} \, u_2^k \big(  h_{(1)} \big) \,
 u_2^{-k} \big( h_{(2)} \big) , & \quad & h \in SU_q (2),
\end{array}
\edm
theorem \ref{BCMii} shows the mappings
\bdm
 \lefteqn{ \Omega_{{\kappa},\iota }^k \: = \: (m\otimes id) \circ
 (id \otimes \tau_{\iota ,{\kappa}}^k \otimes id) \circ (id \otimes \Delta): }
\\
 & & {\cal M} (U_1 \cap U_2 ) \#_{\iota} SU_q  (2) \: \longrightarrow \:
 {\cal M} (U_1 \cap U_2 ) \#_{\kappa} SU_q  (2)
\edm
being homomorphisms. Now by \ref{CQPBTRF} the following is true.
\begin{Theorem}
 For every index $k \in {\Bbb{Z}}$ there exists a quantum principal bundle
 $ {\cal P}^k $ over the $q$-deformed space time $\cal M$
 and with structure quantum group $SU_q(2)$ such that the above defined
 $\tau_{\iota, \kappa}^k$ are its transition functions.
 This quantum principal bundle is the $q$-deformed instanton model
 for the index $k$.
 If $q = 1 $ the $q$-deformed instanton model turns into the
 $SU(2)$-principal fibre bundle with index $k$.
\end{Theorem}
\appendix
\section{Sheaf Theory}
\label{SHTH}
Sheaves provide the natural mathematical language to switch from the local to
the global
and vice versa. Only the definition of sheaves and their morphisms
are given. For further details see the literature, for example Tennison
\cite{ten-1}
or Mac Lane, Moerdijk \cite{mam-1}.

Every topological space $M$ gives rise to the category ${\cal T}_M$
of open sets in $M$ Its morphisms are the inclusion maps
$\, i_V^U  :  V  \longrightarrow  U, \: u  \longmapsto  u \,$,
where $V \subset U \subset M$ open.
Furthermore a continuous mapping $\, F : M \longrightarrow N \,$
between topological spaces defines a contravariant functor
$\, f^{-1} : {\cal T}_N \longrightarrow {\cal T}_M \, $ by
\bdm
\begin{array}{lclrr}
 U & \longmapsto & f^{-1} (U), & U \in
 Obj \big( {\cal T}_N \big), & \\
 i^U_V & \longmapsto & i^{f^{-1} (U)}_{f^{-1} (V)}, &
 \quad V,U \in Obj \big( {\cal T}_N \big), & V \subset U.
\end{array}
\edm
\begin{Definition}
\label{DSH}
 Let $M$ be a topological space, and $\cal K $ a subcategory of the category of
 sets. Then a contravariant functor
\bdm
 {\cal G} : {\cal T}_M \longrightarrow {\cal K}
\edm
 is called a presheaf of $\cal K$-objects over $M$.
 The elements of the set ${\cal G}(U)$ with $U \subset M$ open are the
 sections of $\cal G$ over $U$.
 $\cal G$ is called a sheaf of $\cal K$-objects over $M$, if
 the following conditions are satisfied for each open covering
 $\big( U_{\iota} \big)_{\iota \in I }$ of an open set $U \subset M$:
\begin{enumerate}
\item
 If $s,s' \in {\cal G}(U) $ and
\begin{equation}
 {\cal G}^U_{U_{\iota}} \, (s) \, = \,  {\cal G}^U_{U_{\iota}} \, (s') ,
\end{equation} \noindent
 for all $ \iota \in I $, then $s = s'$.

\item
 Let $\big( s_{\iota} \big)_{\iota \in I } $ be a family of sections
 $s_{\iota} \in {\cal G} \big( U_{\iota} \big)$, such that for
 all $\iota, \kappa \in I $
\begin{equation}
 {\cal G}^{U_{\iota}}_{U_{\iota} \cap U_{\kappa}} \, \big( s_{\iota} \big)
 \, = \,
 {\cal G}^{U_{\kappa}}_{U_{\iota} \cap U_{\kappa}} \, \big( s_{\kappa} \big).
\end{equation} \noindent
 Then there exists a uniquely defined $s \in {\cal G} (U) $, which fulfills
 the relation
\begin{displaymath}
 {\cal G}^U_{U_{\iota}} \, (s) \, = \, s_{\iota}
\end{displaymath} \noindent
 for all $\iota \in I$.
\end{enumerate}
\end{Definition}
The language of sheaves provides for an abstract characterization
of local function algebras. Important examples are given by the sheaf
${\cal C}_M$ (resp.~${\cal C}^{\infty}_M, {\cal C}^{\omega}_M, {\cal O}_M$)
of continuous (resp.~differentiable, analytical, holomorphic) functions
on a topological space (resp.~differentiable, real analytic, complex manifold)
$M$.
\begin{Definition}
 Let $\, f : M \longrightarrow N \,$  be a continuous mapping
 between topological spaces, and let
\bdm
\begin{array}{lclcl}
 {\cal G} & : & {\cal T}_M & \longrightarrow & {\cal K} \\
 {\cal G'} & : & {\cal T}_N & \longrightarrow & {\cal K}
\end{array}
\edm
 be sheaves over  $M$ resp.~$N$ with objects in the category ${\cal K}$.
 A morphism of sheaves from ${\cal G}$ to ${\cal G}'$ over $f$
 is given by a morphism of functors
\bdm
 {\cal F} : {\cal G'} \longrightarrow  {\cal G} \circ f^{-1}.
\edm
\end{Definition}
If $f : M \longrightarrow N $ is a continuous mapping, the pull back
$\, f^{*} : {\cal C}_N \longrightarrow  {\cal C}_M \,$ gives an example
of a morphism of sheaves.
\section{Crossed Products and Smash Products}
\label{CRSMPR}
Crossed products and smash products serve to define a multiplication
on the tensor product of a Hopf algebra and an algebra.
Most of the definitions and theorems are given
according to Blattner, Cohen and Montgomery \cite{bcm-1}.
All algebras are supposed to have a unit.
\begin{Definition}
\label{WEAC}
 Let $H$ be a Hopf algebra and $A$ an algebra over the field k.
 A weak $H$-action on $A$ is a bilinear mapping
\bdm
\begin{array}{lcl}
 H \times A & \longrightarrow & A \\
 (h,a) & \longmapsto & h \cdot a ,
\end{array}
\edm
such that for all $h \in H$ and $a,b \in A$:
\begin{enumerate}
\item
 $ \quad h \cdot ab \, = \, \sum \limits_{(h)} \,
 \big( h_{(1)} \cdot a \big) \,
 \big( h_{(2)} \cdot b \big) $,
\item
 $ \quad h \cdot 1 \, = \, \epsilon (h) \cdot 1 $,
\item
 $ \quad 1 \cdot a \, = \, a $.
\end{enumerate}
 An $H$-action on $A$ is given by a weak action fulfilling the condition
\begin{enumerate}
\setcounter{enumi}{3}
\item
 $ \quad h \cdot ( l \cdot a ) \, = \, (hl) \cdot a $
\end{enumerate}
 for all $h,l \in H$ and $a \in A$.
\end{Definition}
\begin{Theorem}
\label{CRPR}
 Let the Hopf algebra $H$ weakly coact on the algebra $A$.
 Furthermore let $\, \sigma: H \times H \longrightarrow A \,$ be a mapping
 which satisfies:
\begin{enumerate}
\item
 $($normality condition$)$
\bdm
 \lefteqn{
 \sigma (1,h) \, = \, \sigma (h,1) \, = \, \epsilon (h) \quad
 \mbox{for all} \quad h \in H.}
\edm
\item
 $($cocycle condition$)$
\bdm
 \lefteqn{
 \sum \limits_{(h),(l),(m)} \, \big( h_{(1)} \cdot \sigma
 \big( l_{(1)}, m_{(1)} \big) \big) \, \sigma
 \big( h_{(2)} , l_{(2)} m_{(2)} \big) } \\
 & = & \sum  \limits_{(h),(l)} \, \sigma \big( h_{(1)} , l_{(1)} \big)
 \sigma \big( h_{(2)} l_{(2)} , m \big)
 \quad \mbox{for all} \quad h,l,m \in H.
\edm
\item
 $($twisted module condition$)$
\bdm
 \lefteqn{
 \sum \limits_{(h),(l)} \, \big( h_{(1)} \cdot \big( l_{(1)} \cdot a \big)
\big) \,
 \sigma \big( h_{(2)} , l_{(2)} \big) } \\
 & = & \sum \limits_{(h),(l)} \, \sigma \big( h_{(1)} , l_{(1)} \big)
 \, \big( h_{(2)} l_{(2)} \cdot a \big)
 \quad \mbox{for all} \quad h,l \in H, \quad a \in A.
\edm
\end{enumerate}
 Then the equation
\begin{equation}
 (a \otimes h) (b \otimes l) \, = \,
 \sum \limits_{(h),(l)} \, a \big( h_{(1)} \cdot b \big)
 \sigma \big( h_{(2)} , l_{(1)} \big) \otimes
 h_{(3)} l_{(2)}
\end{equation}
 gives a multiplication on $A \otimes H$, which defines an algebra
 $A \#_{\sigma} H$ called a crossed product with unity $1 \otimes 1$.
 The elements $ a \otimes h $ of $A \#_{\sigma} H$ are sometimes written
 in the form $a \# h$.
\end{Theorem}
\bpro
 see Blattner et al.~\cite{bcm-1}, Corollary 4.6.
\epro
\begin{Example}
\label{SMPR}
 Let the Hopf algebra $H$ act on the algebra $A$.
 The bilinear mapping
 $\, \sigma : H \times H \longrightarrow A \,$
 shall be trivial, which means that for all $h,l \in H$
\bdm
 \sigma (h,l) \, = \, \epsilon (h) \epsilon (l) 1.
\edm
 Then the assumptions of theorem $\ref{CRPR}$ are satisfied and the
 crossed product $A \#_{\sigma} H$ is defined. We call it the smash product
 of $H$ and $A$ and simply write $A \# H $.
 The multiplication on the smashed product is given by
\begin{equation}
 (a \# h) (b \# l) \, = \, \sum \limits_{(h)} \, a \, \big( h_{(1)} \cdot b
\big),
 \# h_{(2)} l, \quad (a \# h), (b \# l) \in A \# H.
\end{equation}
\end{Example}
\begin{Theorem}
\label{BCMi}
 Let $H$ weakly act on $A$. Suppose the weak action is defined by an element
 of the convolutuion algebra $Hom (H,A)$, i.~e.~there exists an invertible
 element $u \in Hom (H,A)$ with $ u(1) = 1$ such that for
 all $h \in H$, $a \in A$
\begin{equation}
 h \cdot a \, = \, \sum \limits_{(h)} \, u \big( h_{(1)} \big) \, a \,
 u^{-1} \big( h_{(2)} \big).
\end{equation}
 Define the bilinear mapping $\, \sigma : H \times H \longrightarrow A \,$
 by
\begin{equation}
 \sigma (h,l) \, = \,
 \sum \limits_{(h),(l)} \, u \big( h_{(1)} \big) u \big( l_{(1)} \big)
 u^{-1} \big( h_{(2)} l_{(2)} \big).
\end{equation}
 Then $ \sigma $ is a normal cocycle fulfilling the twisted module condition.
 Therefore the assumptions of theorem $\ref{CRPR}$ are satisfied and the
 crossed product $A {\#}_{\sigma} H$ exists.
\end{Theorem}
\bpro
 see Blattner et al.~\cite{bcm-1}, Example 4.11.
\epro
\begin{Theorem}
\label{BCMii}
 Let $u_{1}, u_{2} \in Hom(H,A)$ be invertible and $u_{1} (1) = u_{2} (1) = 1$.
 For each $i \in {1,2} $ define the weak $H$-action
\bdm
\begin{array}{lclcl}
 \alpha_{i} & : & H \times A & \longrightarrow & A \\
 & & (h,a) & \longmapsto & h \cdot a \, = \, \sum \limits_{(h)} \,
 u_i \big( h_{(1)} \big) \, a \, u_{i}^{-1} \big( h_{(2)} \big).
\end{array}
\edm
 Further construct the bilinear mappings
\bdm
\begin{array}{lclcl}
 \sigma_i & : & H \times H & \longrightarrow & A \\
 & & (h,l) & \longmapsto &
 \sigma_i (h,l) \, = \,
 \sum \limits_{(h),(l)} \, u_i \big( h_{(1)} \big) u_i \big( l_{(1)} \big)
 u_i^{-1} \big( h_{(2)} l_{(2)} \big)
\end{array}
\edm
 and the crossed products $A \#_{\sigma_1} H$ and $A \#_{\sigma_2} H$
 according to theorem $\ref{BCMi}$.
 Then the linear mapping
\bdm
\begin{array}{lclcl}
 \Phi & : & A \#_{\sigma_1} H & \longrightarrow & A \#_{\sigma_2} H \\
 & & (a \# h) & \longmapsto & \sum \limits_{(h)} \,
 a \, u_1 \big( h_{(1)} \big) \, u^{-1}_2 \big( h_{(2)} \big) \# h_{(3)}
\end{array}
\edm
 is a morphism af algebras with unity.
\end{Theorem}
\bpro
 see proposition 1.19, Theorem 5.3 and Corollary 5.4 in
 Blattner et al.~\cite{bcm-1}
\epro
Suppose we are given a Hopf algebra $H$ acting on an algebra $B$ and a
$H$-comodule algebra $A$ with coaction
$\, \phi: A \longrightarrow H \otimes A \,$.
Then define a bilinear mapping
bilineare Abbildung
\bdm
\begin{array}{lclcl}
 m & : & (B \otimes A) \times (B \otimes A) & \longrightarrow & (B \otimes A)
\\
 & & ((f \otimes g),(f' \otimes g')) & \longmapsto &
 (f \otimes g) \cdot (f' \otimes g') \\
 & & & & = \: \sum \limits_{(g)}
 \, f \, \big( g_{(-1)} \cdot f' \big) \otimes g_{(0)} \, g'
\end{array}
\edm
on the tensor product $B \otimes A$. An easy calculation shows $m$ being
associativ.
Additionally we have for $(f \otimes g), (f'\otimes g') \in (B \otimes A) $
\begin{eqnarray}
\begin{array}{lcl}
 (1 \otimes 1) \cdot (f \otimes g) & = & (f \otimes g) \\
 (f'\otimes g') \cdot (1 \otimes 1) & = & (f' \otimes g').
\end{array}
\end{eqnarray}
This proofs the first part of the following theorem.
\begin{Theorem}
\label{SMPRKA}
 Suppose the Hopf algebra $H$ acts on the algebra $B$ and $A$ is an
$H$-comodule algebra.
 Then the mapping
\bdm
\begin{array}{lclcl}
 m & : & (B \otimes A) \times (B \otimes A) & \longrightarrow & (B \otimes A)
\\
 & & ((f \otimes g),(f' \otimes g')) & \longmapsto &
 (f \otimes g) \cdot (f' \otimes g') \\
 & & & & = \: \sum \limits_{(g)}
 \, f \, \big( g_{(-1)} \cdot f' \big) \otimes g_{(0)} \, g'
\end{array}
\edm
turns the vector space $B \otimes A$ into an algebra with unit $\, 1 \otimes 1
$.
This algebra is called the smash product of $B$ and $A$ and will be notated by
$B \# A$.
If $H$ acts only weakly on $B$ and
$\, \sigma : H \times H \longrightarrow {\cal M} \big( U_{\iota} \big) \,$
is a normal cocycle fulfilling the twisted module condition, the term
\bdm
\begin{array}{lclcl}
 m & : & (B \otimes A) \times (B \otimes A) & \longrightarrow & (B \otimes A)
\\
 & & ((f \otimes g),(f' \otimes g')) & \longmapsto &
 (f \otimes g) \cdot (f' \otimes g') \\
\lefteqn{ \hspace{5,0cm} = \: \sum \limits_{(g),(g')}
 \, f \, \big( g_{(-2)} \cdot f' \big) \,
 \sigma \big( g_{(-1)} , g'_{(-1)} \big) \otimes g_{(0)} \, g'_{(0)} }
\end{array}
\edm
 defines an algebra structure on $B \otimes A$ with unity $\, 1 \otimes 1 $.
 The algebra defined in this way is called a crossed product and will be
written
 $B \#_{\sigma} A$.
\end{Theorem}
\bpro
The first part has been shown above, the second one can be proven by
an analog argument.
\epro
\mbox{ } \\
{\bf Note added in proof}:
After having written this paper I received the preprint \cite{brm-1} which is
concerned with a similar matter.
\section*{Acknowledgement}
I would like to thank J.~Wess for introducing me to this interesting topic
and M.~Schottenloher for helpful discussions. I also would like to thank the
Konrad Adenauer Foundation for financial support.

\begin{thebibliography}{99}
%
%
\bibitem{abe-1} E.~Abe: {\em Hopf Algebras}, Cambridge Tracts in Math.~N 74,
                (Cambridge, New York: Cambridge University Press 1980)
%
\bibitem{bcm-1} R.~J.~Blattner, M.~Cohen, S.~Montgomery: {\em Crossed products
and inner actions of Hopf algebras},
                Trans.~Amer.~Math.~Soc. {\bf 298} (1986), 671-711
%
\bibitem{blm-1} R.~J.~Blattner, S.~Montgomery: {\em Crossed Product and Galois
Extensions of Hopf Algebras},
                Pacific Journal of Math.~ Vol.~{\bf 137}, No.~1 (1989), 37-54
%
\bibitem{brm-1} T.~Brzezinski, S.~Majid: {\em Quantum group gauge theory on
quantum spaces},
                preprint, to appear in Comm.~Math.~Physics
%
\bibitem{con-1} A.~Connes: {\em Non-Commutative Differential Geometry},
                Publ.~Math.~IHES {\bf 62} (1985)
%
\bibitem{dav-1} M.~Daniel, C.~M.~Viallet: {\em The geometrical setting of gauge
theories of the Yang-Mills type},
                Rev.~Mod.~Phys.~{\bf 52}, No.~1 (1980)
%
\bibitem{dri-1} V.~Drinfeld: {\em Quantum Groups}, Proc.~of the ICM 1986,
                Berkeley (1986), 798-820
%
\bibitem{ghv-1} W.~Greub, S.~Halperin, R.~Vanstone: {\em Connections, Curvature
and Cohomology, Volume} I
                (New York, London: Academic Press 1972)
%
\bibitem{ghv-2} W.~Greub, S.~Halperin, R.~Vanstone: {\em Connections, Curvature
and Cohomology, Volume} II
                (New York, London: Academic Press 1973)
%
\bibitem{gro-1} A.~Grothendieck: {\em A General Theory of Fibre Spaces with
                Structure Sheaf}, University of Kansas, Department of
Mathematics,
                Report No.~{\bf 4} (Lawrence, Kansas $1965^{2}$)
%
\bibitem{haa-1} R.~Haag: {\em Local Quantum Physics}, Texts and Monographs in
Physics
                (Berlin, Heidelberg, New York: Springer 1992)
%
\bibitem{har-1} R.~Hartshorne: {\em Algebraic Geometry}, Graduate Texts in
Mathematics
                (Berlin, Heidelberg, New York: Springer 1977)
%
\bibitem{itz-1} C.~Itzykson, J.~B.~Zuber: {\em Quantum Field Theory}
                (McGraw-Hill 1980)
%
\bibitem{kas-1} U.~Kasper: {\em Fibre Bundles: An Introduction to Concepts of
Modern Differential Geometry},
                in: {\em Geometry and Theoretical Physics}, eds. J.~Debrus,
A.~C.~Hirshfeld (Berlin, Heidelberg, New York: Springer 1991)
%
\bibitem{kak-1} L.~Kaup, R.~Kaup: {\em Holomorphic Functions of Several
Variables},
                de Gruyter Studies in Mathematics (Berlin, New York 1983)
%
\bibitem{mam-1} S.~Mac Lane, I.~Moerdijk: {\em Sheaves in Geometry and Logic}
Springer-Verlag (Berlin, Heidelberg New York 1992)
%
\bibitem{man-1} Y.~I.~Manin: {\em Quantum Groups and Non-Commutative Geometry},
                Montreal Lecture Notes (1988)
%
\bibitem{man-2} Y.~I.~Manin: {\em Gauge Field Theory and Complex Geometry},
Grundlehren der mathematischen Wissenschaften 289,
                (Berlin, Heidelberg, New York: Springer 1988)
%
\bibitem{pfl-1} M.~J.~Pflaum: {\em Quantengruppen auf Faserb\"undeln},
Diplomarbeit an der Sektion Physik der Universit\"at M\"unchen
                (M\"unchen, 1.~November 1992)
%
\bibitem{stw-1} R.~F.~Streater, A.~S.~Wightman: {\em PCT, Spin and Statistics
and all that}
                (New York, Amsterdam: 1964)
%
\bibitem{ten-1} B.~R.~Tennison: {\em Sheaf Theory}, Cambridge University Press,
                London Math.~Soc.~Lecture Note Series {\bf 20} (Cambridge 1975)
%
\bibitem{web-1} J.~Wess, J.~Bagger: {\em Supersymmetry and supergravity}
                (Princeton, N.~J.: University Press 1983)
%
\bibitem{wez-1} J.~Wess, B.~Zumino: {\em Covariant Differential Calculus on the
                Quantum Hyperplane}, preprint CERN-TH-5697/90 (April 1990)
%
\bibitem{wor-1} S.~L.~Woronowicz: {\em Differential Calculus on Compact Matrix
Pseudogroups},
                Commun.~Math.~Phys.~{\bf 122}, 125-170 (1989)
%
\end{thebibliography}
\end{document}